\documentclass[article,twocolumn,aps,superscriptaddress,longbibliography,floatfix]{revtex4-1}

\usepackage{bm}%
\usepackage{float}
\usepackage{stackengine}
\expandafter\ifx\csname package@font\endcsname\relax\else
 \expandafter\expandafter
 \expandafter\usepackage
 \expandafter\expandafter
 \expandafter{\csname package@font\endcsname}%
\fi

\usepackage{array}
\usepackage{amsmath,amssymb}
\usepackage{MnSymbol}
\usepackage{tikz-feynman,contour} 
\usetikzlibrary{shapes,arrows,positioning,automata,backgrounds,calc,er,patterns}
\usepackage{feynmf}
\tikzfeynmanset{compat=1.0.0} 
\usepackage{mdframed}

\usepackage[export]{adjustbox}
\usepackage{graphicx}
\usepackage[caption=false]{subfig}
\usepackage{mathrsfs}
\usepackage{bm}
\usepackage{mathtools}
\usepackage{epstopdf}
\usepackage{hyperref}
\usepackage{xurl}

\hypersetup{%
   pdfpagemode=None, 
   pdfstartpage=1,
   pdfmenubar=true,
   pdftoolbar=true,
   colorlinks = true,
   linkcolor=blue,
   citecolor=blue,
   urlcolor=blue,
   bookmarksopen=false,
   breaklinks=true
 }
 \usepackage{amsmath}
\usepackage{amsfonts}
\usepackage{mathtools}
\usepackage{graphicx}
\usepackage{fixme}
\usepackage{color}

\usepackage[normalem]{ulem}

\begin{document} 

\title{Microscopic theory of polariton-polariton interactions}




\author{E. R. Christensen}
\affiliation{Center for Complex Quantum Systems, Department of Physics and Astronomy, Aarhus University, Ny Munkegade 120, DK-8000 Aarhus C, Denmark.}
\author{A. Camacho-Guardian}
\affiliation{Instituto de F\'isica, Universidad Nacional Aut\'onoma de M\'exico, Ciudad de M\'exico C.P. 04510, Mexico\looseness=-1}
\author{O. Cotlet}
\affiliation{Institute for Quantum Electronics, ETH Zurich, Zurich, Switzerland}
\author{A. Imamoglu}
\affiliation{Institute for Quantum Electronics, ETH Zurich, Zurich, Switzerland}
\author{M. Wouters}
\affiliation{TQC, Universiteit Antwerpen, Universiteitsplein 1, B-2610 Antwerp, Belgium}

\author{G. M. Bruun}
\affiliation{Center for Complex Quantum Systems, Department of Physics and Astronomy, Aarhus University, Ny Munkegade 120, DK-8000 Aarhus C, Denmark.}

\author{I. Carusotto} 
\affiliation{Pitaevskii BEC Center, CNR-INO and Dipartimento di Fisica,
Universit\`a di Trento, via Sommarive 14, I-38123 Trento, Italy}

\begin{abstract}
We develop a comprehensive theoretical model for the interaction strength between a pair of exciton-polaritons in microcavity devices. Ab initio numerical calculations for dipolar polaritons in one dimension are used as a starting point to build a Born-Oppenheimer theory that generally applies to generic - dipolar or non-dipolar polaritons- in both one and two dimensions. This theory anticipates that the strong coupling to the cavity mode leads to a drastic enhancement of the polariton interactions as compared to bare excitons, and predicts unexpected scaling laws in the interaction strength as a function of system parameters. Comparisons with available experimental data are drawn, and specific suggestions to validate it with new experiments are made. Promising strategies towards the observation of a strong polariton blockade regime are finally sketched.
\end{abstract}

\date{\today}
\maketitle
\section{Introduction}
The dramatic advances in experimental capabilities regarding hybrid light-matter quantum systems have opened the door to engineering systems that possess exotic phases of matter of fundamental interest, which could serve as building blocks for future quantum technology~\cite{Chang2014,Kurizki2015}. In this regard, microcavity exciton-polaritons (polaritons)~\cite{Weisbuch1992,kavokin2003cavity,burstein2012confined}, namely bosonic quasiparticles that emerge from the strong coupling of an optical excitation of the material with a cavity photon mode in a semiconductor microcavity, have proven especially appealing. Characterized by an effective mass four orders of magnitude lower than the electron mass along with interactions inherited from their matter component, polaritons showcase a multitude of interesting collective quantum phenomena such as non-equilibrium Bose-Einstein condensation~\cite{Bloch2022, KPZ, Deng2010} 
and superfluidity~\cite{Carusotto2013,Amo2009}. 
Now the frontier is to enter the regime of strongly correlated fluids of light, which requires strong interactions between polaritons. 

This challenging enterprise calls for a deep understanding of basic mechanisms of polariton-polariton interactions. On the one hand, several pioneering theoretical studies have appeared on this topic, with a special attention towards strategies to reinforce polariton interactions~\cite{Ciuti1998,
Glazov2009, Wouters07,Carusotto_2010,Hu2020,Bleu2020,Li2021,Camacho-Guardian2021,Bastarrachea-Magnani2021,Camacho-Guardian2022}. Meanwhile, many experiments have reported measurements  of the polariton-polariton interaction strength spread over a wide range of values~\cite{Estrecho2019} and first evidences of quantum phenomena arising from strong interactions have been observed~\cite{Delteil2019, Munoz-Matutano2019}.

Among the new avenues being investigated in this context, a special promise is held by the so-called dipolar polaritons -dipolaritons in short- which emerge from the strong coupling of a cavity photon to a dipolar exciton featuring a sizable electric dipole. Such dipolar excitons are being investigated in a number of platforms, from coupled quantum wells~\cite{Christmann2011,Cristofolini2012}, to wide quantum wells~\cite{Rapaport2016}, transition metal dichalchogenide bilayers~\cite{Kremser2020, Li2020, Sun2022}, and electrostatically-defined devices~\cite{Thureja2022}.
For these dipolar excitons, a number of studies have reported strong interactions compared to conventional non-dipolar excitons, leading to strong correlations beyond a mean-field description~\cite{LOZOVIK2007, ZIMMERMANN2007, Astra2007, Shilo2013, Stern2014, Remeika2014, Cohen2016, Misra2018, MazuzHarpaz2019,Lee2009, Schinner2013}. It is then natural to expect that these reinforced interactions between dipolar excitons directly translate to reinforced interactions between dipolaritons. First evidence in this direction have started appearing in the literature~\cite{Tsintzos2018, Togan2018, Rosenberg2018dipolar,datta2022highly,Suarez:PRL2021,Liran:PRX2024}.

Motivated by these exciting developments, in this article we develop a general theory of polariton-polariton interactions, with a special focus on identifying strategies to reinforce them. We start from the case of dipolaritons for which a microscopic Hamiltonian is available and ab initio numerical calculations of the interaction energy can be straightforwardly performed for a one-dimensional geometry. The results of the numerical calculations in this simplest case then provide insight concerning the basic physics of the interaction process in the strong light-matter coupling regime and allow us to build an general and easily tractable Born-Oppenheimer theory describing systems of dipolaritons as well as of non-dipolar contact-interacting polaritons in both one and two dimensions. 
In alternative to the theoretical developments and the diagrammatic calculations in \cite{Bleu2020,Li2021}, our Born-Oppenheimer approach offers direct analytical insight into peculiar and unexpected scalings of the interaction strength as a function of the system parameters.
These may be used to shine light on the available experimental observations and offer well-defined signatures to experimentally assess the accuracy of our theory.

The outline of the paper is as follows. In Sec.~\ref{modelSec}, we present the microscopic model. Section \ref{Resultssec} describes numerical results for the interaction strength of dipolaritons in one-dimensional wire geometries, and we identify the different regimes in terms of the wire length: for long wires, polaritons display a fermionized Tonks-Girardeau-like behaviour, while for short wires they feature a Jaynes-Cummings behavior similar the one of a zero-dimensional quantum dot. In the experimentally most relevant intermediate regime, polaritons display weak inter-particle correlations that can be captured within a mean-field theory. 

In order to understand the value of the interaction strength in this intermediate regime, in Sec.~\ref{BOsec} we develop a Born-Oppenheimer approach: the slow degrees of motion correspond to the spatial separation of the two colliding polaritons; the fast ones correspond to the light-matter dynamics due to the exciton-photon inter-conversion process. The energy landscapes of two-polariton states as a function of their spatial separation provide the effective Born-Oppenheimer potential for polariton-polariton scattering. Analytical formulas for the scaling of the interaction strength between dipolaritons are then derived in Sec.\ref{Lightcouplingsec} as a function of the system parameters. These highlight the key impact of light-matter interaction in determining the interaction strength between polaritons and are shown to well capture the numerical results.  

The agreement of the Born-Oppenheimer theory with the numerical predictions provides a strong motivation to extend it to other configurations of present experimental interest such as two-dimensional microcavity systems and/or non-dipolar contact-interacting polaritons. This is done in in Sec.\ref{extensionsSec}: once again, peculiar scalings as a function of parameters are identified, which can be used to validate the accuracy of our model against experimental measurements in state-of-the-art devices. Critical comparisons with available experimental data for inorganic semiconductor and two-dimensional materials are reported in Sec.\ref{quantitativeSec}, together with a discussion about strategies to further reinforce the interaction strength and enter the polariton blockade regime. Conclusions are finally drawn in Sec.\ref{Concsec}. Additional numerical results for more complex geometries are reported in the two Appendices.

\begin{figure}[htb]
\centering 
\includegraphics[width=0.92\columnwidth]{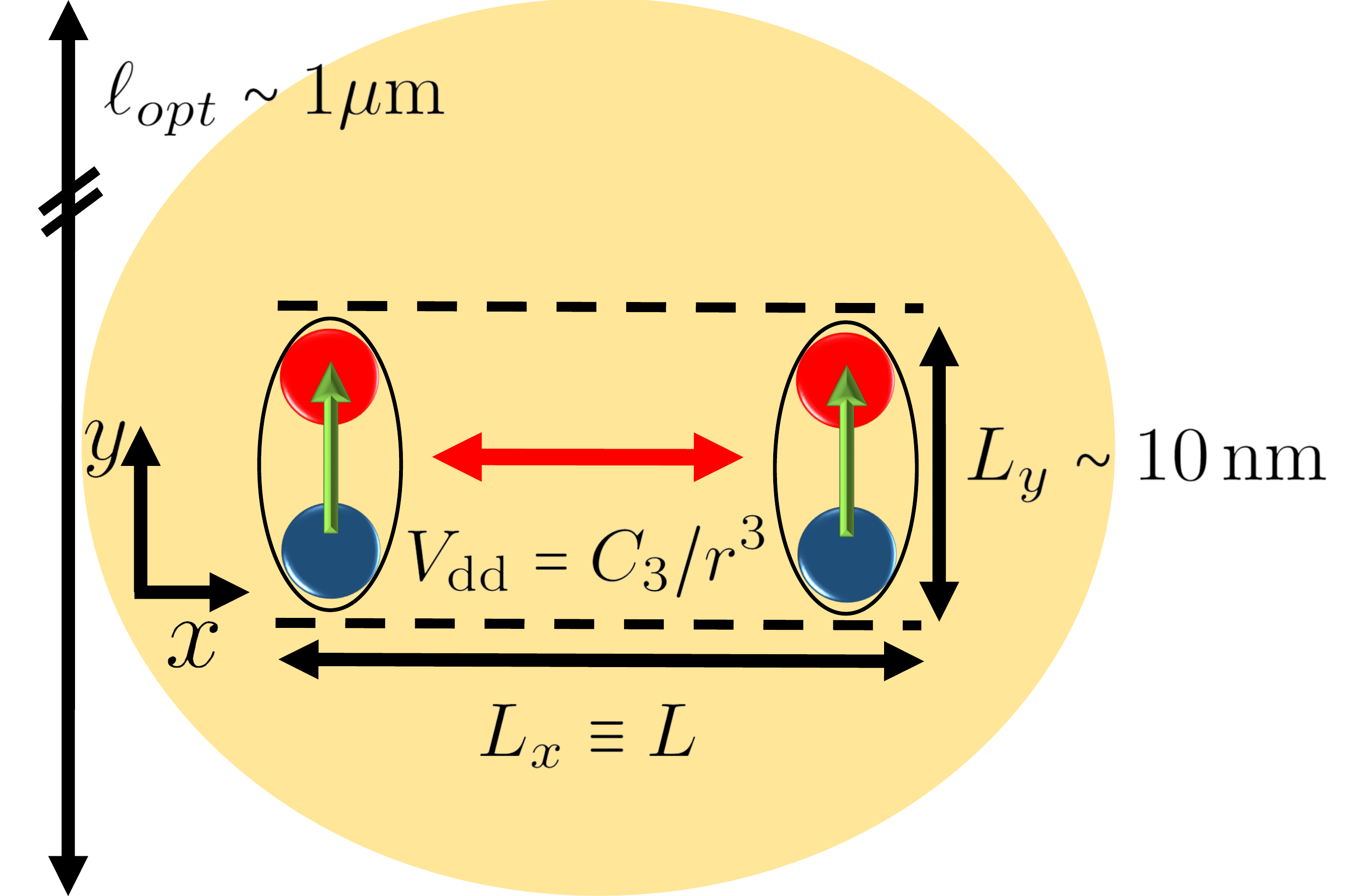}
\llap{\raisebox{-2.85cm}{\includegraphics[height=3.5cm,frame]{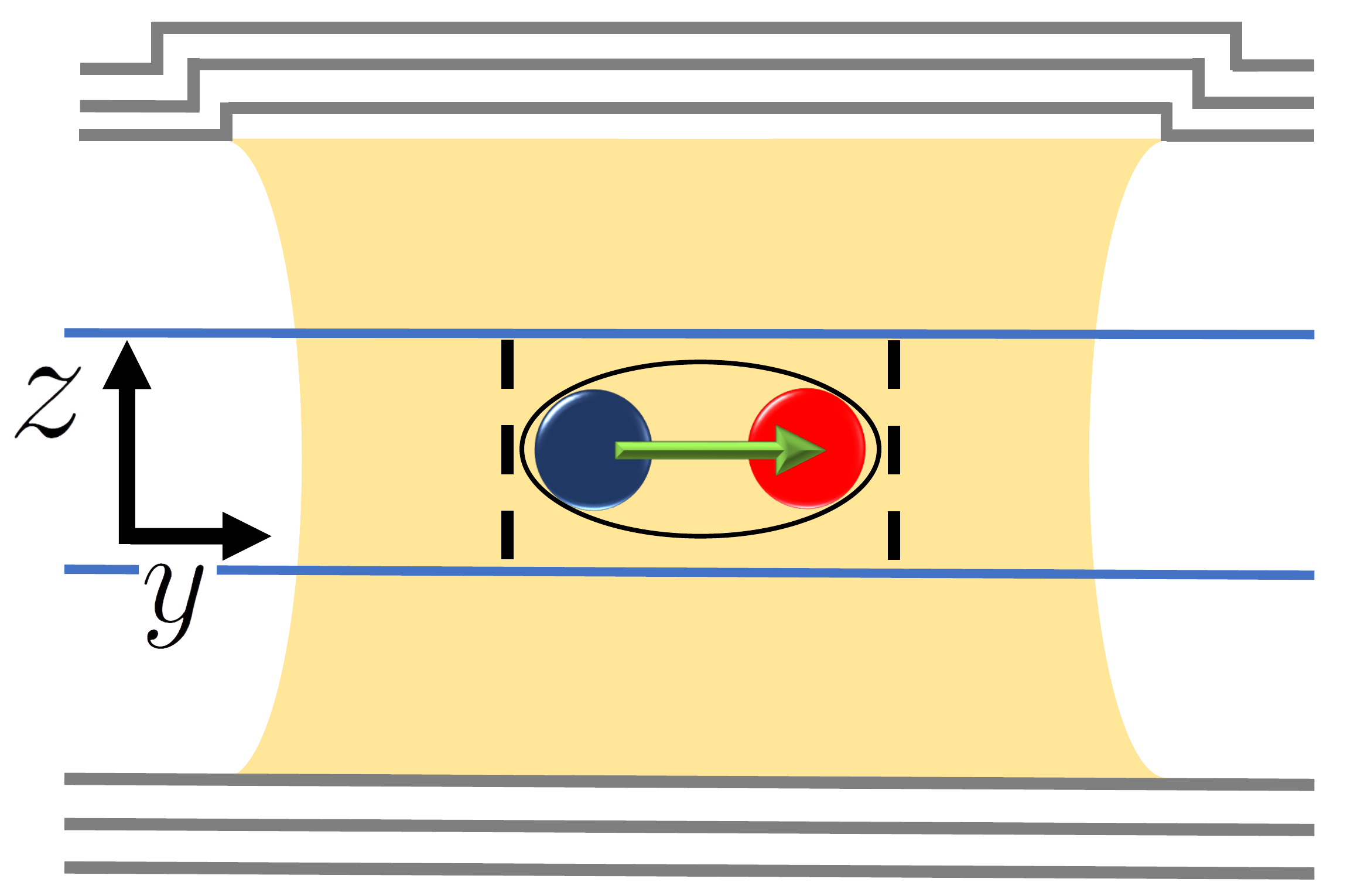}}}
     \caption{
    (Main) Two dipolar excitons in a quantum wire of length $L_x\equiv L$ with a strong transverse confinement along $y$ (illustrated by horizontal dashed lines) as seen from the $z$-direction. The dipole moments of the excitons (green arrows) are perpendicular to the wire leading to a strong repulsion. The excitons are strongly coupled to the one-dimensional mode (yellow ellipse) of a photonic wire cavity. 
    (Lower inset box) The excitons are in a 2D layer (indicated by the blue lines) 
    are further confined along the transverse $y$ direction (as indicated by the dashed lines) so that they can only move in the $x$-direction along the quantum wire axis. The quantum wire is embedded in a photonic wire microcavity and excitons are strongly coupled to a single one-dimensional photonic mode transversally confined within a lateral extent of $\sim 1 \, \mu$m (not drawn to scale), e.g. by means of etched Bragg mirrors (grey lines). 
    }
    \label{fig1}
\end{figure}
\section{One-dimensional dipolar polaritons}
\label{sec:1D_num}
We start our study from the simplest one-dimensional configuration sketched in Fig.~\ref{fig1}. Here, dipolar excitons are confined to a quantum wire and have a static electric dipole moment perpendicular to the allowed direction of motion. The quantum wire is assumed to be embedded in a photonic wire microcavity so that the excitonic transition is in the strong light-matter coupling regime with a one-dimensional optical mode.

\subsection{The physical system and the theoretical model}\label{modelSec}

This system can be described by a one-dimensional model based on the second-quantized Hamiltonian
\begin{gather}
    \hat{H} = \int d x \sum_{i,j=\{C,X\}} \hat{\Psi}_i^\dagger (x) h^{0}_{ij}(x) \hat{\Psi}_j (x) \nonumber\\
    +\frac{1}{2} \int d x_1dx_2 \big[ \hat{\Psi}^\dagger_{X}(x_1)\hat{\Psi}^\dagger_{X}(x_2)\nonumber\\
    \times V_{\text{dd}}(|x_1-x_2|) \hat{\Psi}_{X}(x_1)\hat{\Psi}_{X}(x_2)\big].
    \label{Hamiltonian1}
\end{gather}
where $\hat{\Psi}_{X/C}(x)$ are one-dimensional bosonic field operators describing the exciton and the photons, and $h^0_{ij}$ are the entries of the  matrix 
\begin{align}
    h^0(x) = \begin{pmatrix}
     -\frac{\hbar^2}{2m_C}\frac{\partial^2}{\partial x^2}+\Delta+\delta_C& \Omega \\
     \Omega & -\frac{\hbar^2}{2m_{X} }\frac{\partial^2}{\partial x^2} \end{pmatrix},
     \label{Hamiltonian2}
\end{align}
describing the motion of photons and of excitons along the wire and the coherent exciton-photon coupling of Rabi frequency $\Omega$. Diagonalization of $h^0(x)$ provides the dipolariton states emerging from the coherent mixing of photons and dipolar excitons.

We consider typical experimental values of GaAs-based devices: the exciton and cavity-photon masses are taken as $m_X=2m_e$, i.e. twice the electron mass, and $m_C=10^{-4}\,m_X$, respectively. For the Rabi coupling, a value $\Omega$ = 1 meV is used: while exciton-cavity coupling in planar geometries is on the order of $5\,$meV, the chosen smaller value accounts for the reduced spatial overlap between the excitonic wire (of transverse size of the order of 30~nm) and the one-dimensional optical mode (of transverse size of the order of 1~$\mu$m). ~\cite{Tartakovskii:PRB1998,Lecomte:PRB2013}

Hard-wall boundary conditions at the edges $x_1, x_2 =0, L$, of the wire are imposed. In an actual device, such a configuration is obtained by a suitable design of the photonic wire device and/or an etching of the material hosting the exciton. In order to set the energy of the lowest photon mode on resonance with the exciton one, a detuning
    $\Delta = \left({m_X}^{-1}-{m_C}^{-1}   \right){\hbar^2 \pi^2}/{(2L^2)}$,
is added to the photon energy, so that any further detuning is included in the parameter $\delta_C$. 

The interaction potential between the dipolar excitons has the usual repulsive form for parallel electric dipoles aligned orthogonally to their spatial separation, 
\begin{equation}
    V_{\text{dd}}(x_1,x_2) = \frac{e^2d^2}{4\pi \epsilon |x_1-x_2|^3} \equiv \frac{C_3}{|x_1-x_2|^3}
\label{eq:Vdip}
\end{equation} 
where $e$ is the electron charge, $\epsilon$ is the dielectric constant of the material, and $d$ is the spatial separation between electrons and holes in the dipolar exciton. 

This description is implicitly assuming that dipolar excitons are rigid point-like particles and that their repulsive interactions are strong enough to prevent overlap of their electronic clouds. In particular, the potential \eqref{eq:Vdip} only includes the $r^{-3}$ tail of the interaction potential at inter-particle distances $r>d$ and neglects the distortion of the internal exciton structure under the effect of light-matter coupling~\cite{Levinsen2019b,Li2021} and of the dipolar interaction with neighboring excitons~\cite{Liran:PRX2024}. Unless otherwise specified, in all figures we set $d=5 \,$nm and $\epsilon=10\epsilon_0$. 

A device along these lines was studied in the recent experiment~\cite{Thureja2022} demonstrating strong transverse confinement of neutral excitons in a two-dimensional transition metal dichalcogenide (TMD) monolayer. Here, the strong in-plane confinement along the transverse direction is electrostatically induced by means of a pair of partially-overlapping gates sandwiching the TMD layer, resulting in an effective one-dimensional quantum wire with a transverse confinement length of $\sim 10\,$nm. At the same time, the strong in-plane fields result in a sizable exciton dipole moment, with a dipolar length on the order of a fraction of nm, that is comparable to the exciton radius. Coupling this excitonic system to the one-dimensional photonic mode of a photonic wire cavity is one of the next step of the experiment.

\begin{figure}[htb]
    \centering
\includegraphics[width=1\columnwidth]{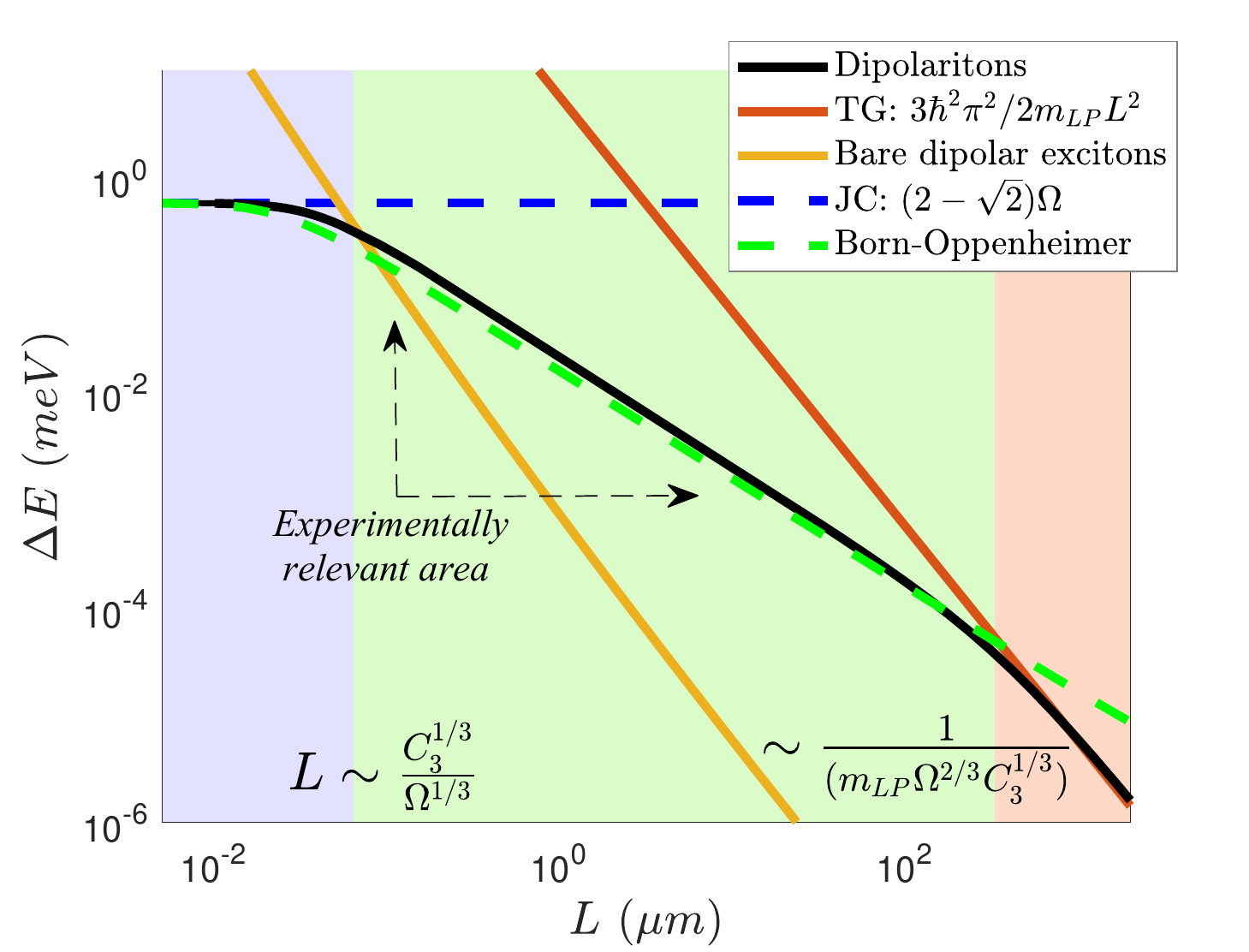}
    \caption{Black solid line: Numerical prediction for the interaction energy of a pair of dipolaritons as a function of the confinement length $L$ obtained by numerically solving Eqs.~\eqref{Hamiltonian1}-\eqref{Hamiltonian2}. The colored lines indicate the approximated theoretical predictions in the different regimes: For large $L$-values (right, red region), the interaction energy tends to the Tonks-Girardeau (TG) limit of two impenetrable bosons of mass $2 m_C$ (red solid line). For intermediate $L$-values (middle, green region), the interaction energy is well-described by a perturbative Born-Oppenheimer (BO) model (green dashed line). For small $L$-values (left, blue region) the interaction energy approaches the Jaynes-Cummings (JC) regime of a zero-dimensional quantum dot (blue dashed horizontal line). For intermediate to large confinement length (green and red regions), the interaction energy  for the polaritons is larger than the one for bare dipolar excitons (solid orange line). The system parameters used in the numerics are summarized in Sec.\ref{modelSec}. {The black dashed arrows indicate experimental limits of state-of-the-art systems, corresponding to $\Delta E \sim \gamma \sim 10^{-3}\,$meV and a minimum confinement of $L \sim 10^2\,$nm. The analytical expression for $L$-values of the cross-overs from TG to Born-Oppenheimer and to JC are indicated in the bottom of the figure.}}
    \label{DEvsL_excgs}
\end{figure}



\begin{figure*}[htb]
 \includegraphics[width=0.8\textwidth]{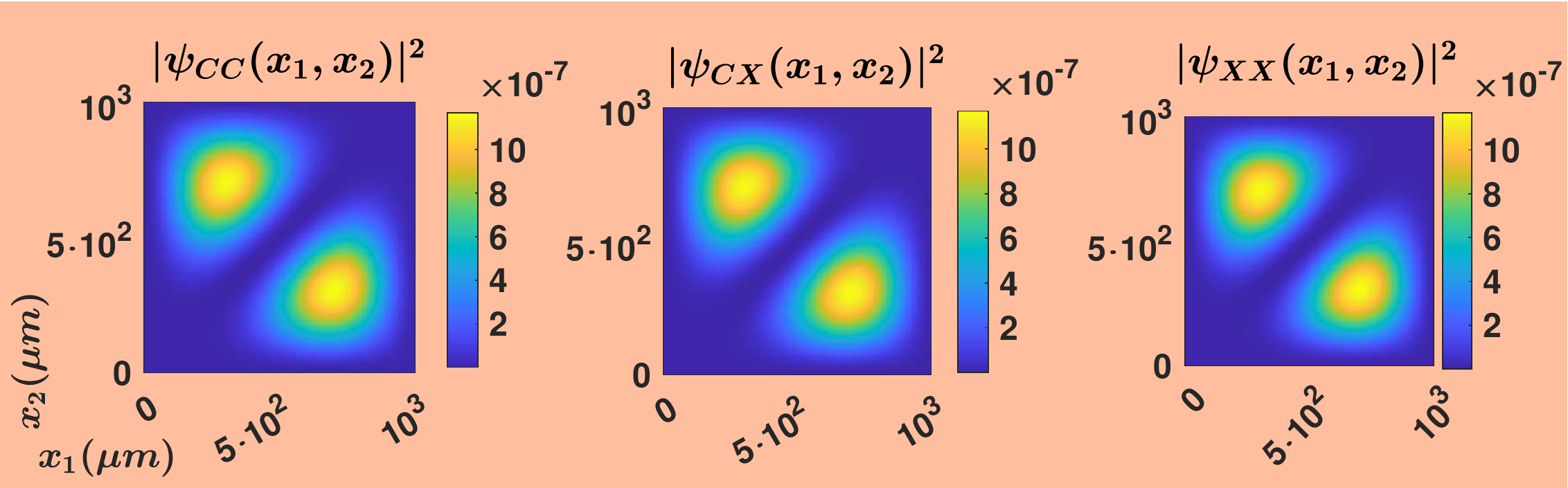} 
 \includegraphics[width=0.8\textwidth]{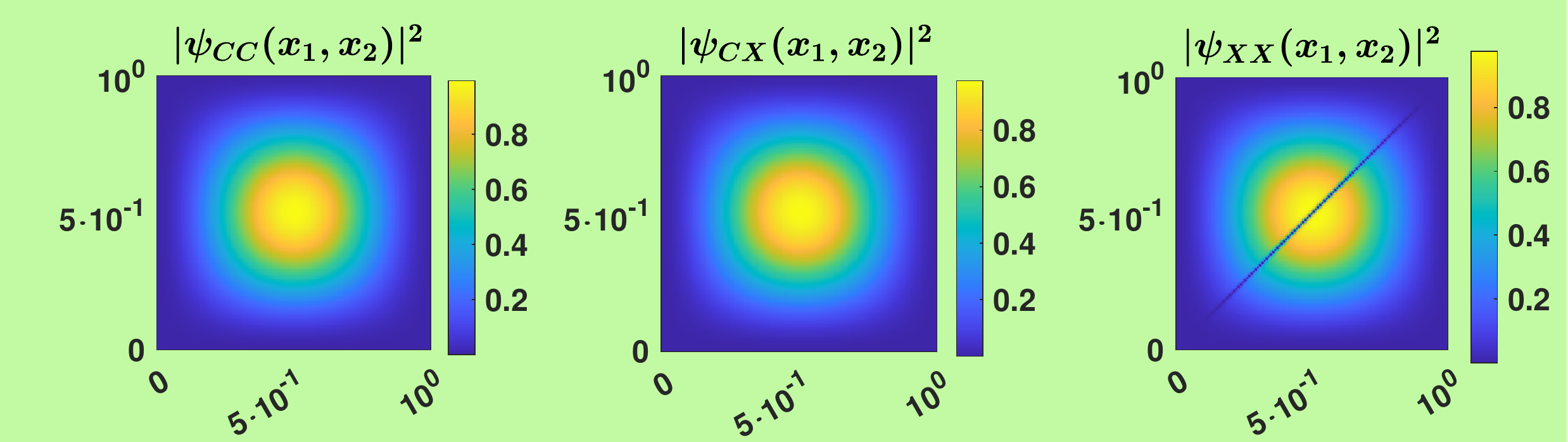}
  \includegraphics[width=0.8\textwidth]{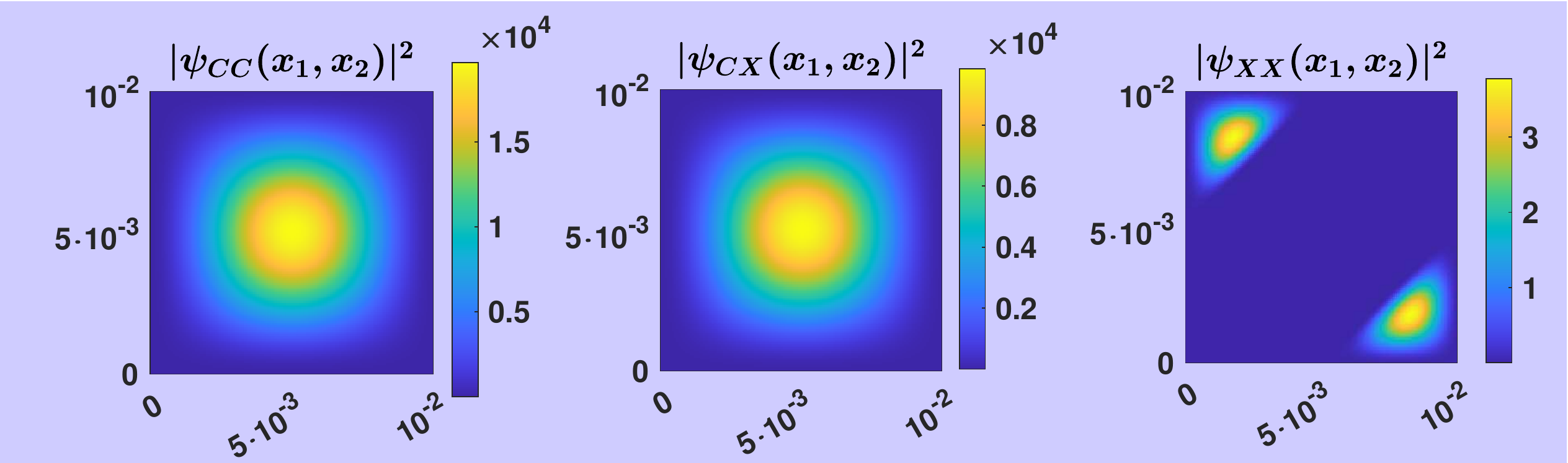}
\caption{Colorplots of the different components of the  two-particle ground state wavefunction for different confinement lengths $L=10^3 \ \mu m$ (top red panel), $L=1 \ \mu m$ (middle green panel), and $L=10^{-2} \ \mu m$ (bottom blue panel). Horizontal and vertical axes show the positions $x_1$ and $x_2$ of the two particles. Left, center, right panels correspond to respectively the $CC$, $XC=CX$, $XX$ components. Same system parameters as in Fig.\ref{DEvsL_excgs}.
}

    \label{threewfs}
\end{figure*}


\subsection{Numerical results for the interaction energy}\label{Resultssec}
In order to characterize the interaction energy between polaritons in the quantum wire configuration, we start from the Hamiltonian \eqref{Hamiltonian1} in the two-body sector and numerically solve the two-body problem to obtain the exact ground state energy and wave function of two interacting dipolar polaritons. 

The two-body wave function has four components $\psi_{i,j}(x_1,x_2)$ with $i,j\in \{C,X\}$
describing the excitonic and photonic amplitudes of the two polaritons, and it is symmetric under particle exchange $\psi_{i,j}(x_1,x_2)=\psi_{j,i}(x_2,x_1)$.
Since we are interested in varying the confinement length $L$ across several orders of magnitude and the dipolar potential displays a strong variation for small separation lengths, it is useful to transform to relative- and center of mass coordinates and adopt a non-uniform grid for the spatial coordinates~\cite{Tan1990}. Because of the large size of the Hamiltonian matrix, we restrict our attention to the lowest eigenvalue, $E_{LP-LP}$ and make use of a Lanczos-type algorithm to determine it and its corresponding eigenstate~\cite{Lanczos1950,Stewart2002}. 

We then define the interaction energy as 
the  difference 
\begin{equation}
\Delta E = E_{LP-LP}-2E_{LP}.
\end{equation}
between the ground state energies of the two-particle system in  the interacting and non-interacting cases respectively. In the latter case, the energy is equal to twice the energy of the one-particle ground state $E_{LP}$. 

As a key numerical result of this work, in Fig.~\ref{DEvsL_excgs} we plot the interaction energy $\Delta E$ as a function of the wire length $L$ for the resonant case $\delta_C=0$ (black line). 
Three regimes can be identified on this curve, that we are going to discuss in the next Subsections.

\subsubsection{Long wire regime: Tonks-Girardeau physics}
For a long wire, i.e.\ large $L$ (right red region in Fig.~\ref{DEvsL_excgs}), the interaction energy (black solid line) shows a $1/L^2$ scaling. This can be interpreted as the Tonks-Girardeau (TG) limit of two impenetrable hard-core bosonic polaritons~\cite{Tonks36,girardeau60}: the dipolar interactions are strong enough to dominate over the kinetic energy and the polaritons get effectively fermionized. 
The physics is then fully determined by the impenetrability condition of two polaritons and does not depend on the specific shape and range of the interaction potential: the interaction energy simply comes from the kinetic energy cost of introducing a node in the two-body wavefunction along the $x_1=x_2$ line.

Quantitatively, the two-body ground state energy $E_{LP-LP}$ is approximately equal to that of two non-interacting \emph{fermionic} particles with the same mass as our polaritons (in the present resonant case, $m_{LP}=2 m_C$), which occupy the lowest ($n=1$) and second lowest ($n=2)$ single-particle energy levels of the infinite square well of length $L$~\cite{Lieb63}. The interaction energy is obtained by subtracting from this value $E_{LP-LP}=5 \hbar^2 \pi^2/[2 m_{LP} L^2]$ twice the energy of the lowest $n=1$ single particle ground state. This gives the result 
\begin{equation}
    \Delta E_{TG}= \frac{3 \hbar^2 \pi^2}{2 m_{LP} L^2},
    \label{TGenergy}
\end{equation} 
which is plotted as a red solid line in Fig.~\ref{DEvsL_excgs}. The agreement with the numerical results is excellent in the long wire region highlighted in red in Fig.\ref{DEvsL_excgs}.

To further corroborate this interpretation, in the top row of Fig.~\ref{threewfs} we  show the square modulus of the different components of the (normalized) wavefunction $\psi_{i,j}(x_1,x_2)$: in this large $L$ regime, the strong light-matter coupling makes all four components of the wavefunction to  have comparable  weights and, most importantly, to vanish along the $x_1\simeq x_2$ contact line. 
This repulsive dip originates from the interaction between excitons acting on the $\psi_{XX}$ component. Thanks to the small value of the kinetic energy compared to the Rabi coupling, this dip also extends to the other components $\psi_{CX}$ and $\psi_{CC}$.


\subsubsection{Short wire regime: Jaynes-Cummings physics}
In the opposite limit of a short wire highlighted in blue in Fig.~\ref{DEvsL_excgs}, we observe a saturation of the interaction energy to a finite value. This can be understood as a Jaynes-Cummings (JC) regime~\cite{JCM63,Tian92}, where the short quantum wire behaves as a zero dimensional quantum dot and can only host a single exciton. 

This physical observation allows to analytically estimate the interaction energy. Setting the $\psi_{XX}$ component to zero, we can restrict to a basis where the spatial profile of the remaining $\psi_{CC}(x_1,x_2)$ and $\psi_{XC}(x_1,x_2)$ components is fixed (for each component $x_{1,2}$) to the spatial profile of the lowest single-particle energy level  $\psi_{LP}(x)=[2/L^{1/2}]\,\sin(\pi x/L)$. Diagonalization of light-matter interaction in this basis gives the energy $-\sqrt2\Omega$ for the two-particle ground state state. Subtracting twice energy of a single lower polariton $-\Omega$, one straightforwardly obtains an interaction energy
\begin{equation}
    \Delta E_{JC}=(2-\sqrt{2})\Omega.
    \label{JCenergy}
\end{equation}
This value is plotted in Fig.~\ref{DEvsL_excgs} as a horizontal blue dashed line. We see that the agreement with the numerical results is excellent in the short wire region highlighted in blue.

These physical features of the JC regime are confirmed looking at the different components of the wavefunction that are shown in the bottom panels of Fig. \ref{threewfs}. Here, the components $\psi_{CC}(x_1,x_2)$ and $\psi_{XC}(x_1,x_2)$ indeed follow the lowest single-particle energy level, while the amplitude of  $\psi_{XX}(x_1,x_2)$  is suppressed by several orders of magnitude because of the repulsive interactions.

\subsubsection{Intermediate length regime: weakly interacting polaritons}
\label{subsubsec:intermediate}
A different behaviour is observed in the intermediate wire length $L$ regime highlighted in green in Fig.~\ref{DEvsL_excgs}: here the interaction energy displays a $1/L$ scaling, to be contrasted to the $1/L^2$ scaling (constant value) of the impenetrable boson (Jaynes-Cummings) cases discussed above for the large (small) $L$ regimes. This $1/L$ scaling is typical of weakly interacting bosons in one dimension~\cite{Lieb63,PitaevskiiStringari2003book}: building a theoretical model to understand this behavior will be the goal of the next Section.



\begin{figure}
\includegraphics[width=\columnwidth]{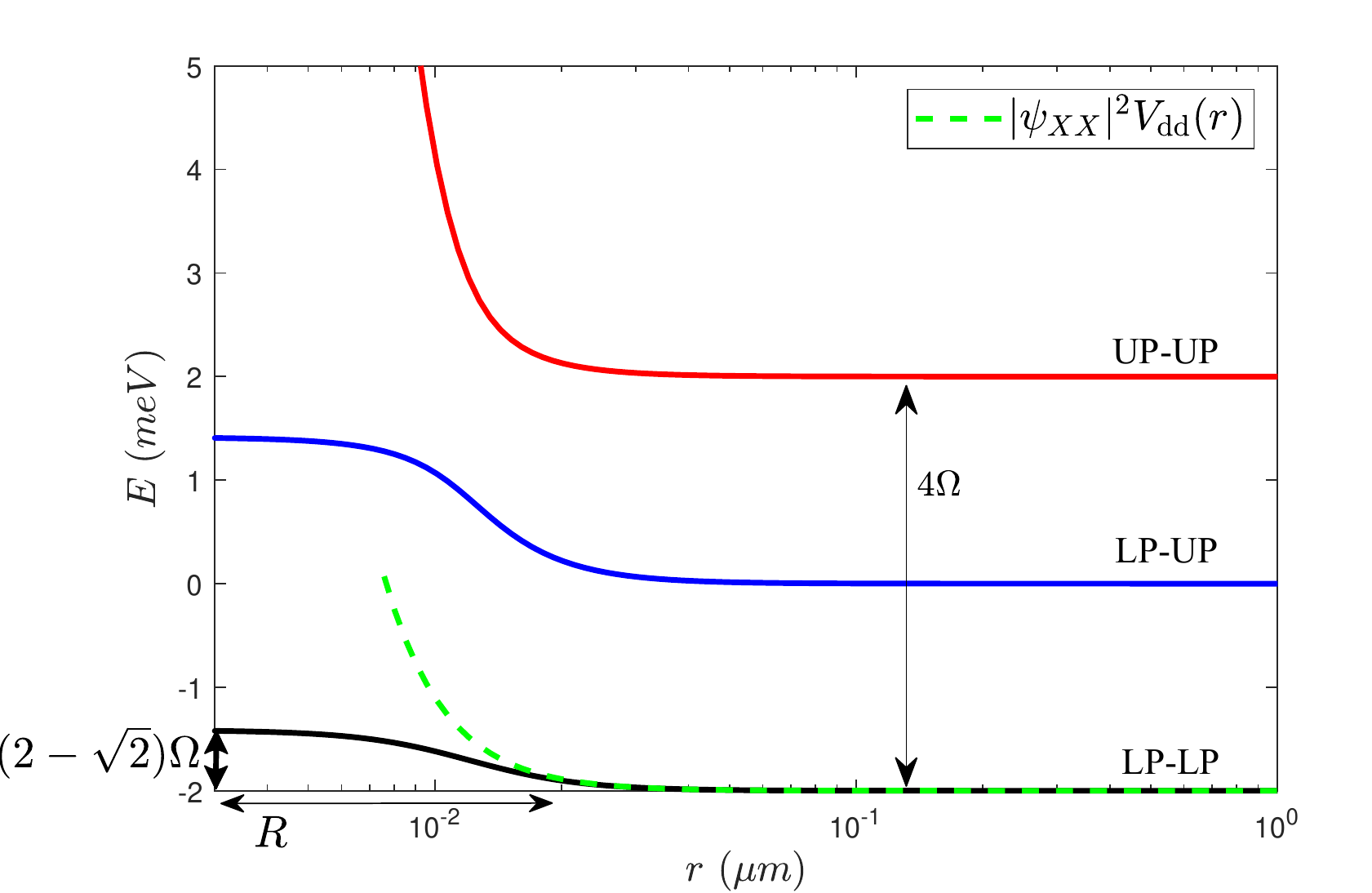}
\caption{Born-Oppenheimer energy landscapes as a function of the separation $r$ between the polaritons. The green dashed line indicates the dipolar potential between bare excitons weighted by the Hopfield coefficient. We have used the same system parameters as in Fig.\ref{DEvsL_excgs}. }
\label{DEvsL_excgs2}
\end{figure}

\section{Born-Oppenheimer theory}\label{BOsec}

In this Section we develop a theoretical description of polariton-polariton scattering in terms of Born-Oppenheimer (BO) energy landscapes as a function of the inter-polariton distance. 
The first goal of this theory will be to understand the physics of one-dimensional dipolar polaritons in the intermediate length $L$ regime of Sec.\ref{subsubsec:intermediate}. In the next Sections we will see that it has a much wider application range and provides deep insight also in generic configurations beyond the one-dimensional and dipolar-interacting case.

Our BO theory takes inspiration from molecular physics, where the speed of motion of nuclei is typically slow enough for an adiabatic approximation to be valid for the electronic dynamics, so that the electronic ground state energy provides an effective energy landscape for the quantum motion of the nuclei~\cite{BO1927}.
In the polariton context, the key idea of the BO approach is to separate out the fast dynamics due to the light-matter coupling whose energy scale is set by the Rabi coupling $\Omega$ from the slow spatial motion of the polaritons. For each value of the inter-polariton distance $r$, the eigenenergies of the fast light-matter dynamics then provide the $r$-dependent BO energy landscapes for a pair of interacting polaritons. The collisional properties of polaritons are then straightforwardly obtained as a standard scattering problem under the BO interaction potential.

\subsection{General formalism and Born-Oppenheimer energy landscapes}
\label{subsec:BOgeneral}
In formal terms, the fast light-matter dynamics is determined by a Hamiltonian that in the $CC,XC,CX,XX$ basis reads
\begin{align}
\label{eq:hBO4}
    h_{BO}=\begin{pmatrix}
    2\delta_C & \Omega & \Omega & 0\\
    \Omega & \delta_C & 0 & \Omega \\
    \Omega & 0 & \delta_C & \Omega \\
    0 & \Omega & \Omega &  V_{\text{dd}}(r) 
    \end{pmatrix}\,.
\end{align}
Here, $\delta_C$ is the detuning of the photonic mode and $V_{\text{dd}}(r)$ is the dipolar potential at a separation distance $r$. By bosonic symmetry, the $XC$ and $CX$ states have identical coupling to the other states, so they only enter via their symmetric  $(XC+CX)/\sqrt{2}$ combination; the anti-symmetric one $(XC-CX)/\sqrt{2}$ is instead decoupled and can be safely neglected in what follows. This results in a simplified 3-by-3 Hamiltonian matrix of the form
\begin{equation}
    h_{BO}=\begin{pmatrix} 
    2\delta_C & \sqrt{2} \Omega & 0 \\
    \sqrt{2} \Omega & \delta_C & \sqrt{2}\Omega \\
    0 & \sqrt{2}\Omega & V_{\rm dd}(r)
    \end{pmatrix}.
    \label{eq:hBO}
\end{equation}
An example of the BO energy landscapes $V_{BO,i}(r)$ for the resonant $\delta_C=0$ case is shown in Fig.~\ref{DEvsL_excgs2}.  At infinitely large distances, the dipolar potential is negligible and the three landscapes tend to the rest (i.e. zero momentum) energies of a pair of non-interacting polaritons in the different channels, namely two lower polaritons (LP-LP, black), one lower and one upper polariton (LP-UP, blue), and two upper polaritons (UP-UP, red).

For large but finite distances, the dipolar interaction energy induces a perturbative shift of the three energy landscapes, proportional to the dipolar potential $V_{\rm dd}(r)$ weighted by the excitonic weight of the polariton pair, given by the product of the squared excitonic Hopfield coefficients $|u_X|^2$ of the two polaritons in the pair. 
In Fig.~\ref{DEvsL_excgs2}, the perturbatively shifted shape of the lowest landscape (corresponding to a pair of lower polaritons) is shown as a green dashed line: at large distances it recovers the full BO calculation (black line) but strong deviations start being visible as the polaritons approach and non-perturbative effects start dominating.

At short distances, the $XX$ state is pushed away to high energies by the dipolar potential, so that the last row and column of \eqref{eq:hBO} are effectively decoupled. As a result, the weight on the $XX$ state is concentrated in the highest energy eigenstate corresponding to a pair of colliding UPs which displays the $1/r^3$ divergence of the dipolar potential. On the other hand, the lowest BO energy landscape saturates to the Jaynes-Cummings value of the interaction energy, equal, in the considered $\delta_C=0$ case, to $(2-\sqrt{2})\Omega$.  

The spatial size $R$ of the energy bump corresponding to this saturation plateau can be estimated as the location where the dipolar interaction energy $V_{\rm dd}(r)$ equals the Rabi coupling $\Omega$, that is $V_{\text{dd}}(R) = {C_3}/{R^3}=\Omega$, which gives a value
\begin{equation}
\label{eq:dipolarlength}
R=(C_3/\Omega)^{1/3}\,.
\end{equation}
for the dipolar radius. In the plot of the interaction energy as a function of the confinement length shown in Fig.~\ref{DEvsL_excgs}, this value sets the position of the crossover between the intermediate and the JC regimes. For the specific choice of parameters used in the Figures, the dipolar radius $R\simeq 15\,$nm: this relatively large value suppresses overlap of the electronic clouds; furthermore, the fact that $R>d$ also guarantees consistency of our model based on the $1/r^3$ tail of the dipolar interaction potential.


\subsection{Weakly-interacting one-dimensional dipolar polaritons: perturbative BO theory}
\label{subsec:weaklyinter}
The BO picture can be used to obtain a prediction for the ground state energy of the two-body problem by reducing the multi-component problem of \eqref{Hamiltonian1} and \eqref{Hamiltonian2} to a simpler single-component scattering problem on the BO energy landscape. If the interaction energy stemming from the BO energy landscape is weak enough, one can perform the additional approximation of treating it at a perturbative level. 
This immediately leads to the perturbative BO prediction
\begin{multline}
   \Delta E_{\rm int} =\int_0^L \!\!\! \int_0^L dx_1\,dx_2\,V^{(BO)}
   (|x_1-x_2|)\\ \times |\psi_{LP}(x_1)|^2|\psi_{LP}(x_2)|^2\, ,
   \label{eq:BOpert}
\end{multline}
where $|\psi_{LP}(x)|^2=\frac{2}{L}\,\sin^2(\pi x/L)$ is the squared modulus of the wavefunction of the lowest single-particle state in the wire and $V^{(BO)}(r)$ is the BO energy landscape corresponding to the polariton pair under consideration, e.g. the lowest one for a pair of colliding lower polaritons. 

If the characteristic size $R$ of the interaction potential is much smaller than the wire length $L$, we can neglect the integration limits and further simplify \eqref{eq:BOpert} to the form
\begin{equation}
   \Delta E_{\rm int} \simeq \frac{3}{2L}\,\int\,dx\,V^{(BO)}
   (x),
   \label{eq:BOpert2}
\end{equation}
where the $3/2$ factor takes into account the spatial inhomogeneity of the single-particle wavefunction in the presence of the hard-wall boundary conditions and the $1/L$ scaling typical of a weak interaction regime is apparent. Based on the qualitative shape of the lowest BO energy landscape discussed above, the magnitude of the interaction energy \eqref{eq:BOpert} can be then estimated to be on the order of $\Delta E_{\rm int}\sim \Omega R/L$. As our calculations so far assume a constant number of particles equal to two, the wire length $L$ is physically equivalent to a sort of inter-particle distance, that is the inverse of the one-dimensional density $n_{1D}$.

A quantitative numerical evaluation of the perturbative BO prediction \eqref{eq:BOpert} for the interaction energy of a lower polariton pair is plotted in Fig.~\ref{DEvsL_excgs} as a dashed green line. In the intermediate-$L$ range (green region in the figure), this line indeed well approximates the full numerical calculation. The slight deviations that one observes for decreasing values of $L$ can be traced to corrections to the BO picture due to the breakdown of the adiabaticity condition. Of course, the $1/L$ scaling of the interaction energy becomes completely inaccurate when $L$ becomes of the order of the dipolar length $R$ and one enters the JC regime (blue region in Fig.~\ref{DEvsL_excgs}). 

It is important to note that the expression \eqref{eq:BOpert} is perturbative in the effective Born-Oppenheimer interaction potential $V^{(BO)}(r)$, but, in contrast to much previous work on polariton-polariton interactions~\cite{Ciuti1998,Rochat2000, Glazov2009, Estrecho2019}, remains non-perturbative in the underlying exciton-exciton interaction potential. As such, it safely applies also to the typical physically relevant cases where the bare exciton-exciton interactions is very strong but polaritons remain in a weakly interacting regime. 

On the other hand, our perturbative treatment of the BO interaction energy underlying the estimate \eqref{eq:BOpert} of the interaction energy is expected to be accurate provided the wire is short enough for the energy gap to the next single-particle energy level to well exceed the interaction energy,
\begin{equation}
    \frac{\hbar^2}{m_{LP} L^2} \gg \Delta E_{\rm int} \,.
    \label{eq:BOpert_cond}
\end{equation}
From a physical standpoint, this condition 
corresponds to the boundary between the weakly interacting and the TG regimes. Using the estimate $\Delta E_{\rm int}\sim R\Omega/L$ for the perturbative BO interaction energy one obtains a value $L\sim 1/(m_{LP}\Omega^{2/3}C_3^{1/3})$ for the location of the crossover indicated in Fig.\ref{DEvsL_excgs}. 
Of course, we expect that a more refined application of the BO framework fully including the quantum mechanical motion in the two-body potential arising from BO energy landscape would be able to provide a quantitative description of the two-body problem also in the TG regime.




Further insight on the physics of the weakly-interacting-polariton regime 
is provided by the plots of the different components of the two-body ground state wavefunction $\psi_{ij}(x_1,x_2)$ shown in the middle row of Fig.~\ref{threewfs} for an intermediate value $L=1 \ \mu$m. 
This wavefunction closely resembles the product of free-particle sine-shaped wavefunctions at all spatial positions, except for the narrow dip that is present only in the $XX$-component along the contact line $x_1=x_2$.
This is in agreement with the BO picture, whose starting point is to approximate the full two-body wavefunction as the product of the slowly varying envelope times the local eigenvector of the fast light-matter dynamics \eqref{eq:hBO}: the fact that the depth of the  dip in the $XX$ component is almost complete confirms how the BO framework of \eqref{eq:BOpert} is indeed able to accurately capture non-perturbative features in the underlying dipolar interaction potential. 



\section{Scaling laws and role of strong coupling with light}\label{Lightcouplingsec}
After having presented our numerical results for one-dimensional dipolar polaritons and having laid the grounds of the theoretical BO framework, in this Section we proceed with a systematic study of the dependence of the interaction energy for a pair of colliding lower polaritons on the different system parameters, in particular the strength $C_3$ of the dipolar interactions, the Rabi coupling $\Omega$, and the exciton-photon detuning $\delta_C$. 

Even though the analysis of this Section will be focused on the case of one-dimensional dipolar polaritons for which comparison with numerics is available, its conclusions will provide general insight into the crucial role of light-matter coupling in the polariton interaction process. As such, the results of this Section will be propedeutical to the investigation of other configurations of experimental interest that we are going to present in the following Sections.

\subsection{Polaritons vs. bare excitons}
As a first step, we compare the interaction energy for a pair of dipolaritons with the one of a pair of bare dipolar excitons. Naively~\cite{Carusotto2013}, one might expect that the effect of light-matter coupling is to reduce the interaction between the polaritons compared to the one between bare excitons by a factor $|u_X|^4$ where $u_X$ is the excitonic Hopfield coefficient of the polariton. Based on this argument, one expects to find a stronger interaction energy for excitons than for polaritons.

As it is shown by a solid orange line in Fig.\ref{DEvsL_excgs}, our numerics find a completely different physics. In particular, we see that the interaction energy between bare excitons is much smaller than the one between polaritons for intermediate to long wire lengths. Furthermore, the interaction energy of excitons shows the $1/L^2$ scaling typical of the TG regime for all considered values of $L$ with no trace of a weak interaction regime. 

These findings can be understood in terms of the far heavier mass of exciton, which strongly reduces the energy cost of inserting a node in the two-body wavefunction and thus stabilizes a TG state. 
A quantitative comparison of the interaction energies of excitons and of polaritons can be performed in the large $L$ regime (red region in Fig.\ref{DEvsL_excgs}),  where both display a TG behaviour: according to \eqref{TGenergy}, the far heavier mass of excitons provides a $m_X/m_{LP} \sim 0.5 \times 10^4$ suppression factor of the interaction energy compared to polaritons.

This ordering of the curves persists at smaller $L$ values, albeit with a smaller magnitude: while in the intermediate-$L$ regime (green region) the polaritons transition into a weakly interacting regime with a slower $1/L$ scaling of the interaction energy, the larger mass of excitons allows them to maintain the $1/L^2$ scaling of the TG regime down to small lengths $L$. 
\begin{figure}[t]
       \includegraphics[width=0.85\columnwidth]{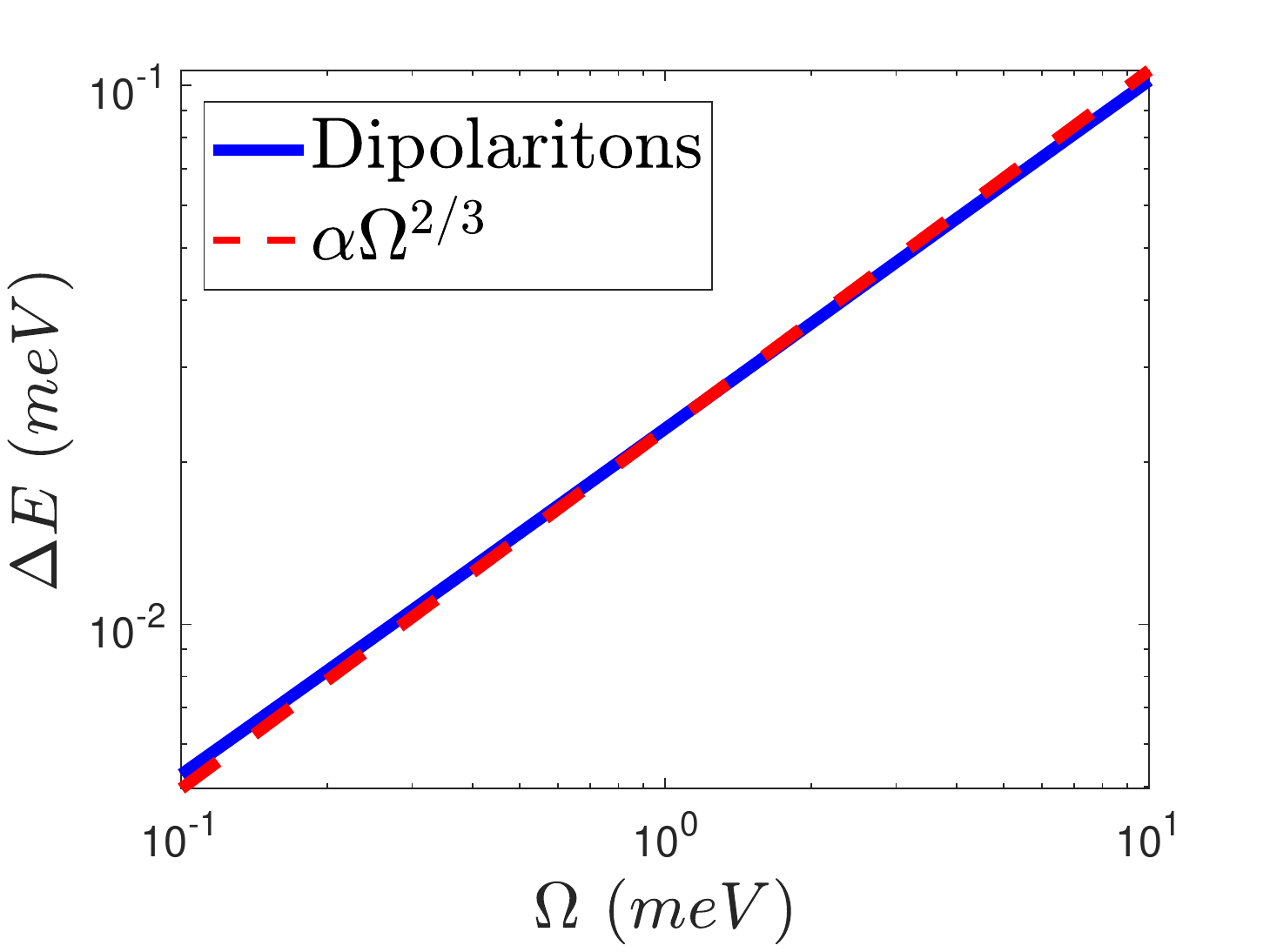}\\
    \includegraphics[width=0.85\columnwidth]{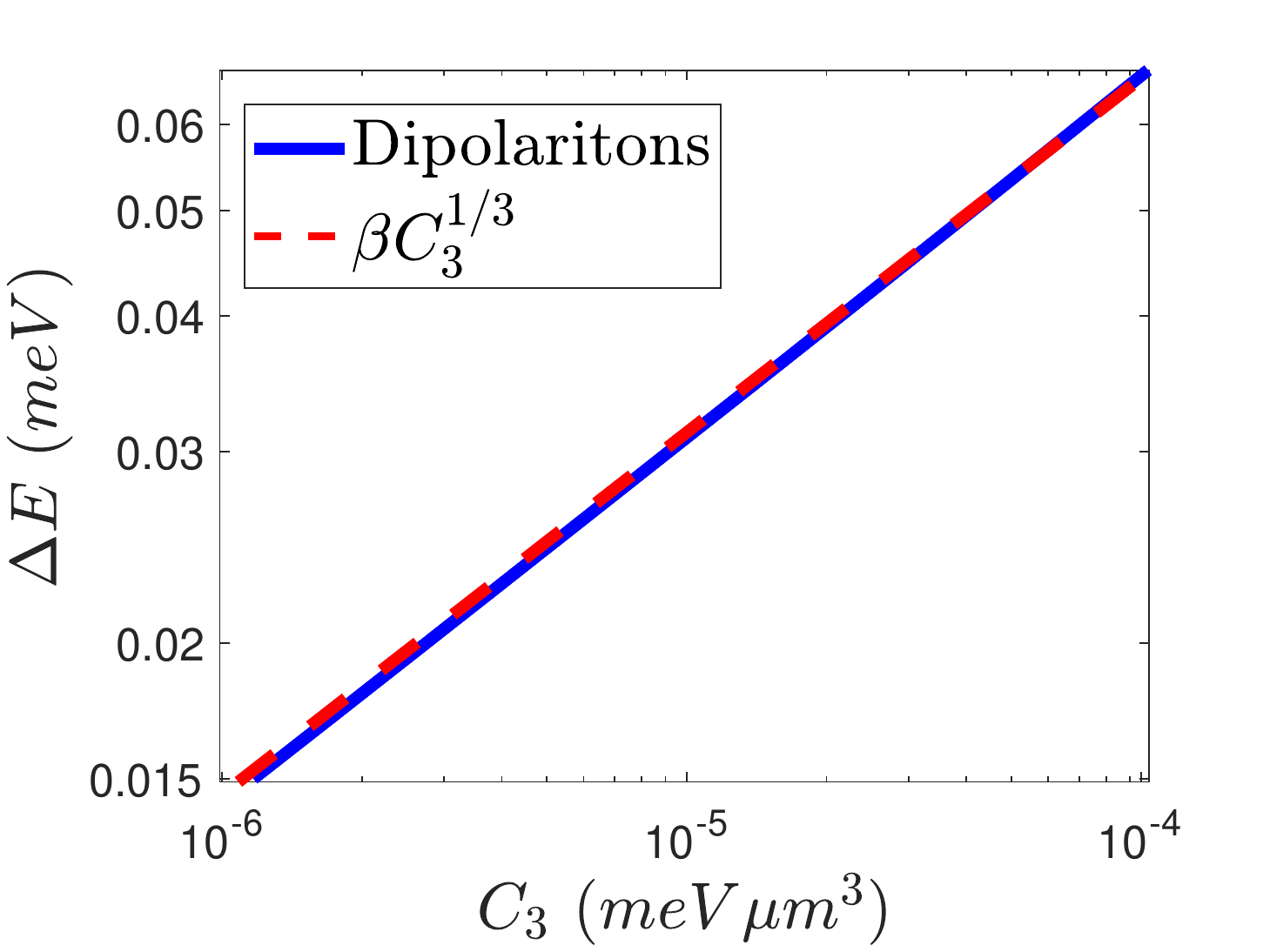}
    \caption{Numerical calculation (solid blue line) and power-law analytical fit (red dashed line) of the dipolariton interaction energy as function of the Rabi coupling $\Omega$ (left) and the dipolar strength $C_3$ (right) at a fixed $L=1\,\mu$m. Same system parameters as in Fig.\ref{DEvsL_excgs}. 
} 
    \label{parameters_scale}
\end{figure}

\subsection{Resonant case $\delta_L=0$}
\label{subsec:resonant}
Focusing our attention on the experimentally most relevant regime of weakly interacting lower polaritons in intermediate-$L$ configurations, we now make use of the BO framework to identify the functional dependence of the polariton interaction energy on the different system parameters. 

Let us start from the $\delta_C=0$ case, for which the BO energy landscape was plotted as a black line in Fig.\ref{DEvsL_excgs2}: at short distances, $E_{LP-LP}(r)$ displays a bump of height $(2-\sqrt{2})\Omega$ that was straightforwardly interpreted understood in terms of JC physics. Its spatial extension is set by the dipolar radius $R=(C_3/\Omega)^{1/3}$ defined in \eqref{eq:dipolarlength}. Under the perturbative BO approximation, Eq.\eqref{eq:BOpert2} then predicts a scaling
\begin{equation}\label{eq:BOpert_scaling}
    \Delta E_{\rm int} = a_{1D}\,\frac{\Omega R}{L}=a_{1D}\,\frac{\Omega^{2/3}C_3^{1/3}}{L}
\end{equation} 
of the polariton interaction energy with the system paraeters. 

Numerical integration of the BO energy landscape provides a value $a_{1D}\sim 3/2$ for the constant prefactor, in agreement with the numerical results in Fig.\ref{DEvsL_excgs}.
The functional form of the scaling is validated in Fig.~\ref{parameters_scale}, where we plot the numerical prediction for the interaction energy as a function of the Rabi coupling $\Omega$ (left) and of the dipolar interaction strength $C_3$ (right) for a fixed value of the wire length $L=1\,\mu$m well within the weakly interacting polariton regime.
The curves in both panels  confirm the quite non-trivial $C_3^{1/3}$ and $\Omega^{2/3}$ behaviors predicted by \eqref{eq:BOpert_scaling}. 

These results highlight the crucial role of the light-matter coupling in the polariton interaction process: the naive prediction~\cite{Carusotto2013} of an interaction energy proportional to the $|u_X|^4$ Hopfield coefficient would in fact lead to a linear dependence on the dipolar potential $C_3$ and no dependence on $\Omega$.
As a physically most salient feature, our prediction of a $\Omega^{2/3}$ scaling of the interaction energy implies that, for fixed values of the dipolar length $d$, 
more tightly-bound dipolar excitons with a stronger oscillator strength are favorable in view of achieving strong interactions between dipolaritons. 

\begin{figure*}[htbp]
    \includegraphics[width=0.95\columnwidth]{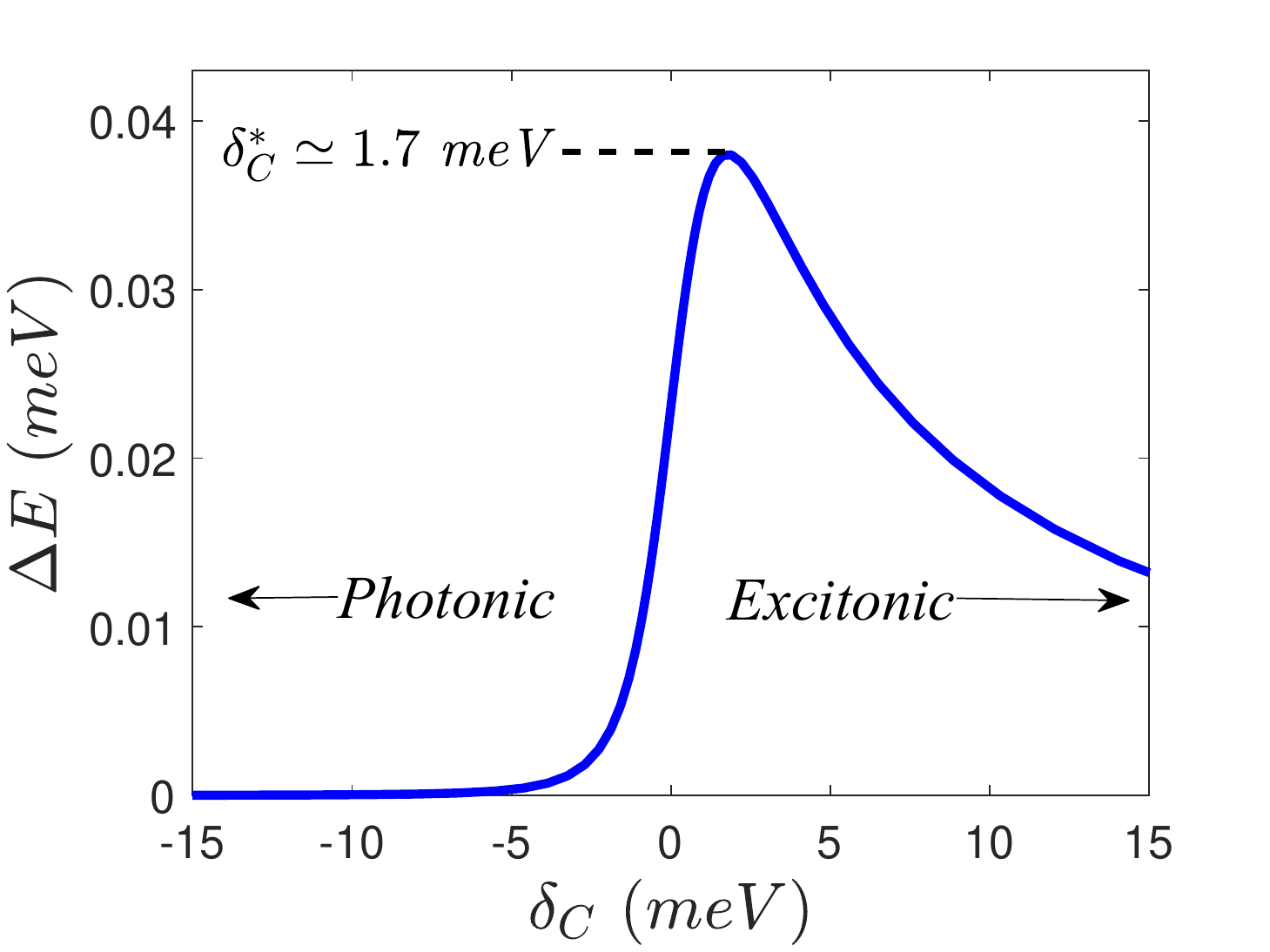}\hfill
    \includegraphics[width=0.95\columnwidth]{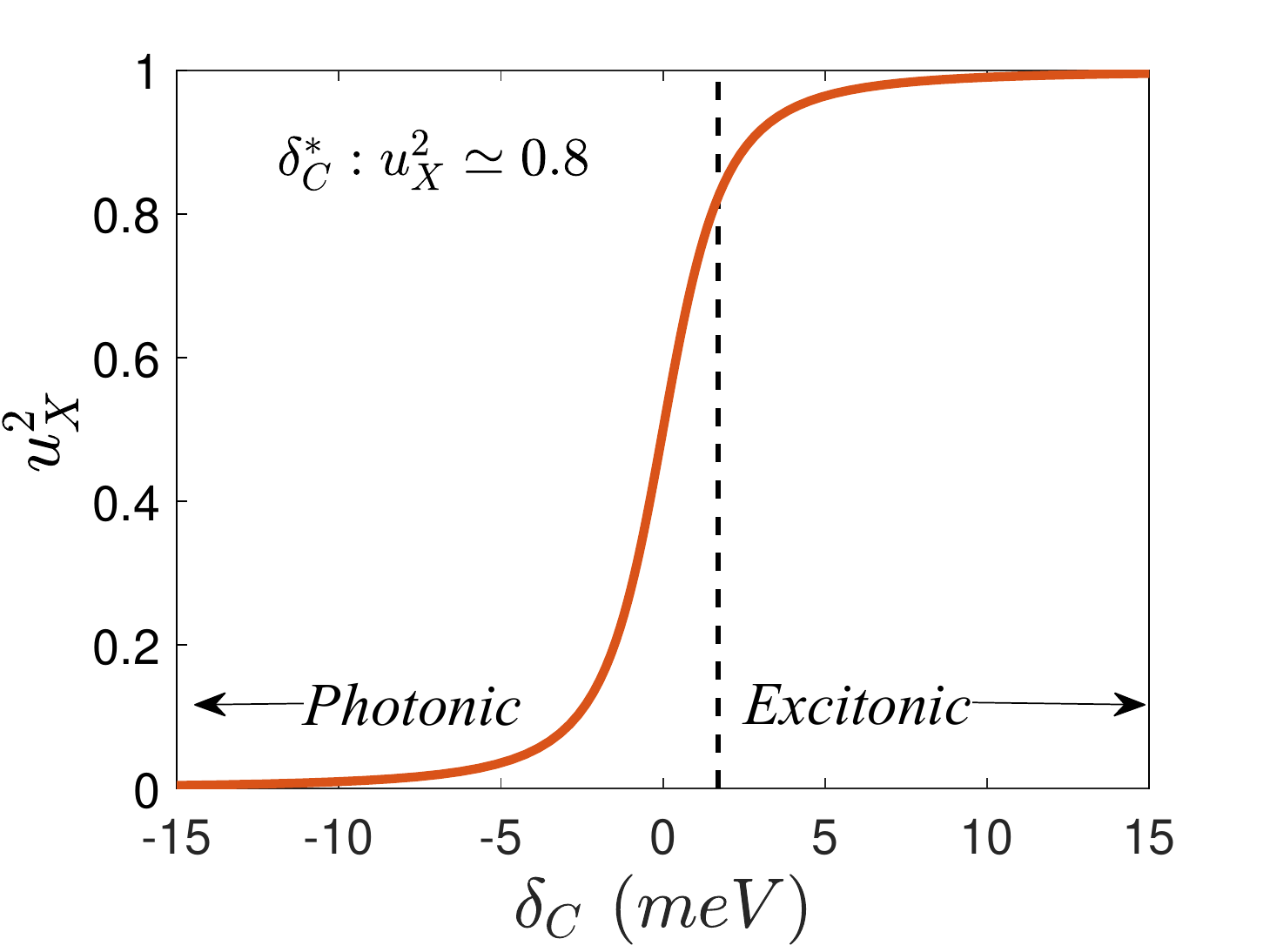}
    \includegraphics[width=0.95\columnwidth]{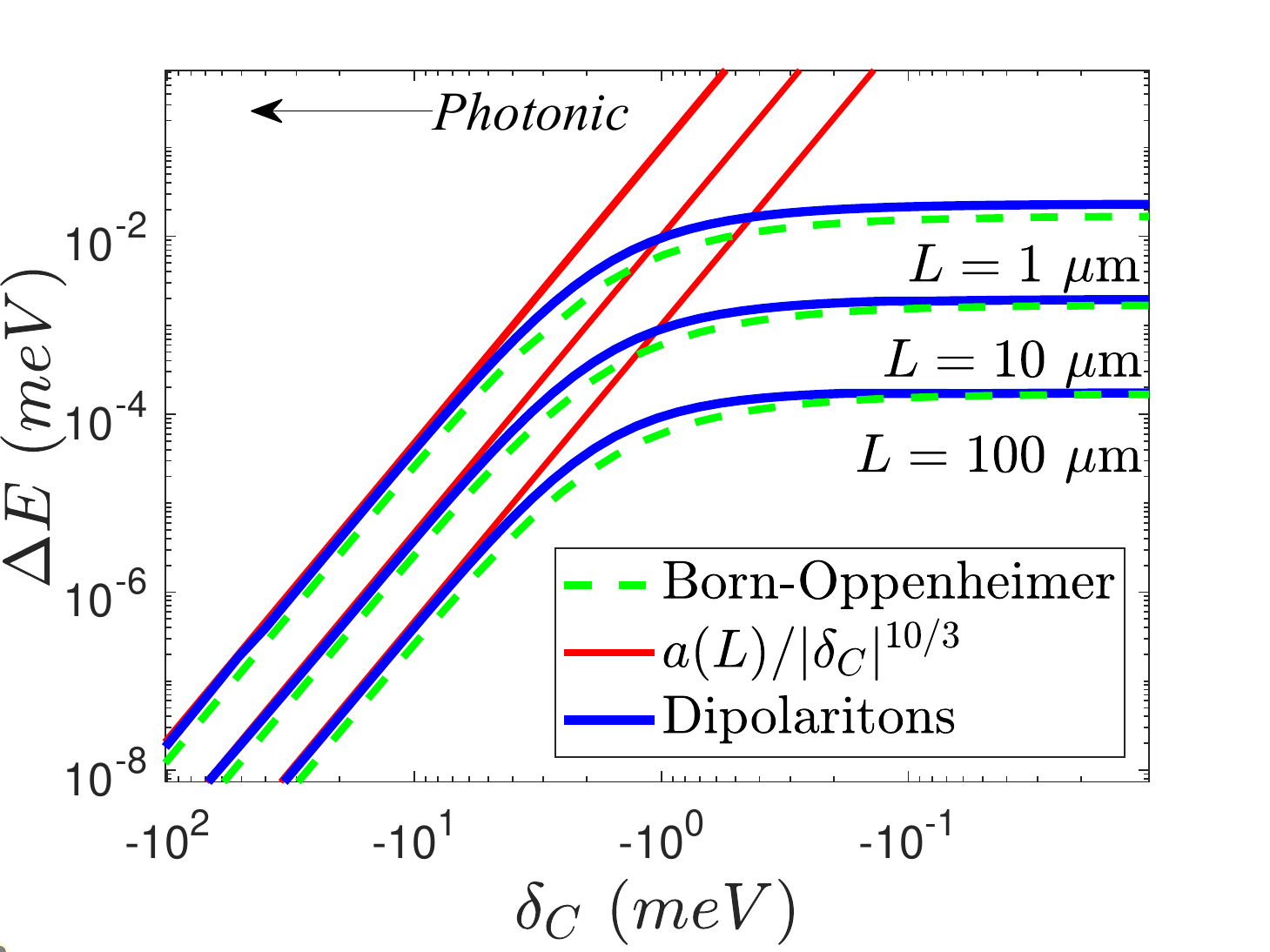}\hfill
    \includegraphics[width=0.95\columnwidth]{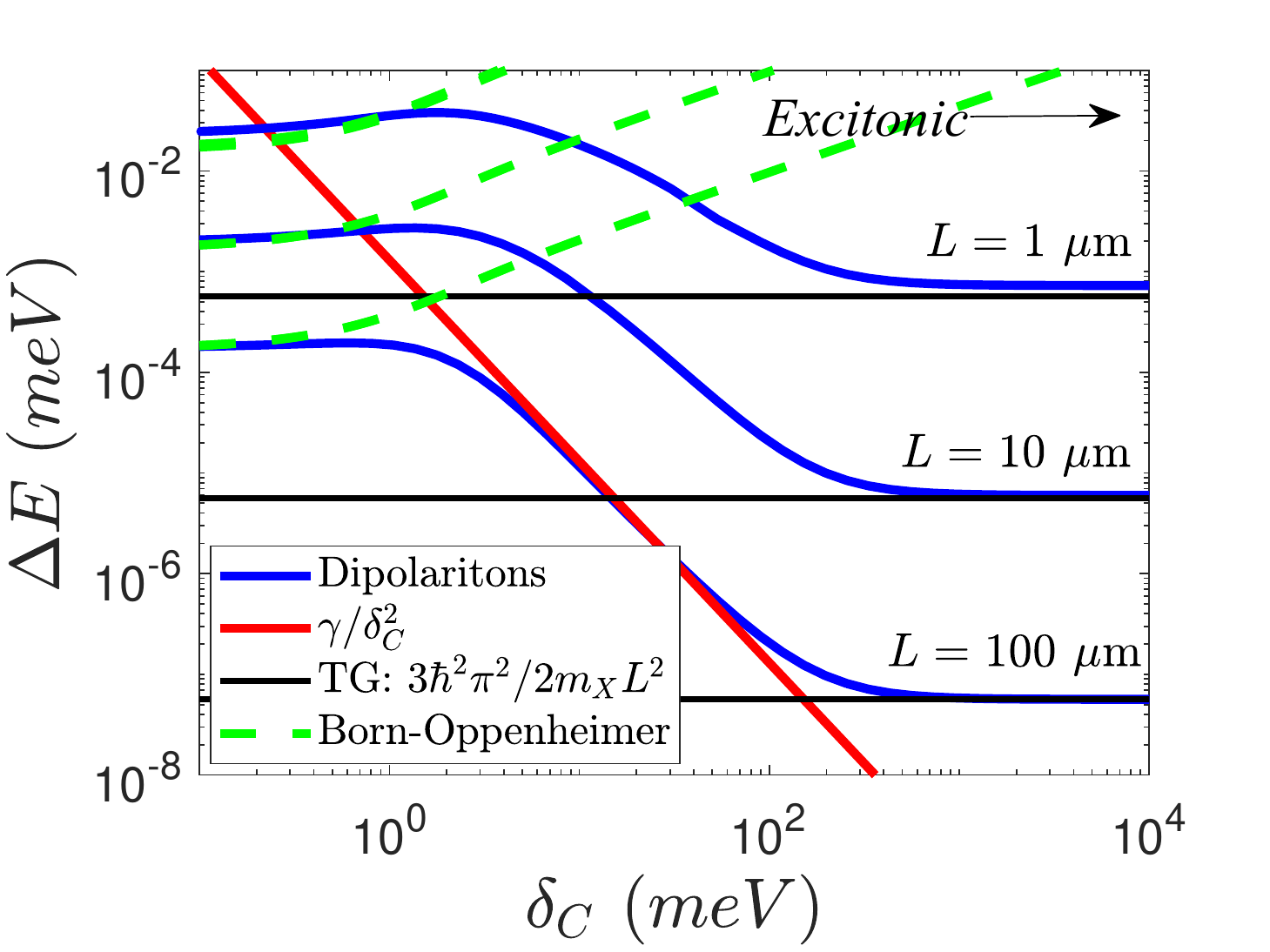}
    \caption{(Top left) Numerical calculation of the interaction energy, $\Delta E$, as function of detuning, $\delta_C$, for fixed values of the length $L=1\, \mu$m and of the Rabi coupling $\Omega=1$. The maximum of the interaction energy occurs at a finite detuning $\delta_C^*>0$ on the excitonic side. (Top right) excitonic content of the lower polariton as function of detuning $\delta_C$. Dashed vertical line indicates the point of maximum interaction energy, corresponding to $u_X^2 \simeq 0.8$. Bottom panels show log-log scale plots of the interaction energy as function of detuning on the photonic side (bottom left) and on the excitonic side (bottom right) for different values of the length $L$. The blue solid lines are full numerics, whereas the solid red and green dashed lines indicate the analytical power-law fit and the prediction of the perturbative BO approximation, respectively.  Same system parameters as in Fig.\ref{DEvsL_excgs}.}
    \label{DEvsdetun}
\end{figure*}
\subsection{Dependence on the exciton-photon detuning $\delta_C$}
\label{subsec:deltaC}
The non-trivial role of the light-matter coupling is further highlighted in the plot of the numerically calculated interaction energy as a function of the exciton-photon detuning $\delta_C$ shown in the upper-left panel of Fig.~\ref{DEvsdetun} for fixed values of the wire length $L=1 \ \mu$m and of the Rabi coupling $\Omega=1$ meV. For this choice of parameters, at the resonant $\delta_C=0$ point the system is in the weakly interacting polariton regime.

The most apparent feature of this plot is that the interaction energy features a pronounced peak at a specific value $\delta_C^*>0$ of the detuning on the excitonic side. 
For the specific parameters of the Figure, the peak is located at a detuning value around $\delta_C^*\simeq1.7\,$ meV, which corresponds (upper-right panel) to an exciton content of $u_X^2 \simeq 0.8$.
Once again, this result is to be contrasted with the naive prediction of the standard theory~\cite{Carusotto2013} that $\Delta E \sim |u_X|^4$ would be a monotonically growing funcion of $\delta_C$ and, thus, would attain its maximum value on the far excitonic side at large and positive $\delta_C$. 

\subsubsection{Excitonic regime at $\delta_C>0$}

A closer view on the excitonic side is given in the bottom right panel of Fig.~\ref{DEvsdetun}, where we show a log-log plot of the interaction energy as a function of (positive) detuning $\delta_C$ for a few different values of $L$. 
Most of its features can be understood in terms of the $\delta_C$-dependence of the polariton mass $m_{LP}$: following the reduction of the photonic character of the polariton for growing $\delta_C$, the polariton mass goes from a value equal to $2m_C$ at the resonance point $\delta_C=0$ towards the much heavier value $m_X$ on the far excitonic side $\delta_C\gg \Omega$.
As a consequence of the growing $m_{LP}$, the weak interaction condition \eqref{eq:BOpert_cond} gets eventually broken at large detunings and the system eventually recovers a TG behaviour. 

For small values of $\delta_C$, the polariton mass is close to its resonant value $2m_C$ and 
the polaritons are in a weakly interacting regime. This is confirmed by the fact that the interaction energy is well captured by the perturbative BO prediction \eqref{eq:BOpert} shown as a dashed green line. The main dependence of the interaction energy comes from the Hopfield coefficient $|u_X|^4$ which grows on the excitonic side from $1/4$ towards $1$ giving rise to a corresponding increase of the interaction energy. In the BO framework, this reinforced value of $|u_X|^4$ corresponds to a higher and wider bump in the BO energy landscape for a pair of LPs.

The increase of the interaction energy with $\delta_C$ stops when the increased polariton mass makes the system to transition to a TG regime and, thus, strongly deviate from the perturbative BO prediction. In agreement with \eqref{eq:BOpert_cond}, this transition occurs at a lower value of the mass (in our case of the detuning $\delta_C$) for lower values of the density (in our case, the inverse wire length $L^{-1}$). The following decrease of the interaction energy with $\delta_C$ in the TG regime of \eqref{TGenergy} is then responsible for the appearance of a maximum of the interaction energy at a finite value $\delta_C^*$ of the detuning.


At very large $\delta_C$, the polariton mass tends to the excitonic one $m_X$, so the interaction energy tends towards the small value $3\hbar^2 \pi^2/(2m_X L^2)$ expected for two bare excitons in the TG regime and indicated by the horizontal black lines. The slight deviations that are visible for small values of the wire length $L$ are a consequence of the long-range nature of the dipole interaction.

On the other hand, for large $L$, it is possible to clearly isolate the effect of the photonic component of the polariton, which gives a correction to the inverse polariton mass of the form 
\begin{equation}
    m_{LP}^{-1}\simeq m_{X}^{-1}+m_C^{-1}\frac{\Omega^2}{\delta_C^2}\,.
\end{equation} 
In the plot, this $\delta_C$-dependence of the mass is visible as an intermediate region where the interaction energy follows a $1/\delta_C^2$ slope (red line) before tending to the asymptotically constant value set by the exciton mass $m_X$. For smaller $L$, this behavior is made harder to isolate by the incipient transition to the weak interaction regime.

\subsubsection{Photonic regime at $\delta_C<0$}
\label{subsubsec:deltaneg}

For negative detunings $\delta_C<0$, polaritons have a mostly photonic character and the interaction energy monotonically drops to zero for large $|\delta_C|$.
To pin down the $\delta_C$-dependence on the negative side, we display in the bottom left panel of Fig.~\ref{DEvsdetun} a log-log plot of the interaction energy as a function of (negative) detuning $\delta_C$ for different values of wire length $L$. Quite interestingly, for our choice of parameters the perturbative BO approximation well captures the physics at play for all values of $\delta_C<0$.

For large and negative $\delta_C$ such that $|\delta_C|\gg \Omega$, the dependence displays an unexpected and very non-trivial $|\delta_C|^{-10/3}$ scaling (green dashed lines).
This dependence can be analytically understood in terms of the perturbative BO picture by calculating the energy landscape $E_{LP-LP}(r)$ of two colliding lower polaritons as the lowest eigenvalue of the 3-by-3 BO Hamiltonian \eqref{eq:hBO}, taken in the limit of large and negative $\delta_C$. 

In this regime, the height of the energy bump around $r=0$ is obtained by comparing the asymptotic value of $E_{LP,LP}(r)$ for $r\to \infty$ with its value at $r=0$. 
At $r=0$, the dipolar potential is infinite, so we can restrict \eqref{eq:hBO} to the first and second rows and columns, obtaining, in the large and negative $\delta_C$ limit
\begin{equation}
    E_{LP-LP}(r=0)\simeq 2\delta_C +\frac{2\Omega^2}{\delta_C}-\frac{4\Omega^4}{\delta_C^3} 
\end{equation}
At $r=\infty$, particles are instead non-interacting, so the asymptotic energy is equal to twice the single LP energy, $E_{r=\infty}=2E_{LP}$ with
\begin{equation}
    E_{LP}= \delta_C +\frac{\Omega^2}{\delta_C}-\frac{\Omega^4}{\delta_C^3}
\end{equation}
Taking the difference of the two quantities gives the following $\delta_C$-dependence
\begin{equation}
    E_{LP-LP}(r=0)-E_{LP-LP}(r=\infty) \simeq \frac{2\Omega^4}{|\delta_C|^3}
    \label{peak_delta}
\end{equation}
for the height of the potential bump.
In order to estimate the spatial extension $R_{\delta_C}$ of the energy bump, one has to note that for large $|\delta_C|\gg \Omega$, the relevant energy scale appearing on the diagonal of the matrix \eqref{eq:hBO} is now $\delta_C$ instead of $\Omega$ as in the resonant case. This means that
\begin{equation}
R_{\delta_C}\sim \left(\frac{C_3}{|\delta_C|}\right)^{1/3}\,.
\label{eq:peak_ext}
\end{equation}
Mutltiplying \eqref{peak_delta} and \eqref{eq:peak_ext}, one obtains a dependence
\begin{equation}
    \Delta E_{\rm int}\propto\frac{\Omega^4 \, (C_3)^{1/3}}{|\delta_C|^{10/3}}
    \label{eq:deltaC10/3}
\end{equation}
that indeed recovers the behavior that was numerically observed in the bottom left panel of Fig.~\ref{DEvsdetun}. 

In particular, note how this $\sim |\delta_C|^{-10/3}$ dependence is different from the prediction of the naive perturbation theory, where the $|u_X|^4$ scaling on the Hopfield coefficient would translate into a $(\Omega/\delta_C)^4$ dependence on the detuning~\cite{Carusotto2013}.

\section{Two-dimensional and/or contact-interacting polaritons} 
\label{extensionsSec}
The studies reported in the previous Sections started from a numerical study of the simplest case of dipolar polaritons in a one-dimensional geometry and led us to formulate an analytically tractable model having a much wider applicability range and providing an intuitive understanding of the physics at play.

In this Section, we take advantage of this physical insight to analytically extend the theory to other configurations of great experimental importance, in particular polaritons emerging from non-dipolar excitons and/or in two-dimensional geometries. 
This extension will allow us to compare and validate our theoretical predictions against available experimental observations in state-of-the-art two-dimensional systems as well as diagrammatic microscopic theories~\cite{Li2021}.

Throughout this Section, we keep focusing on a hard-wall wire geometry of length $L$ or, in the two-dimensional case, on a square geometry of side $L$. This result  straightforwardly extends to spatially homogeneous geometries  by replacing $L$ (resp. $L^2$) with $2/n_{1D}$ (resp. $2/n_{2D}$) in one-dimension (resp. two-dimensions) where the factor $2$ accounts for the $\sin^2$ shape of the ground state wavefunction of the spatially confined problem.

\subsection{Two-dimensional geometries}
\label{subsec:2D}

A full numerical calculation of polariton interactions in two dimensions along the lines of Sec.\ref{sec:1D_num} would be  numerically cumbersome. 
On the other hand, 
the Born-Oppenheimer theory developed in Sec.\ref{BOsec} is not specific to the one-dimensional case and can be applied in higher dimensions as well. 

In particular, the BO energy landscapes directly translate to higher dimensions, so one can obtain a prediction for the interaction energy in  two dimensions with the dipoles perpendicular to the planes,  by means of a straightforward generalization of \eqref{eq:BOpert} and \eqref{eq:BOpert_scaling} to two-dimensions.  In  the resonant $\delta_C=0$ case, this gives
\begin{equation}
\Delta E_{\rm int}=a_{2D} \Omega \, \frac{R^2}{L^2}=a_{2D} \frac{\Omega^{1/3} C_3^{2/3}}{L^2}\,.
\label{eq:BOpert_scaling2D}
\end{equation}
with a prefactor $a_{2D}$ for which a numerical integration of the BO energy landscape predicts a value $a_{2D}\sim 5$~\footnote{It is interesting to note how the scaling \eqref{eq:BOpert_scaling2D} strongly differs from the prediction of the so-called capacitor formula, $\Delta E_{\rm int}^{\rm (cap)} = (de^2/\epsilon) n_X$ often used to describe gases of dipolar excitons, $n_X=L^{-2}$ being the exciton density~\cite{Liran:PRX2024}. Several physical reasons justify this disagreement, including the key role of the light-matter interaction and the perturbative nature of the capacitor formula in the exciton-exciton interaction potential. In contrast to this, our theory assumes in fact from the outset that interaction between excitons are strongly repulsive so that $R>d$ and excitons are not able to overlap.}.
Numerical evidence in support to this approach is provided in App.~\ref{2Dapp}, where we successfully benchmark the predictions of the BO model on a numerically tractable geometry mimicking a fully two-dimensional one, namely a coupled-wire system where dipolaritons can tunnel between a pair of parallel wires.


A key difference of two-dimensional systems compared to one-dimensional ones is the different scaling of the characteristic scales of interaction vs. kinetic energy with the system size. In the one-dimensional case, we have seen in Sec.\ref{BOsec} that the characteristic kinetic energy scales as $1/L^2$ while the interaction one scales as $1/L$. As such, perturbation theory on the BO energy landscape underlying \eqref{eq:BOpert_scaling} is accurate for moderate values of $L$ but fails for large $L$ when polaritons enter the TG regime.
In two-dimensional systems on the other hand, the kinetic and the interaction energies display the same scaling as $1/L^2$. As a result, the validity of the perturbation theory underlying 
Eq.~\eqref{eq:BOpert_scaling2D} does not depend on $L$ and is expressed in terms of the system parameters as $\hbar^2/(m_{LP} R^2 \Omega) \gg 1$ (or, equivalently $\hbar^2/(m_{LP} \Omega^{1/3} C_3^{2/3})\gg 1$) in the $\delta_C=0$ case: for the chosen values of the system parameters, the LHS is of the order of $100$, which guarantees the validity of the perturbative BO approach and the weakly interacting nature of two-dimensional dipolar polaritons in the resonant $\delta_C=0$ case. 

Regimes with significant polariton-polariton correlations can be achieved in other parameter regimes, e.g. by working in a strongly excitonic $\delta_C\gg \Omega$ regime where the polariton mass is much heavier and the system may recover the behaviour of bare dipolar excitons~\cite{LOZOVIK2007,Astra2007}. 
On the other hand, in the strongly photonic regime $\delta_C<0$ with $|\delta_C|\gg \Omega$, the perturbative BO theory used to obtain \eqref{eq:deltaC10/3} can be straightforwardly extended to the two dimensional case, which gives a peculiar scaling
\begin{equation}
    \Delta E_{\rm int}\propto\frac{\Omega^4 \, (C_3)^{2/3}}{|\delta_C|^{11/3}}
    \label{eq:deltaC11/3}
\end{equation}
for the interaction energy of the weakly interacting polariton system.

 \subsection{Contact-interacting polaritons}
Even though this work was focused on the case of dipolaritons arising from dipolar excitons, our theoretical approach can be extended to the case of non-dipolar excitons with only short-range interactions. A quantitatively accurate analysis would require a detailed knowledge of the exciton-exciton interaction potential, which combines a hard-core repulsive part of radius $a_{\rm exc}$ on the order of the exciton Bohr radius $a_B$, and an attractive tail due, in particular, to $1/r^6$ van der Waals interactions. 
Postponing a complete study to a future work, we assume here for simplicity that the hard-core repulsion dominates the polariton-polariton interactions and neglect the $1/r^6$ tail.

Within the BO framework developed in Sec.\ref{BOsec}, it is easy to see that the hard core potential between excitons translates into a BO energy landscape for a pair of colliding lower polaritons which is piecewise flat, the potential step for $r<a_{\rm exc}$ having a height $(2-\sqrt{2})\Omega$ in the $\delta_C=0$ resonant case. At the level of the perturbative BO approximation of \eqref{eq:BOpert_scaling} and \eqref{eq:BOpert_scaling2D}, this gives scalings of the form
\begin{eqnarray}
    \Delta E_{non-dip}=2\,(2-\sqrt{2})\Omega\, \frac{2\,a_{\rm exc}}{L}  \label{eq:DeltaEnon-dip1}\\
    \Delta E_{non-dip}=2\,(2-\sqrt{2})\Omega\, \frac{\pi a_{\rm exc}^2}{L^2}
        \label{eq:DeltaEnon-dip2}
\end{eqnarray}
in respectively $D=1,2$ spatial dimensions, where the factor $2$ in front accounts for the $\sin^2$ form of the ground state wavefunction. This formula again highlights the key role of the light-matter coupling $\Omega$, as also found in the diagrammatic calculation of~\cite{Li2021}.
As a main difference with respect to dipolaritons, the interaction strength is set here by the radius of the effective hard-core potential rather than the (typically larger) dipolar radius $R$.

The typical $\Omega\sim a_{\rm exc}^{-D/2}$ scaling of the exciton Rabi coupling with exciton radius~\cite{yu-cardona} translates via Eqs.~(\ref{eq:DeltaEnon-dip1}-\ref{eq:DeltaEnon-dip2}) into a $\Delta E_{non-dip}\propto a_{\rm exc}^{D/2}$
scaling of the interaction energy between polaritons. This confirms the experimentally established fact that Wannier excitons with a large Bohr radius display stronger interactions. This result is in stark contrast with our predictions for dipolar polaritons in Sec.\ref{subsec:resonant} where, for a given dipolar length $d$, interactions are stronger for tightly-bound dipolar excitons with stronger $\Omega$. 

Additional remarkable features are found concerning the dependence of the interaction strength on the exciton-photon detuning $\delta_C$, also discussed in~\cite{Li2021}. This can be evaluated using the same perturbative BO approach used in Sec.\ref{subsec:deltaC}.
On the excitonic $\delta_C>0$ side, the initial reinforcement of the interaction energy due to the larger excitonic weight predicted by the perturbative BO theory is eventually compensated at large $\delta_C$ by corrections to the perturbative BO theory due to the much heavier polariton mass deep in the excitonic regime. As in the one-dimensional dipolar case of Sec.\ref{subsec:deltaC}, we may conjecture that, for sufficiently large $L$, the interaction energy will drop at large $\delta_C$ to a constant and relatively small value proportional to the excitonic kinetic energy $\hbar^2/(m_X L^2)$. 

On the photonic $\delta_C<0$ side, the polariton mass tends to the photonic mass and we can make use of the perturbative BO theory. This predicts that the scaling \eqref{eq:deltaC10/3} of the interaction energy is modified by the fact that the radius of the potential bump is now fixed by the exciton radius, which resuls into a slightly slower way as $ \Delta E_{non-dip} \sim |\delta_C|^{-3}$ independently of the dimensionality $D=1,2$. 
It is interesting to note that this scaling differs again from the $|\delta_C|^{-4}$ scaling of the naive standard theory for repulsive exciton-exciton interactions, but coincides instead with the one predicted by this same theory for the nonlinearity stemming from the saturation of the excitonic oscillator strength, proportional to the $u_X^3 u_C$ Hopfield coefficients~\cite{Carusotto2013}. This coincidence with the saturation model appears natural when seen in terms of the BO energy landscapes: the repulsive hard-core potential results in fact in an effective decoupling of the XX state, closely analogous to the effect of saturation. 

\subsection{Several quantum wires/wells}\label{Sec:SeveralLayers}
Typical samples host several quantum wires/wells arranged within the cavity in a stacked geometry. In the case of non-dipolar polaritons, this is anticipated to give a reduction of the effective interaction by the number $N$ of wells/wires~\cite{Carusotto2013}. 

Within our formalism, having several $N>1$ wells/wires means that single excitation states are delocalized among the $N$ wells/wires,
\begin{equation}
|X\rangle=\frac{1}{\sqrt{N}} \sum_{j=1}^N |X_j\rangle
\label{eq:delocal}
\end{equation}
where $|X_j\rangle$ are the excitonic states in each well/wire and the delocalized excitation $|X\rangle$ has an $\sqrt{N}$ times larger effective Rabi frequency, $\Omega_N=\sqrt{N}\,\Omega$. 

If we restrict to the simplest case of a hard-core potential of radius $a_{\rm exc}$, the two-exciton state for $r<a_{\rm exc}$ is restricted to pairs of excitons sitting in different wells/wires,
\begin{equation}
|XX\rangle=\frac{1}{\sqrt{N(N-1)}} \sum_{j\neq j', j,j'=1}^N |X_j X_{j'}\rangle\,.
\end{equation}
As a result, the Hamiltonian matrix \eqref{eq:hBO} is replaced by 
\begin{equation}
    h_{BO}=\begin{pmatrix} 
    2\delta_C & \sqrt{2N} \Omega & 0 \\
    \sqrt{2N} \Omega & \delta_C & \sqrt{2(N-1)}\Omega \\
    0 & \sqrt{2(N-1)}\Omega & 0
    \end{pmatrix}
    \label{eq:hBON}
\end{equation}
where the effect of the exciton interactions has been included in the smaller value of the oscillator strength to the two-exciton state. A perturbative analytical diagonalization of \eqref{eq:hBON} in the limit of large $N$ gives 
\begin{equation}
    \Delta E_{non-dip}\simeq \frac{1}{2\sqrt{N}}\,\Omega\, \frac{2\pi a_{\rm exc}^2}{L^2}=\frac{\Omega_N}{N}\,\, \frac{\pi a_{\rm exc}^2}{L^2}
\end{equation}
in the two-dimensional case (a similar formula holds in the one-dimensional one).
Interestingly, this formula confirms the usual $1/N$ scaling at a given value of the effective Rabi frequency $\Omega_N$, that is the experimentally measured polariton splitting. However, given that $\Omega_N$ scales itself as $\sqrt{N}$, the actual reduction of the interaction strength predicted by our theory is only as $1/\sqrt{N}$, in qualitative agreement with~\cite{Bleu2020,Li2021}.

A different behaviour can arise from the spatially long-range shape of the dipolar interactions. Provided the QWs are located close enough along the growth direction, excitons interact  with approximately the same dipolar potential $V_{\rm dd}(r)$ independently on the specific well in which they are sitting. As a result, the delocalized excitations defined in \eqref{eq:delocal} will interact via $V_{\rm dd}(r)$, which brings us to the original BO Hamiltonian Eq.~\eqref{eq:hBO} albeit with the increased Rabi frequency $\Omega_N=\sqrt{N}\,\Omega$. Based on \eqref{eq:BOpert_scaling} and \eqref{eq:BOpert_scaling2D}, this results in a (slow) $N$-dependent increase of the interaction energy as $N^{(1-D/3)/2}$ for dipolaritons in respectively $D=1,2$ dimensional geometries.

\subsection{Dipolaritons stemming from coupled direct and indirect excitons}
\label{sec:IXDX}
Starting from the pioneering work~\cite{agranovich1998excitons}, coherent coupling of different species of excitons with different features has provided a successful strategy to combine seemingly contradictory properties such as a strong oscillator strength and a large exciton radius. In Ref.~\cite{Cristofolini2012}, a related idea was explored to combine a strong coupling to light and a sizable electric dipole moment by mixing direct (DX) and indirect (IX) excitons in a pair of quantum wells via inter-well tunneling processes. The resulting enhancement of polariton interactions was then experimentally reported by many authors~\cite{Byrnes2014, Rapaport2016, Tsintzos2018, Togan2018, Rosenberg2018dipolar,datta2022highly,Suarez:PRL2021,Liran:PRX2024}. 

While full numerical calculations in a one dimensional geometry are reported in App.\ref{app:DXIX}, in this Subsection we extend the perturbative BO approach of \eqref{eq:BOpert} to this configuration to obtain analytical insight on the interaction energy between such mixed polaritons. Instead of the $4\times4$ matrix given by Eq.~\eqref{eq:hBO4}, 
we need to consider a $9\times 9$ Hamiltonian where each particle can sit in 3 states, namely $\{C, DX, IX\}$. 
The BO Hamiltonian reads
\begin{widetext}
\begin{align}
\label{eq:hBO9} h_{BO}=\begin{pmatrix}
    2\delta_C & \Omega & 0 & \Omega & 0 & 0 & 0 & 0 & 0\\
    \Omega & \delta_C & J & 0 & \Omega & 0 &0 &0 & 0\\
    0 & J & \delta_C+\delta_{IX} & 0 & 0 & \Omega & 0 &0 & 0\\
    \Omega & 0 & 0 & \delta_C & \Omega &  0 & J & 0 & 0 \\
    0 & \Omega & 0 & \Omega & 0 & J & 0 & J &0 \\
0 & 0 & \Omega & 0 & J & \delta_{IX} & 0 & 0 & J \\    
    0& 0 & 0 & J & 0 & 0 & \delta_{C} + \delta_{IX} & 
    \Omega & 0 \\
    0 & 0 & 0 & 0 &J & 0 & \Omega & \delta_{IX} & J \\
    0 & 0 & 0 & 0 & 0 & J & 0 & J & 2 \delta_{IX}+ V_{\text{dd}}(r) 
    \end{pmatrix}
\end{align}
in the two-particle basis $\{ C,C;\, C,DX; \, C,IX; \, DX,C; \,DX,DX; \,DX,IX; \,IX,C; \,IX,DX; \,IX,IX\}$, 
\end{widetext}
where the detunings $\delta_C$ and $\delta_{IX}$ are measured from the direct DX exciton, the coupling $J$ between direct and indirect excitons stems from inter-well tunneling, and $\Omega$ is the usual Rabi frequency of the direct-exciton/photon coupling. For simplicity, we have neglected here the interactions between direct excitons and the ones between direct and indirect excitons. Diagonalization of \eqref{eq:hBO9} provides the BO energy landscapes, whose numerical integration offers a prediction of the effective interaction strength between polaritons. 

For instance, in the specific $J=\Omega$ case for which the relative photonic, direct-excitonic and indirect-excitonic amplitudes are proportional to $(1,\sqrt{2},1)$, one obtains for resonant $\delta_C=0$ detuning 
and two-dimensions the same scaling \eqref{eq:BOpert_scaling2D} as the simpler photon/indirect exciton case,
\begin{equation}
\Delta E_{\rm int}=a'_{2D} \Omega \, \frac{R^2}{L^2}=a'_{2D} \frac{\Omega^{1/3} C_3^{2/3}}{L^2}\,.
\label{eq:BOpert_scaling2Dbis}
\end{equation}
except for a slightly reduced value of the prefactor $a'_{2D}\sim 1.25$ as compared to $a_{2D}\sim 5$ found in Sec.\ref{subsec:2D}. This suppression can be qualitatively understood in connection with the reduced weight of the polariton on the dipolar-interacting indirect exciton component. 

\subsection{Summary}

In this and the previous Section, we have capitalized on the numerical results obtained for the one-dimensional geometry to build a general theory of polariton interactions that provides concrete predictions for a variety of different regimes. While diagrammatic theories have already hinted at some of these trends~\cite{Li2021}, in this work we have provided physical insight on the underlying mechanisms and well-defined scaling law to be compared to experiments.
\begin{itemize}
\item Comparison with the bare excitons shows the key role played by the coupling to light, mostly via the dramatically reduced mass of polaritons. The correspondingly increased kinetic energy prevents polaritons from forming a Tonks-Girardeau state where interactions would be suppressed by the reduced spatial overlap of particles. Indeed, it  instead stabilizes a weakly correlated polariton state with a strongly reinforced interaction energy compared to the bare exciton case. 

\item Even though our theory is based on interactions between the excitonic component, the scaling of the interaction strength on the photonic side follows a slower decay than the $|\delta_C|^{-4}$ decay of standard theory. In particular a $|\delta_C|^{-10/3}$ vs. $|\delta_C|^{-11/3}$  dependence is found for dipolaritons in one vs. two dimensions, while a $|\delta_C|^{-3}$ dependence is found for contact-interacting polaritons independently of the dimensionality $D=1,2$, analogous to the effect of a saturation of the excitonic oscillator strength.

\item For a given value of the single well/wire Rabi frequency $\Omega$, we predict a decrease of the interaction strength as $1/\sqrt{N}$ instead of the usual $1/N$ one for spatially well separated wells/wires. This behaviour changes into a $\sqrt{N}$ increase if the wells/wires are located very close in space, at distances smaller than the dipolar radius.

\item For given exciton properties, the polariton interaction strength depends on the photonic environment via the Rabi frequency $\Omega$ according to a power-law with a configuration-dependent exponent, namely $\Omega$ for contact-interacting polaritons and $\Omega^{2/3}$ ($\Omega^{1/3}$) for dipolaritons in one- (two-)dimensions. This feature could be experimentally probed by placing the same quantum well
structure in cavities of different thickness $L_z$ , so to vary
$\Omega \sim \sqrt{1/L_z}$ while keeping the internal structure of the
exciton (i.e. $E_B$ and $a_B$) constant. 

\end{itemize}
We anticipate that experimental verification of these trends is possible in state-of-the-art samples. While this will provide the ultimate verification of our model, in the next Section we will provide some preliminary comparisons of our predictions with experimental data that are already available in the literature.

\section{Comparison with experiments}
\label{quantitativeSec}

In this final Section we provide a first validation of our framework by critically comparing its predictions with the available experimental data. We also sketch the promising perspectives of our work in view of achieving the polariton blockade regime.

\subsection{Non-dipolar polaritons}

Even though no specific studies of the scaling of the polariton interaction strength across different materials are available in the literature, 
the general, well-established trend is that interactions are stronger for spatially extended Wannier excitons in inorganic-semiconductor-based devices~\cite{Estrecho2019} than for tightly-bound ones in, e.g., two-dimensional materials~\cite{barachati2018interacting} or perovskites~\cite{fieramosca2019two}. As we have mentioned above, this basic feature is well accounted for by our theory of Eqs.(\ref{eq:DeltaEnon-dip1}-\ref{eq:DeltaEnon-dip2}).

Using typical values of inorganic semiconductor materials $\Omega=4\,$meV and taking as an educated guess for the effective hard-core potential radius $a_{\rm exc}$ a value equal to the exciton Bohr radius $a_B=10\,nm$, the overall value of the interaction constant extracted from \eqref{eq:DeltaEnon-dip2} turns out to be on the order of $g_{non-dip}=\Delta E_{non-dip}\,(L^2/2)\sim 1\,\mu\textrm{eV}\,\mu\textrm{m}^2$, well within the range of available experimental data~\cite{Estrecho2019}. Also for two-dimensional materials, using the reported values for the light-matter coupling $\Omega=21\,$meV and Bohr radius $a_B=1.7\,$nm one gets a value on the order of $g_{non-dip}\sim 0.1\,\mu\textrm{eV}\,\mu\textrm{m}^2$ for the interaction constant which, in contrast to the standard perturbative theory, is in rough agreement with the experiment~\cite{barachati2018interacting}.

Whereas many other discussions assume the $|u_X|^4$ scaling of the standard theory~\cite{Carusotto2013}, a central role of saturation effects in polariton-polariton interactions was highlighted by a recent experiment in polariton fluids~\cite{frerot2023bogoliubov}. A reanalysis of the points in Fig.3a of Ref.~\cite{Togan2018} may also suggest a similarly slower dependence on $u_X$ and indicate the importance of saturation effects. Experimental work to put this key issue on solid grounds is presently on-going, including an additional study of interactions between lower and upper polaritons that have not been covered here.

Besides this peculiar dependence on $\delta_C$, it is important to note that our theory also predicts a dependence of the interaction strength on the electromagnetic environment via the $\Omega$ factor. Such a dependence was absent in the standard perturbative theory~\cite{Ciuti1998} where the interaction energy only depends on intrinsic properties of the exciton such as the binding energy $E_B$ and Bohr radius $a_B$. 
In general, the more complex dependence of the polariton interaction strength on the system parameters anticipated by our theory could be responsible for the large spread of experimentally reported values~\cite{Estrecho2019}.

\subsection{Dipolar polaritons}
Experimental investigation of the reinforced interactions between dipolaritons were recently reported by several authors~\cite{Tsintzos2018, Togan2018, Rosenberg2018dipolar,datta2022highly,Suarez:PRL2021,Liran:PRX2024}.
In particular, a quantitative comparison with non-dipolar polaritons was carried out in~\cite{Togan2018} by varying the amplitude of the static electric field used to induce the electric dipole, finding an increase of the effective interaction strength by a factor around $5$.

It is interesting to see whether our theory is able to reproduce such an observation. Since in the experiment the Rabi frequency per quantum well $\Omega=1.33\,$meV 
is not far from the tunneling $J=1.75\,$meV (note the different convention used in \cite{Togan2018}), we can approximately use our theoretical prediction \eqref{eq:BOpert_scaling2Dbis}. The large value of the electron-hole separation $d\sim 21\,$nm gives via \eqref{eq:dipolarlength} a large dipolar radius $R\sim 36\,$nm. Inserting this value of $R$ and approximating once again the hard-core potential radius $a_{\rm exc}$ with the Bohr radius $a_B=10\,$nm, 
the dipolar enhancement  
\begin{equation}
\eta_{\rm dip}=    \frac{a'_{2D}}{2(2-\sqrt{2})} \frac{R^2}{\pi a_B^2}
\label{eq:etadip}
\end{equation}
set by the ratio of \eqref{eq:BOpert_scaling2Dbis} to \eqref{eq:DeltaEnon-dip2} gives a value around $5$. Our theory thus 
roughly recovers the dipolariton enhancement factor observed in the experiment~\cite{Togan2018}. 
Another recent experiment based on a bilayer of two-dimensional materials~\cite{datta2022highly} has reported a similar value around $\eta_{\rm dip}\simeq 10$.
On the other hand, our model does not predict the quite larger enhancement claimed in~\cite{Rosenberg2018dipolar}, which could stem from other effects beyond our simplified model of rigid dipolar excitons~\cite{Liran:PRX2024}. 


The expression \eqref{eq:etadip} might suggest that the dipolar enhancement factor can reach very high values for tightly-bound excitons with small $a_B$.
To properly assess the actual limitations of this enhancement strategy, it is useful to rewrite \eqref{eq:etadip} in terms of the dipolar length $d$ and the exciton radius $a_B$. This gives a scaling $\eta_{\rm dip}$ proportional to $(d/a_B)^{4/3}$. As the very different Coulomb attraction between electron and holes sitting on the same or different wells may give very different shapes for the direct and indirect excitons, we anticipate that the dipolar enhancement factor is limited by the maximum electron-hole separation distances $d$ up to which tunneling between direct and indirect excitons can remain efficient, a quantity that can hardly be orders of magnitude larger than the exciton radius $a_B$.

\subsection{Towards polariton blockade}

Fig.~\ref{DEvsL_excgs} suggests that the TG regime is hardly observable for polaritons in state-of-the-art wire devices. Realistic state-of-art limits on linewidth and polariton confinement are indicated in the figure by the dashed arrow, in particular the polariton linewidth $\gamma$ is typically on the $\mu$eV range at the very best~\cite{SunY2017}. The situation would of course be different for a system of bare excitons (or strongly excitonic polaritons), where the exciton lifetime can be orders of magnitude longer than the typical polariton one. Still, given the much heavier value of the exciton mass, difficulties will likely arise from the reduced characteristic kinetic energy scale (orange line in Fig.~\ref{DEvsL_excgs}), to be compared to the finite temperature of the system.

On the contrary, the same Fig.~\ref{DEvsL_excgs} suggests that an effective polariton blockade $\Delta E>\gamma$ is within reach of state-of-the-art experiments. In addition to configurations similar to the pioneering works~\cite{Munoz-Matutano2019,Delteil2019}, a promising alternative consists of the transversally confined one-dimensional wire geometries as in the experiment~\cite{Thureja2022}, with a $\mu$m-range length. 

\section{Conclusions and outlook}\label{Concsec}
In this article we have developed a microscopic model of the interactions between polaritons in solid-state photonic devices. Starting from exact numerical calculations for dipolar polaritons in a one-dimensional
geometry, we have built a general theory based on a Born-Oppenheimer approximation that provides predictions for different geometries and different forms of the interaction potentials. 
These predictions are amenable to experimental verification using state-of-the-art photonic devices as well as to comparison with fully microscopic calculations~\cite{Li2021}, and suggest promising strategies towards observing a full-fledged polariton blockade effect in a semiconductor device.

In particular, our theory highlights 
the key role of coupling to light in reinforcing interactions as compared to bare excitons. 
Whereas available experimental data are recovered, new peculiar scalings of the interaction strength on the system parameters are predicted. These scalings differ in a marked way from the predictions of standard theories and may contribute to explaining the spread that still exists in the experimental values of the polariton interaction constants~\cite{Estrecho2019}.

Furthermore, the generality of our framework based on Born-Oppenheimer energy landscapes makes it applicable to a wide range of physical systems. Motivated by on-going experiments, straightforward extensions of the theory will include interactions between lower and upper polaritons and between polaritons and dark excitons, as well as field-induced (rather than rigid) dipolar excitons~\cite{Liran:PRX2024}. Longer-term work will then address application to more exotic systems of Rydberg polaritons in atomic gases~\cite{peyronel2012quantum,Gorshkov2011,firstenberg2013attractive} or solid-state materials~\cite{Walther2018,Orfanakis2022}, that are presently attracting a great interest as an alternative platform for strongly interacting polaritons. To this purpose, a promising model is provided by the three-state theory discussed in Sec.\ref{sec:IXDX}: the interaction potential between the Rydberg-atom component of the polariton corresponds in our language to the interaction potential between indirect excitons, the intermediate atomic state corresponds to the direct exciton, and the coherent optical coupling corresponds to the tunneling. 


On the long term, our results will provide useful guidance towards designing systems with reinforced polariton-polariton interactions so to experimentally realize new many-body states of the polariton fluid. Such a result would open a new era in the study of strongly correlated fluids of light using state-of-the-art semiconductor optics technology.

\section*{Acknowledgments} We acknowledge financial support from
 the Danish National Research Foundation through the Center of Excellence “CCQ” (Grant agreement no.: DNRF156) and the U.S. Army CCDC Atlantic Basic and Applied Research via Grant No. W911NF-19-1-0403. The work at ETH Zurich was supported
by the Swiss National Science Foundation (SNSF) under Grant Number 200020\_207520. I. C. acknowledges financial support from the European Union H2020-FETFLAG-2018-2020 project ``PhoQuS'' (n.820392), from the Provincia Autonoma di Trento, from the Q@TN initiative, and from the National Quantum Science and Technology Institute through the PNRR MUR Project under Grant PE0000023-NQSTI, co-funded by the European Union - NextGeneration EU.  A. C. G. acknowledges financial support from Grant UNAM DGAPA PAPIIT No. IN108620, PAPIIT No. IA101923, and PIIF23. 
We also thank Deepnakur Thureja, Puneet Murty, Daniele De Bernardis and Thomas Pohl for many discussions.

During revision of this manuscript, theoretical work extending our calculations in the many-body direction was reported~\cite{knorzer2024fermionization} as well as experimental advances in the direction of tight in-plane confinement of excitons in two-dimensional materials~\cite{hu2024quantum,thureja2024electrically}. 
\bibliography{references}

\begin{thebibliography}{77}%
\makeatletter
\providecommand \@ifxundefined [1]{%
 \@ifx{#1\undefined}
}%
\providecommand \@ifnum [1]{%
 \ifnum #1\expandafter \@firstoftwo
 \else \expandafter \@secondoftwo
 \fi
}%
\providecommand \@ifx [1]{%
 \ifx #1\expandafter \@firstoftwo
 \else \expandafter \@secondoftwo
 \fi
}%
\providecommand \natexlab [1]{#1}%
\providecommand \enquote  [1]{``#1''}%
\providecommand \bibnamefont  [1]{#1}%
\providecommand \bibfnamefont [1]{#1}%
\providecommand \citenamefont [1]{#1}%
\providecommand \href@noop [0]{\@secondoftwo}%
\providecommand \href [0]{\begingroup \@sanitize@url \@href}%
\providecommand \@href[1]{\@@startlink{#1}\@@href}%
\providecommand \@@href[1]{\endgroup#1\@@endlink}%
\providecommand \@sanitize@url [0]{\catcode `\\12\catcode `\$12\catcode
  `\&12\catcode `\#12\catcode `\^12\catcode `\_12\catcode `\%12\relax}%
\providecommand \@@startlink[1]{}%
\providecommand \@@endlink[0]{}%
\providecommand \url  [0]{\begingroup\@sanitize@url \@url }%
\providecommand \@url [1]{\endgroup\@href {#1}{\urlprefix }}%
\providecommand \urlprefix  [0]{URL }%
\providecommand \Eprint [0]{\href }%
\providecommand \doibase [0]{http://dx.doi.org/}%
\providecommand \selectlanguage [0]{\@gobble}%
\providecommand \bibinfo  [0]{\@secondoftwo}%
\providecommand \bibfield  [0]{\@secondoftwo}%
\providecommand \translation [1]{[#1]}%
\providecommand \BibitemOpen [0]{}%
\providecommand \bibitemStop [0]{}%
\providecommand \bibitemNoStop [0]{.\EOS\space}%
\providecommand \EOS [0]{\spacefactor3000\relax}%
\providecommand \BibitemShut  [1]{\csname bibitem#1\endcsname}%
\let\auto@bib@innerbib\@empty
\bibitem [{\citenamefont {Chang}\ \emph {et~al.}(2014)\citenamefont {Chang},
  \citenamefont {Vuleti{\'{c}}},\ and\ \citenamefont {Lukin}}]{Chang2014}%
  \BibitemOpen
  \bibfield  {author} {\bibinfo {author} {\bibfnamefont {Darrick~E.}\
  \bibnamefont {Chang}}, \bibinfo {author} {\bibfnamefont {Vladan}\
  \bibnamefont {Vuleti{\'{c}}}}, \ and\ \bibinfo {author} {\bibfnamefont
  {Mikhail~D.}\ \bibnamefont {Lukin}},\ }\bibfield  {title} {\enquote {\bibinfo
  {title} {Quantum nonlinear optics -- photon by photon},}\ }\href {\doibase
  10.1038/nphoton.2014.192} {\bibfield  {journal} {\bibinfo  {journal} {Nature
  Photonics}\ }\textbf {\bibinfo {volume} {8}},\ \bibinfo {pages} {685--694}
  (\bibinfo {year} {2014})}\BibitemShut {NoStop}%
\bibitem [{\citenamefont {Kurizki}\ \emph {et~al.}(2015)\citenamefont
  {Kurizki}, \citenamefont {Bertet}, \citenamefont {Kubo}, \citenamefont
  {M{\o}lmer}, \citenamefont {Petrosyan}, \citenamefont {Rabl},\ and\
  \citenamefont {Schmiedmayer}}]{Kurizki2015}%
  \BibitemOpen
  \bibfield  {author} {\bibinfo {author} {\bibfnamefont {Gershon}\ \bibnamefont
  {Kurizki}}, \bibinfo {author} {\bibfnamefont {Patrice}\ \bibnamefont
  {Bertet}}, \bibinfo {author} {\bibfnamefont {Yuimaru}\ \bibnamefont {Kubo}},
  \bibinfo {author} {\bibfnamefont {Klaus}\ \bibnamefont {M{\o}lmer}}, \bibinfo
  {author} {\bibfnamefont {David}\ \bibnamefont {Petrosyan}}, \bibinfo {author}
  {\bibfnamefont {Peter}\ \bibnamefont {Rabl}}, \ and\ \bibinfo {author}
  {\bibfnamefont {Jörg}\ \bibnamefont {Schmiedmayer}},\ }\bibfield  {title}
  {\enquote {\bibinfo {title} {Quantum technologies with hybrid systems},}\
  }\href {\doibase 10.1073/pnas.1419326112} {\bibfield  {journal} {\bibinfo
  {journal} {Proceedings of the National Academy of Sciences}\ }\textbf
  {\bibinfo {volume} {112}},\ \bibinfo {pages} {3866--3873} (\bibinfo {year}
  {2015})}\BibitemShut {NoStop}%
\bibitem [{\citenamefont {Weisbuch}\ \emph {et~al.}(1992)\citenamefont
  {Weisbuch}, \citenamefont {Nishioka}, \citenamefont {Ishikawa},\ and\
  \citenamefont {Arakawa}}]{Weisbuch1992}%
  \BibitemOpen
  \bibfield  {author} {\bibinfo {author} {\bibfnamefont {C.}~\bibnamefont
  {Weisbuch}}, \bibinfo {author} {\bibfnamefont {M.}~\bibnamefont {Nishioka}},
  \bibinfo {author} {\bibfnamefont {A.}~\bibnamefont {Ishikawa}}, \ and\
  \bibinfo {author} {\bibfnamefont {Y.}~\bibnamefont {Arakawa}},\ }\bibfield
  {title} {\enquote {\bibinfo {title} {Observation of the coupled
  exciton-photon mode splitting in a semiconductor quantum microcavity},}\
  }\href {\doibase 10.1103/PhysRevLett.69.3314} {\bibfield  {journal} {\bibinfo
   {journal} {Phys. Rev. Lett.}\ }\textbf {\bibinfo {volume} {69}},\ \bibinfo
  {pages} {3314--3317} (\bibinfo {year} {1992})}\BibitemShut {NoStop}%
\bibitem [{\citenamefont {Kavokin}\ and\ \citenamefont
  {Malpuech}(2003)}]{kavokin2003cavity}%
  \BibitemOpen
  \bibfield  {author} {\bibinfo {author} {\bibfnamefont {Alexey}\ \bibnamefont
  {Kavokin}}\ and\ \bibinfo {author} {\bibfnamefont {Guillaume}\ \bibnamefont
  {Malpuech}},\ }\href@noop {} {\emph {\bibinfo {title} {Cavity polaritons}}}\
  (\bibinfo  {publisher} {Elsevier},\ \bibinfo {year} {2003})\BibitemShut
  {NoStop}%
\bibitem [{\citenamefont {Burstein}\ and\ \citenamefont
  {Weisbuch}(2012)}]{burstein2012confined}%
  \BibitemOpen
  \bibfield  {author} {\bibinfo {author} {\bibfnamefont {Elias}\ \bibnamefont
  {Burstein}}\ and\ \bibinfo {author} {\bibfnamefont {Claude}\ \bibnamefont
  {Weisbuch}},\ }\href@noop {} {\emph {\bibinfo {title} {Confined electrons and
  photons: New physics and applications}}},\ Vol.\ \bibinfo {volume} {340}\
  (\bibinfo  {publisher} {Springer Science \& Business Media},\ \bibinfo {year}
  {2012})\BibitemShut {NoStop}%
\bibitem [{\citenamefont {Bloch}\ \emph {et~al.}(2022)\citenamefont {Bloch},
  \citenamefont {Carusotto},\ and\ \citenamefont {Wouters}}]{Bloch2022}%
  \BibitemOpen
  \bibfield  {author} {\bibinfo {author} {\bibfnamefont {Jacqueline}\
  \bibnamefont {Bloch}}, \bibinfo {author} {\bibfnamefont {Iacopo}\
  \bibnamefont {Carusotto}}, \ and\ \bibinfo {author} {\bibfnamefont {Michiel}\
  \bibnamefont {Wouters}},\ }\bibfield  {title} {\enquote {\bibinfo {title}
  {Non-equilibrium bose--einstein condensation in photonic systems},}\ }\href
  {\doibase 10.1038/s42254-022-00464-0} {\bibfield  {journal} {\bibinfo
  {journal} {Nature Reviews Physics}\ }\textbf {\bibinfo {volume} {4}},\
  \bibinfo {pages} {470--488} (\bibinfo {year} {2022})}\BibitemShut {NoStop}%
\bibitem [{\citenamefont {Fontaine}\ \emph {et~al.}(2022)\citenamefont
  {Fontaine}, \citenamefont {Squizzato}, \citenamefont {Baboux}, \citenamefont
  {Amelio}, \citenamefont {Lema{\^i}tre}, \citenamefont {Morassi},
  \citenamefont {Sagnes}, \citenamefont {Le~Gratiet}, \citenamefont {Harouri},
  \citenamefont {Wouters}, \citenamefont {Carusotto}, \citenamefont {Amo},
  \citenamefont {Richard}, \citenamefont {Minguzzi}, \citenamefont {Canet},
  \citenamefont {Ravets},\ and\ \citenamefont {Bloch}}]{KPZ}%
  \BibitemOpen
  \bibfield  {author} {\bibinfo {author} {\bibfnamefont {Quentin}\ \bibnamefont
  {Fontaine}}, \bibinfo {author} {\bibfnamefont {Davide}\ \bibnamefont
  {Squizzato}}, \bibinfo {author} {\bibfnamefont {Florent}\ \bibnamefont
  {Baboux}}, \bibinfo {author} {\bibfnamefont {Ivan}\ \bibnamefont {Amelio}},
  \bibinfo {author} {\bibfnamefont {Aristide}\ \bibnamefont {Lema{\^i}tre}},
  \bibinfo {author} {\bibfnamefont {Martina}\ \bibnamefont {Morassi}}, \bibinfo
  {author} {\bibfnamefont {Isabelle}\ \bibnamefont {Sagnes}}, \bibinfo {author}
  {\bibfnamefont {Luc}\ \bibnamefont {Le~Gratiet}}, \bibinfo {author}
  {\bibfnamefont {Abdelmounaim}\ \bibnamefont {Harouri}}, \bibinfo {author}
  {\bibfnamefont {Michiel}\ \bibnamefont {Wouters}}, \bibinfo {author}
  {\bibfnamefont {Iacopo}\ \bibnamefont {Carusotto}}, \bibinfo {author}
  {\bibfnamefont {Alberto}\ \bibnamefont {Amo}}, \bibinfo {author}
  {\bibfnamefont {Maxime}\ \bibnamefont {Richard}}, \bibinfo {author}
  {\bibfnamefont {Anna}\ \bibnamefont {Minguzzi}}, \bibinfo {author}
  {\bibfnamefont {L{\'e}onie}\ \bibnamefont {Canet}}, \bibinfo {author}
  {\bibfnamefont {Sylvain}\ \bibnamefont {Ravets}}, \ and\ \bibinfo {author}
  {\bibfnamefont {Jacqueline}\ \bibnamefont {Bloch}},\ }\bibfield  {title}
  {\enquote {\bibinfo {title} {Kardar--parisi--zhang universality in a
  one-dimensional polariton condensate},}\ }\href {\doibase
  10.1038/s41586-022-05001-8} {\bibfield  {journal} {\bibinfo  {journal}
  {Nature}\ }\textbf {\bibinfo {volume} {608}},\ \bibinfo {pages} {687--691}
  (\bibinfo {year} {2022})}\BibitemShut {NoStop}%
\bibitem [{\citenamefont {Deng}\ \emph {et~al.}(2010)\citenamefont {Deng},
  \citenamefont {Haug},\ and\ \citenamefont {Yamamoto}}]{Deng2010}%
  \BibitemOpen
  \bibfield  {author} {\bibinfo {author} {\bibfnamefont {Hui}\ \bibnamefont
  {Deng}}, \bibinfo {author} {\bibfnamefont {Hartmut}\ \bibnamefont {Haug}}, \
  and\ \bibinfo {author} {\bibfnamefont {Yoshihisa}\ \bibnamefont {Yamamoto}},\
  }\bibfield  {title} {\enquote {\bibinfo {title} {Exciton-polariton
  bose-einstein condensation},}\ }\href {\doibase 10.1103/RevModPhys.82.1489}
  {\bibfield  {journal} {\bibinfo  {journal} {Rev. Mod. Phys.}\ }\textbf
  {\bibinfo {volume} {82}},\ \bibinfo {pages} {1489--1537} (\bibinfo {year}
  {2010})}\BibitemShut {NoStop}%
\bibitem [{\citenamefont {Carusotto}\ and\ \citenamefont
  {Ciuti}(2013)}]{Carusotto2013}%
  \BibitemOpen
  \bibfield  {author} {\bibinfo {author} {\bibfnamefont {Iacopo}\ \bibnamefont
  {Carusotto}}\ and\ \bibinfo {author} {\bibfnamefont {Cristiano}\ \bibnamefont
  {Ciuti}},\ }\bibfield  {title} {\enquote {\bibinfo {title} {Quantum fluids of
  light},}\ }\href {\doibase 10.1103/RevModPhys.85.299} {\bibfield  {journal}
  {\bibinfo  {journal} {Rev. Mod. Phys.}\ }\textbf {\bibinfo {volume} {85}},\
  \bibinfo {pages} {299--366} (\bibinfo {year} {2013})}\BibitemShut {NoStop}%
\bibitem [{\citenamefont {Amo}\ \emph {et~al.}(2009)\citenamefont {Amo},
  \citenamefont {Lefr\`ere}, \citenamefont {Pigeon}, \citenamefont {Adrados},
  \citenamefont {Ciuti}, \citenamefont {Carusotto}, \citenamefont {Houdr\'e},
  \citenamefont {Giacobino},\ and\ \citenamefont {Bramati}}]{Amo2009}%
  \BibitemOpen
  \bibfield  {author} {\bibinfo {author} {\bibfnamefont {Alberto}\ \bibnamefont
  {Amo}}, \bibinfo {author} {\bibfnamefont {J\'er\^ome}\ \bibnamefont
  {Lefr\`ere}}, \bibinfo {author} {\bibfnamefont {Simon}\ \bibnamefont
  {Pigeon}}, \bibinfo {author} {\bibfnamefont {Claire}\ \bibnamefont
  {Adrados}}, \bibinfo {author} {\bibfnamefont {Cristiano}\ \bibnamefont
  {Ciuti}}, \bibinfo {author} {\bibfnamefont {Iacopo}\ \bibnamefont
  {Carusotto}}, \bibinfo {author} {\bibfnamefont {Romuald}\ \bibnamefont
  {Houdr\'e}}, \bibinfo {author} {\bibfnamefont {Elisabeth}\ \bibnamefont
  {Giacobino}}, \ and\ \bibinfo {author} {\bibfnamefont {Alberto}\ \bibnamefont
  {Bramati}},\ }\bibfield  {title} {\enquote {\bibinfo {title} {Superfluidity
  of polaritons in semiconductor microcavities},}\ }\href
  {https://doi.org/10.1038/nphys1364} {\bibfield  {journal} {\bibinfo
  {journal} {Nature Physics}\ }\textbf {\bibinfo {volume} {5}},\ \bibinfo
  {pages} {805 EP --} (\bibinfo {year} {2009})}\BibitemShut {NoStop}%
\bibitem [{\citenamefont {Ciuti}\ \emph {et~al.}(1998)\citenamefont {Ciuti},
  \citenamefont {Savona}, \citenamefont {Piermarocchi}, \citenamefont
  {Quattropani},\ and\ \citenamefont {Schwendimann}}]{Ciuti1998}%
  \BibitemOpen
  \bibfield  {author} {\bibinfo {author} {\bibfnamefont {C.}~\bibnamefont
  {Ciuti}}, \bibinfo {author} {\bibfnamefont {V.}~\bibnamefont {Savona}},
  \bibinfo {author} {\bibfnamefont {C.}~\bibnamefont {Piermarocchi}}, \bibinfo
  {author} {\bibfnamefont {A.}~\bibnamefont {Quattropani}}, \ and\ \bibinfo
  {author} {\bibfnamefont {P.}~\bibnamefont {Schwendimann}},\ }\bibfield
  {title} {\enquote {\bibinfo {title} {Role of the exchange of carriers in
  elastic exciton-exciton scattering in quantum wells},}\ }\href {\doibase
  10.1103/PhysRevB.58.7926} {\bibfield  {journal} {\bibinfo  {journal} {Phys.
  Rev. B}\ }\textbf {\bibinfo {volume} {58}},\ \bibinfo {pages} {7926--7933}
  (\bibinfo {year} {1998})}\BibitemShut {NoStop}%
\bibitem [{\citenamefont {Glazov}\ \emph {et~al.}(2009)\citenamefont {Glazov},
  \citenamefont {Ouerdane}, \citenamefont {Pilozzi}, \citenamefont {Malpuech},
  \citenamefont {Kavokin},\ and\ \citenamefont {D'Andrea}}]{Glazov2009}%
  \BibitemOpen
  \bibfield  {author} {\bibinfo {author} {\bibfnamefont {M.~M.}\ \bibnamefont
  {Glazov}}, \bibinfo {author} {\bibfnamefont {H.}~\bibnamefont {Ouerdane}},
  \bibinfo {author} {\bibfnamefont {L.}~\bibnamefont {Pilozzi}}, \bibinfo
  {author} {\bibfnamefont {G.}~\bibnamefont {Malpuech}}, \bibinfo {author}
  {\bibfnamefont {A.~V.}\ \bibnamefont {Kavokin}}, \ and\ \bibinfo {author}
  {\bibfnamefont {A.}~\bibnamefont {D'Andrea}},\ }\bibfield  {title} {\enquote
  {\bibinfo {title} {Polariton-polariton scattering in microcavities: A
  microscopic theory},}\ }\href {\doibase 10.1103/PhysRevB.80.155306}
  {\bibfield  {journal} {\bibinfo  {journal} {Phys. Rev. B}\ }\textbf {\bibinfo
  {volume} {80}},\ \bibinfo {pages} {155306} (\bibinfo {year}
  {2009})}\BibitemShut {NoStop}%
\bibitem [{\citenamefont {Wouters}(2007)}]{Wouters07}%
  \BibitemOpen
  \bibfield  {author} {\bibinfo {author} {\bibfnamefont {Michiel}\ \bibnamefont
  {Wouters}},\ }\bibfield  {title} {\enquote {\bibinfo {title} {Resonant
  polariton-polariton scattering in semiconductor microcavities},}\ }\href
  {\doibase 10.1103/PhysRevB.76.045319} {\bibfield  {journal} {\bibinfo
  {journal} {Phys. Rev. B}\ }\textbf {\bibinfo {volume} {76}},\ \bibinfo
  {pages} {045319} (\bibinfo {year} {2007})}\BibitemShut {NoStop}%
\bibitem [{\citenamefont {Carusotto}\ \emph {et~al.}(2010)\citenamefont
  {Carusotto}, \citenamefont {Volz},\ and\ \citenamefont
  {Imamo{\u{g}}lu}}]{Carusotto_2010}%
  \BibitemOpen
  \bibfield  {author} {\bibinfo {author} {\bibfnamefont {I.}~\bibnamefont
  {Carusotto}}, \bibinfo {author} {\bibfnamefont {T.}~\bibnamefont {Volz}}, \
  and\ \bibinfo {author} {\bibfnamefont {A.}~\bibnamefont {Imamo{\u{g}}lu}},\
  }\bibfield  {title} {\enquote {\bibinfo {title} {Feshbach blockade:
  Single-photon nonlinear optics using resonantly enhanced cavity polariton
  scattering from biexciton states},}\ }\href {\doibase
  10.1209/0295-5075/90/37001} {\bibfield  {journal} {\bibinfo  {journal} {{EPL}
  (Europhysics Letters)}\ }\textbf {\bibinfo {volume} {90}},\ \bibinfo {pages}
  {37001} (\bibinfo {year} {2010})}\BibitemShut {NoStop}%
\bibitem [{\citenamefont {Hu}\ \emph {et~al.}(2020)\citenamefont {Hu},
  \citenamefont {Deng},\ and\ \citenamefont {Liu}}]{Hu2020}%
  \BibitemOpen
  \bibfield  {author} {\bibinfo {author} {\bibfnamefont {Hui}\ \bibnamefont
  {Hu}}, \bibinfo {author} {\bibfnamefont {Hui}\ \bibnamefont {Deng}}, \ and\
  \bibinfo {author} {\bibfnamefont {Xia-Ji}\ \bibnamefont {Liu}},\ }\bibfield
  {title} {\enquote {\bibinfo {title} {Polariton-polariton interaction beyond
  the born approximation: A toy model study},}\ }\href {\doibase
  10.1103/PhysRevA.102.063305} {\bibfield  {journal} {\bibinfo  {journal}
  {Phys. Rev. A}\ }\textbf {\bibinfo {volume} {102}},\ \bibinfo {pages}
  {063305} (\bibinfo {year} {2020})}\BibitemShut {NoStop}%
\bibitem [{\citenamefont {Bleu}\ \emph {et~al.}(2020)\citenamefont {Bleu},
  \citenamefont {Li}, \citenamefont {Levinsen},\ and\ \citenamefont
  {Parish}}]{Bleu2020}%
  \BibitemOpen
  \bibfield  {author} {\bibinfo {author} {\bibfnamefont {Olivier}\ \bibnamefont
  {Bleu}}, \bibinfo {author} {\bibfnamefont {Guangyao}\ \bibnamefont {Li}},
  \bibinfo {author} {\bibfnamefont {Jesper}\ \bibnamefont {Levinsen}}, \ and\
  \bibinfo {author} {\bibfnamefont {Meera~M.}\ \bibnamefont {Parish}},\
  }\bibfield  {title} {\enquote {\bibinfo {title} {Polariton interactions in
  microcavities with atomically thin semiconductor layers},}\ }\href {\doibase
  10.1103/PhysRevResearch.2.043185} {\bibfield  {journal} {\bibinfo  {journal}
  {Phys. Rev. Research}\ }\textbf {\bibinfo {volume} {2}},\ \bibinfo {pages}
  {043185} (\bibinfo {year} {2020})}\BibitemShut {NoStop}%
\bibitem [{\citenamefont {Li}\ \emph {et~al.}(2021)\citenamefont {Li},
  \citenamefont {Parish},\ and\ \citenamefont {Levinsen}}]{Li2021}%
  \BibitemOpen
  \bibfield  {author} {\bibinfo {author} {\bibfnamefont {Guangyao}\
  \bibnamefont {Li}}, \bibinfo {author} {\bibfnamefont {Meera~M.}\ \bibnamefont
  {Parish}}, \ and\ \bibinfo {author} {\bibfnamefont {Jesper}\ \bibnamefont
  {Levinsen}},\ }\bibfield  {title} {\enquote {\bibinfo {title} {Microscopic
  calculation of polariton scattering in semiconductor microcavities},}\ }\href
  {\doibase 10.1103/PhysRevB.104.245404} {\bibfield  {journal} {\bibinfo
  {journal} {Phys. Rev. B}\ }\textbf {\bibinfo {volume} {104}},\ \bibinfo
  {pages} {245404} (\bibinfo {year} {2021})}\BibitemShut {NoStop}%
\bibitem [{\citenamefont {Camacho-Guardian}\ \emph {et~al.}(2021)\citenamefont
  {Camacho-Guardian}, \citenamefont {Bastarrachea-Magnani},\ and\ \citenamefont
  {Bruun}}]{Camacho-Guardian2021}%
  \BibitemOpen
  \bibfield  {author} {\bibinfo {author} {\bibfnamefont {A.}~\bibnamefont
  {Camacho-Guardian}}, \bibinfo {author} {\bibfnamefont {M.~A.}\ \bibnamefont
  {Bastarrachea-Magnani}}, \ and\ \bibinfo {author} {\bibfnamefont {G.~M.}\
  \bibnamefont {Bruun}},\ }\bibfield  {title} {\enquote {\bibinfo {title}
  {Mediated interactions and photon bound states in an exciton-polariton
  mixture},}\ }\href {\doibase 10.1103/PhysRevLett.126.017401} {\bibfield
  {journal} {\bibinfo  {journal} {Phys. Rev. Lett.}\ }\textbf {\bibinfo
  {volume} {126}},\ \bibinfo {pages} {017401} (\bibinfo {year}
  {2021})}\BibitemShut {NoStop}%
\bibitem [{\citenamefont {Bastarrachea-Magnani}\ \emph
  {et~al.}(2021)\citenamefont {Bastarrachea-Magnani}, \citenamefont
  {Camacho-Guardian},\ and\ \citenamefont {Bruun}}]{Bastarrachea-Magnani2021}%
  \BibitemOpen
  \bibfield  {author} {\bibinfo {author} {\bibfnamefont {Miguel~A.}\
  \bibnamefont {Bastarrachea-Magnani}}, \bibinfo {author} {\bibfnamefont
  {Arturo}\ \bibnamefont {Camacho-Guardian}}, \ and\ \bibinfo {author}
  {\bibfnamefont {Georg~M.}\ \bibnamefont {Bruun}},\ }\bibfield  {title}
  {\enquote {\bibinfo {title} {Attractive and repulsive exciton-polariton
  interactions mediated by an electron gas},}\ }\href {\doibase
  10.1103/PhysRevLett.126.127405} {\bibfield  {journal} {\bibinfo  {journal}
  {Phys. Rev. Lett.}\ }\textbf {\bibinfo {volume} {126}},\ \bibinfo {pages}
  {127405} (\bibinfo {year} {2021})}\BibitemShut {NoStop}%
\bibitem [{\citenamefont {Camacho-Guardian}\ \emph {et~al.}(2022)\citenamefont
  {Camacho-Guardian}, \citenamefont {Bastarrachea-Magnani}, \citenamefont
  {Pohl},\ and\ \citenamefont {Bruun}}]{Camacho-Guardian2022}%
  \BibitemOpen
  \bibfield  {author} {\bibinfo {author} {\bibfnamefont {A.}~\bibnamefont
  {Camacho-Guardian}}, \bibinfo {author} {\bibfnamefont {M.}~\bibnamefont
  {Bastarrachea-Magnani}}, \bibinfo {author} {\bibfnamefont {T.}~\bibnamefont
  {Pohl}}, \ and\ \bibinfo {author} {\bibfnamefont {G.~M.}\ \bibnamefont
  {Bruun}},\ }\bibfield  {title} {\enquote {\bibinfo {title} {Strong photon
  interactions from weakly interacting particles},}\ }\href {\doibase
  10.1103/PhysRevB.106.L081302} {\bibfield  {journal} {\bibinfo  {journal}
  {Phys. Rev. B}\ }\textbf {\bibinfo {volume} {106}},\ \bibinfo {pages}
  {L081302} (\bibinfo {year} {2022})}\BibitemShut {NoStop}%
\bibitem [{\citenamefont {Estrecho}\ \emph {et~al.}(2019)\citenamefont
  {Estrecho}, \citenamefont {Gao}, \citenamefont {Bobrovska}, \citenamefont
  {Comber-Todd}, \citenamefont {Fraser}, \citenamefont {Steger}, \citenamefont
  {West}, \citenamefont {Pfeiffer}, \citenamefont {Levinsen}, \citenamefont
  {Parish}, \citenamefont {Liew}, \citenamefont {Matuszewski}, \citenamefont
  {Snoke}, \citenamefont {Truscott},\ and\ \citenamefont
  {Ostrovskaya}}]{Estrecho2019}%
  \BibitemOpen
  \bibfield  {author} {\bibinfo {author} {\bibfnamefont {E.}~\bibnamefont
  {Estrecho}}, \bibinfo {author} {\bibfnamefont {T.}~\bibnamefont {Gao}},
  \bibinfo {author} {\bibfnamefont {N.}~\bibnamefont {Bobrovska}}, \bibinfo
  {author} {\bibfnamefont {D.}~\bibnamefont {Comber-Todd}}, \bibinfo {author}
  {\bibfnamefont {M.~D.}\ \bibnamefont {Fraser}}, \bibinfo {author}
  {\bibfnamefont {M.}~\bibnamefont {Steger}}, \bibinfo {author} {\bibfnamefont
  {K.}~\bibnamefont {West}}, \bibinfo {author} {\bibfnamefont {L.~N.}\
  \bibnamefont {Pfeiffer}}, \bibinfo {author} {\bibfnamefont {J.}~\bibnamefont
  {Levinsen}}, \bibinfo {author} {\bibfnamefont {M.~M.}\ \bibnamefont
  {Parish}}, \bibinfo {author} {\bibfnamefont {T.~C.~H.}\ \bibnamefont {Liew}},
  \bibinfo {author} {\bibfnamefont {M.}~\bibnamefont {Matuszewski}}, \bibinfo
  {author} {\bibfnamefont {D.~W.}\ \bibnamefont {Snoke}}, \bibinfo {author}
  {\bibfnamefont {A.~G.}\ \bibnamefont {Truscott}}, \ and\ \bibinfo {author}
  {\bibfnamefont {E.~A.}\ \bibnamefont {Ostrovskaya}},\ }\bibfield  {title}
  {\enquote {\bibinfo {title} {Direct measurement of polariton-polariton
  interaction strength in the thomas-fermi regime of exciton-polariton
  condensation},}\ }\href {\doibase 10.1103/PhysRevB.100.035306} {\bibfield
  {journal} {\bibinfo  {journal} {Phys. Rev. B}\ }\textbf {\bibinfo {volume}
  {100}},\ \bibinfo {pages} {035306} (\bibinfo {year} {2019})}\BibitemShut
  {NoStop}%
\bibitem [{\citenamefont {Delteil}\ \emph {et~al.}(2019)\citenamefont
  {Delteil}, \citenamefont {Fink}, \citenamefont {Schade}, \citenamefont
  {H{\"o}fling}, \citenamefont {Schneider},\ and\ \citenamefont
  {{\.{I}}mamo{\u{g}}lu}}]{Delteil2019}%
  \BibitemOpen
  \bibfield  {author} {\bibinfo {author} {\bibfnamefont {Aymeric}\ \bibnamefont
  {Delteil}}, \bibinfo {author} {\bibfnamefont {Thomas}\ \bibnamefont {Fink}},
  \bibinfo {author} {\bibfnamefont {Anne}\ \bibnamefont {Schade}}, \bibinfo
  {author} {\bibfnamefont {Sven}\ \bibnamefont {H{\"o}fling}}, \bibinfo
  {author} {\bibfnamefont {Christian}\ \bibnamefont {Schneider}}, \ and\
  \bibinfo {author} {\bibfnamefont {Ata{\c{c}}}\ \bibnamefont
  {{\.{I}}mamo{\u{g}}lu}},\ }\bibfield  {title} {\enquote {\bibinfo {title}
  {Towards polariton blockade of confined exciton--polaritons},}\ }\href
  {\doibase 10.1038/s41563-019-0282-y} {\bibfield  {journal} {\bibinfo
  {journal} {Nature Materials}\ }\textbf {\bibinfo {volume} {18}},\ \bibinfo
  {pages} {219--222} (\bibinfo {year} {2019})}\BibitemShut {NoStop}%
\bibitem [{\citenamefont {Mu{\~n}oz-Matutano}\ \emph
  {et~al.}(2019)\citenamefont {Mu{\~n}oz-Matutano}, \citenamefont {Wood},
  \citenamefont {Johnsson}, \citenamefont {Vidal}, \citenamefont {Baragiola},
  \citenamefont {Reinhard}, \citenamefont {Lema{\^\i}tre}, \citenamefont
  {Bloch}, \citenamefont {Amo}, \citenamefont {Nogues}, \citenamefont {Besga},
  \citenamefont {Richard},\ and\ \citenamefont {Volz}}]{Munoz-Matutano2019}%
  \BibitemOpen
  \bibfield  {author} {\bibinfo {author} {\bibfnamefont {Guillermo}\
  \bibnamefont {Mu{\~n}oz-Matutano}}, \bibinfo {author} {\bibfnamefont
  {Andrew}\ \bibnamefont {Wood}}, \bibinfo {author} {\bibfnamefont {Mattias}\
  \bibnamefont {Johnsson}}, \bibinfo {author} {\bibfnamefont {Xavier}\
  \bibnamefont {Vidal}}, \bibinfo {author} {\bibfnamefont {Ben~Q.}\
  \bibnamefont {Baragiola}}, \bibinfo {author} {\bibfnamefont {Andreas}\
  \bibnamefont {Reinhard}}, \bibinfo {author} {\bibfnamefont {Aristide}\
  \bibnamefont {Lema{\^\i}tre}}, \bibinfo {author} {\bibfnamefont {Jacqueline}\
  \bibnamefont {Bloch}}, \bibinfo {author} {\bibfnamefont {Alberto}\
  \bibnamefont {Amo}}, \bibinfo {author} {\bibfnamefont {Gilles}\ \bibnamefont
  {Nogues}}, \bibinfo {author} {\bibfnamefont {Benjamin}\ \bibnamefont
  {Besga}}, \bibinfo {author} {\bibfnamefont {Maxime}\ \bibnamefont {Richard}},
  \ and\ \bibinfo {author} {\bibfnamefont {Thomas}\ \bibnamefont {Volz}},\
  }\bibfield  {title} {\enquote {\bibinfo {title} {Emergence of quantum
  correlations from interacting fibre-cavity polaritons},}\ }\href {\doibase
  10.1038/s41563-019-0281-z} {\bibfield  {journal} {\bibinfo  {journal} {Nature
  Materials}\ }\textbf {\bibinfo {volume} {18}},\ \bibinfo {pages} {213--218}
  (\bibinfo {year} {2019})}\BibitemShut {NoStop}%
\bibitem [{\citenamefont {Christmann}\ \emph {et~al.}(2011)\citenamefont
  {Christmann}, \citenamefont {Askitopoulos}, \citenamefont {Deligeorgis},
  \citenamefont {Hatzopoulos}, \citenamefont {Tsintzos}, \citenamefont
  {Savvidis},\ and\ \citenamefont {Baumberg}}]{Christmann2011}%
  \BibitemOpen
  \bibfield  {author} {\bibinfo {author} {\bibfnamefont {Gabriel}\ \bibnamefont
  {Christmann}}, \bibinfo {author} {\bibfnamefont {Alexis}\ \bibnamefont
  {Askitopoulos}}, \bibinfo {author} {\bibfnamefont {George}\ \bibnamefont
  {Deligeorgis}}, \bibinfo {author} {\bibfnamefont {Zacharias}\ \bibnamefont
  {Hatzopoulos}}, \bibinfo {author} {\bibfnamefont {Simeon~I.}\ \bibnamefont
  {Tsintzos}}, \bibinfo {author} {\bibfnamefont {Pavlos~G.}\ \bibnamefont
  {Savvidis}}, \ and\ \bibinfo {author} {\bibfnamefont {Jeremy~J.}\
  \bibnamefont {Baumberg}},\ }\bibfield  {title} {\enquote {\bibinfo {title}
  {Oriented polaritons in strongly-coupled asymmetric double quantum well
  microcavities},}\ }\href {\doibase 10.1063/1.3559909} {\bibfield  {journal}
  {\bibinfo  {journal} {Applied Physics Letters}\ }\textbf {\bibinfo {volume}
  {98}},\ \bibinfo {pages} {081111} (\bibinfo {year} {2011})},\ \Eprint
  {http://arxiv.org/abs/https://doi.org/10.1063/1.3559909}
  {https://doi.org/10.1063/1.3559909} \BibitemShut {NoStop}%
\bibitem [{\citenamefont {Cristofolini}\ \emph {et~al.}(2012)\citenamefont
  {Cristofolini}, \citenamefont {Christmann}, \citenamefont {Tsintzos},
  \citenamefont {Deligeorgis}, \citenamefont {Konstantinidis}, \citenamefont
  {Hatzopoulos}, \citenamefont {Savvidis},\ and\ \citenamefont
  {Baumberg}}]{Cristofolini2012}%
  \BibitemOpen
  \bibfield  {author} {\bibinfo {author} {\bibfnamefont {Peter}\ \bibnamefont
  {Cristofolini}}, \bibinfo {author} {\bibfnamefont {Gabriel}\ \bibnamefont
  {Christmann}}, \bibinfo {author} {\bibfnamefont {Simeon~I.}\ \bibnamefont
  {Tsintzos}}, \bibinfo {author} {\bibfnamefont {George}\ \bibnamefont
  {Deligeorgis}}, \bibinfo {author} {\bibfnamefont {George}\ \bibnamefont
  {Konstantinidis}}, \bibinfo {author} {\bibfnamefont {Zacharias}\ \bibnamefont
  {Hatzopoulos}}, \bibinfo {author} {\bibfnamefont {Pavlos~G.}\ \bibnamefont
  {Savvidis}}, \ and\ \bibinfo {author} {\bibfnamefont {Jeremy~J.}\
  \bibnamefont {Baumberg}},\ }\bibfield  {title} {\enquote {\bibinfo {title}
  {Coupling quantum tunneling with cavity photons},}\ }\href {\doibase
  10.1126/science.1219010} {\bibfield  {journal} {\bibinfo  {journal}
  {Science}\ }\textbf {\bibinfo {volume} {336}},\ \bibinfo {pages} {704--707}
  (\bibinfo {year} {2012})}\BibitemShut {NoStop}%
\bibitem [{\citenamefont {Rosenberg}\ \emph {et~al.}(2016)\citenamefont
  {Rosenberg}, \citenamefont {Mazuz-Harpaz}, \citenamefont {Rapaport},
  \citenamefont {West},\ and\ \citenamefont {Pfeiffer}}]{Rapaport2016}%
  \BibitemOpen
  \bibfield  {author} {\bibinfo {author} {\bibfnamefont {Itamar}\ \bibnamefont
  {Rosenberg}}, \bibinfo {author} {\bibfnamefont {Yotam}\ \bibnamefont
  {Mazuz-Harpaz}}, \bibinfo {author} {\bibfnamefont {Ronen}\ \bibnamefont
  {Rapaport}}, \bibinfo {author} {\bibfnamefont {Kenneth}\ \bibnamefont
  {West}}, \ and\ \bibinfo {author} {\bibfnamefont {Loren}\ \bibnamefont
  {Pfeiffer}},\ }\bibfield  {title} {\enquote {\bibinfo {title} {Electrically
  controlled mutual interactions of flying waveguide dipolaritons},}\ }\href
  {\doibase 10.1103/PhysRevB.93.195151} {\bibfield  {journal} {\bibinfo
  {journal} {Phys. Rev. B}\ }\textbf {\bibinfo {volume} {93}},\ \bibinfo
  {pages} {195151} (\bibinfo {year} {2016})}\BibitemShut {NoStop}%
\bibitem [{\citenamefont {Kremser}\ \emph {et~al.}(2020)\citenamefont
  {Kremser}, \citenamefont {Brotons-Gisbert}, \citenamefont {Kn{\"o}rzer},
  \citenamefont {G{\"u}ckelhorn}, \citenamefont {Meyer}, \citenamefont
  {Barbone}, \citenamefont {Stier}, \citenamefont {Gerardot}, \citenamefont
  {M{\"u}ller},\ and\ \citenamefont {Finley}}]{Kremser2020}%
  \BibitemOpen
  \bibfield  {author} {\bibinfo {author} {\bibfnamefont {Malte}\ \bibnamefont
  {Kremser}}, \bibinfo {author} {\bibfnamefont {Mauro}\ \bibnamefont
  {Brotons-Gisbert}}, \bibinfo {author} {\bibfnamefont {Johannes}\ \bibnamefont
  {Kn{\"o}rzer}}, \bibinfo {author} {\bibfnamefont {Janine}\ \bibnamefont
  {G{\"u}ckelhorn}}, \bibinfo {author} {\bibfnamefont {Moritz}\ \bibnamefont
  {Meyer}}, \bibinfo {author} {\bibfnamefont {Matteo}\ \bibnamefont {Barbone}},
  \bibinfo {author} {\bibfnamefont {Andreas~V.}\ \bibnamefont {Stier}},
  \bibinfo {author} {\bibfnamefont {Brian~D.}\ \bibnamefont {Gerardot}},
  \bibinfo {author} {\bibfnamefont {Kai}\ \bibnamefont {M{\"u}ller}}, \ and\
  \bibinfo {author} {\bibfnamefont {Jonathan~J.}\ \bibnamefont {Finley}},\
  }\bibfield  {title} {\enquote {\bibinfo {title} {Discrete interactions
  between a few interlayer excitons trapped at a mose2--wse2
  heterointerface},}\ }\href {\doibase 10.1038/s41699-020-0141-3} {\bibfield
  {journal} {\bibinfo  {journal} {npj 2D Materials and Applications}\ }\textbf
  {\bibinfo {volume} {4}},\ \bibinfo {pages} {8} (\bibinfo {year}
  {2020})}\BibitemShut {NoStop}%
\bibitem [{\citenamefont {Li}\ \emph {et~al.}(2020)\citenamefont {Li},
  \citenamefont {Lu}, \citenamefont {Dubey}, \citenamefont {Devenica},\ and\
  \citenamefont {Srivastava}}]{Li2020}%
  \BibitemOpen
  \bibfield  {author} {\bibinfo {author} {\bibfnamefont {Weijie}\ \bibnamefont
  {Li}}, \bibinfo {author} {\bibfnamefont {Xin}\ \bibnamefont {Lu}}, \bibinfo
  {author} {\bibfnamefont {Sudipta}\ \bibnamefont {Dubey}}, \bibinfo {author}
  {\bibfnamefont {Luka}\ \bibnamefont {Devenica}}, \ and\ \bibinfo {author}
  {\bibfnamefont {Ajit}\ \bibnamefont {Srivastava}},\ }\bibfield  {title}
  {\enquote {\bibinfo {title} {Dipolar interactions between localized
  interlayer excitons in van der waals heterostructures},}\ }\href {\doibase
  10.1038/s41563-020-0661-4} {\bibfield  {journal} {\bibinfo  {journal} {Nature
  Materials}\ }\textbf {\bibinfo {volume} {19}},\ \bibinfo {pages} {624--629}
  (\bibinfo {year} {2020})}\BibitemShut {NoStop}%
\bibitem [{\citenamefont {Sun}\ \emph {et~al.}(2022)\citenamefont {Sun},
  \citenamefont {Zhu}, \citenamefont {Qin}, \citenamefont {Liu}, \citenamefont
  {Tang}, \citenamefont {L{\"u}}, \citenamefont {Rahman}, \citenamefont
  {Yildirim},\ and\ \citenamefont {Lu}}]{Sun2022}%
  \BibitemOpen
  \bibfield  {author} {\bibinfo {author} {\bibfnamefont {Xueqian}\ \bibnamefont
  {Sun}}, \bibinfo {author} {\bibfnamefont {Yi}~\bibnamefont {Zhu}}, \bibinfo
  {author} {\bibfnamefont {Hao}\ \bibnamefont {Qin}}, \bibinfo {author}
  {\bibfnamefont {Boqing}\ \bibnamefont {Liu}}, \bibinfo {author}
  {\bibfnamefont {Yilin}\ \bibnamefont {Tang}}, \bibinfo {author}
  {\bibfnamefont {Tieyu}\ \bibnamefont {L{\"u}}}, \bibinfo {author}
  {\bibfnamefont {Sharidya}\ \bibnamefont {Rahman}}, \bibinfo {author}
  {\bibfnamefont {Tanju}\ \bibnamefont {Yildirim}}, \ and\ \bibinfo {author}
  {\bibfnamefont {Yuerui}\ \bibnamefont {Lu}},\ }\bibfield  {title} {\enquote
  {\bibinfo {title} {Enhanced interactions of interlayer excitons in
  free-standing heterobilayers},}\ }\href {\doibase 10.1038/s41586-022-05193-z}
  {\bibfield  {journal} {\bibinfo  {journal} {Nature}\ }\textbf {\bibinfo
  {volume} {610}},\ \bibinfo {pages} {478--484} (\bibinfo {year}
  {2022})}\BibitemShut {NoStop}%
\bibitem [{\citenamefont {Thureja}\ \emph {et~al.}(2022)\citenamefont
  {Thureja}, \citenamefont {Imamoglu}, \citenamefont {Smole{\'{n}}ski},
  \citenamefont {Amelio}, \citenamefont {Popert}, \citenamefont {Chervy},
  \citenamefont {Lu}, \citenamefont {Liu}, \citenamefont {Barmak},
  \citenamefont {Watanabe}, \citenamefont {Taniguchi}, \citenamefont {Norris},
  \citenamefont {Kroner},\ and\ \citenamefont {Murthy}}]{Thureja2022}%
  \BibitemOpen
  \bibfield  {author} {\bibinfo {author} {\bibfnamefont {Deepankur}\
  \bibnamefont {Thureja}}, \bibinfo {author} {\bibfnamefont {Atac}\
  \bibnamefont {Imamoglu}}, \bibinfo {author} {\bibfnamefont {Tomasz}\
  \bibnamefont {Smole{\'{n}}ski}}, \bibinfo {author} {\bibfnamefont {Ivan}\
  \bibnamefont {Amelio}}, \bibinfo {author} {\bibfnamefont {Alexander}\
  \bibnamefont {Popert}}, \bibinfo {author} {\bibfnamefont {Thibault}\
  \bibnamefont {Chervy}}, \bibinfo {author} {\bibfnamefont {Xiaobo}\
  \bibnamefont {Lu}}, \bibinfo {author} {\bibfnamefont {Song}\ \bibnamefont
  {Liu}}, \bibinfo {author} {\bibfnamefont {Katayun}\ \bibnamefont {Barmak}},
  \bibinfo {author} {\bibfnamefont {Kenji}\ \bibnamefont {Watanabe}}, \bibinfo
  {author} {\bibfnamefont {Takashi}\ \bibnamefont {Taniguchi}}, \bibinfo
  {author} {\bibfnamefont {David~J.}\ \bibnamefont {Norris}}, \bibinfo {author}
  {\bibfnamefont {Martin}\ \bibnamefont {Kroner}}, \ and\ \bibinfo {author}
  {\bibfnamefont {Puneet~A.}\ \bibnamefont {Murthy}},\ }\bibfield  {title}
  {\enquote {\bibinfo {title} {Electrically tunable quantum confinement of
  neutral excitons},}\ }\href {\doibase 10.1038/s41586-022-04634-z} {\bibfield
  {journal} {\bibinfo  {journal} {Nature}\ }\textbf {\bibinfo {volume} {606}},\
  \bibinfo {pages} {298--304} (\bibinfo {year} {2022})}\BibitemShut {NoStop}%
\bibitem [{\citenamefont {Lozovik}\ \emph {et~al.}(2007)\citenamefont
  {Lozovik}, \citenamefont {Kurbakov}, \citenamefont {Astrakharchik},
  \citenamefont {Boronat},\ and\ \citenamefont {Willander}}]{LOZOVIK2007}%
  \BibitemOpen
  \bibfield  {author} {\bibinfo {author} {\bibfnamefont {Yurii~E.}\
  \bibnamefont {Lozovik}}, \bibinfo {author} {\bibfnamefont {I.L.}\
  \bibnamefont {Kurbakov}}, \bibinfo {author} {\bibfnamefont {G.E.}\
  \bibnamefont {Astrakharchik}}, \bibinfo {author} {\bibfnamefont
  {J.}~\bibnamefont {Boronat}}, \ and\ \bibinfo {author} {\bibfnamefont
  {Magnus}\ \bibnamefont {Willander}},\ }\bibfield  {title} {\enquote {\bibinfo
  {title} {Strong correlation effects in 2d bose–einstein condensed dipolar
  excitons},}\ }\href {\doibase https://doi.org/10.1016/j.ssc.2007.07.041}
  {\bibfield  {journal} {\bibinfo  {journal} {Solid State Communications}\
  }\textbf {\bibinfo {volume} {144}},\ \bibinfo {pages} {399--404} (\bibinfo
  {year} {2007})},\ \bibinfo {note} {spontaneous coherence in exciton
  systems}\BibitemShut {NoStop}%
\bibitem [{\citenamefont {Zimmermann}\ and\ \citenamefont
  {Schindler}(2007)}]{ZIMMERMANN2007}%
  \BibitemOpen
  \bibfield  {author} {\bibinfo {author} {\bibfnamefont {Roland}\ \bibnamefont
  {Zimmermann}}\ and\ \bibinfo {author} {\bibfnamefont {Christoph}\
  \bibnamefont {Schindler}},\ }\bibfield  {title} {\enquote {\bibinfo {title}
  {Exciton–exciton interaction in coupled quantum wells},}\ }\href {\doibase
  https://doi.org/10.1016/j.ssc.2007.07.044} {\bibfield  {journal} {\bibinfo
  {journal} {Solid State Communications}\ }\textbf {\bibinfo {volume} {144}},\
  \bibinfo {pages} {395--398} (\bibinfo {year} {2007})},\ \bibinfo {note}
  {spontaneous coherence in exciton systems}\BibitemShut {NoStop}%
\bibitem [{\citenamefont {Astrakharchik}\ \emph {et~al.}(2007)\citenamefont
  {Astrakharchik}, \citenamefont {Boronat}, \citenamefont {Kurbakov},\ and\
  \citenamefont {Lozovik}}]{Astra2007}%
  \BibitemOpen
  \bibfield  {author} {\bibinfo {author} {\bibfnamefont {G.~E.}\ \bibnamefont
  {Astrakharchik}}, \bibinfo {author} {\bibfnamefont {J.}~\bibnamefont
  {Boronat}}, \bibinfo {author} {\bibfnamefont {I.~L.}\ \bibnamefont
  {Kurbakov}}, \ and\ \bibinfo {author} {\bibfnamefont {Yu.~E.}\ \bibnamefont
  {Lozovik}},\ }\bibfield  {title} {\enquote {\bibinfo {title} {Quantum phase
  transition in a two-dimensional system of dipoles},}\ }\href {\doibase
  10.1103/PhysRevLett.98.060405} {\bibfield  {journal} {\bibinfo  {journal}
  {Phys. Rev. Lett.}\ }\textbf {\bibinfo {volume} {98}},\ \bibinfo {pages}
  {060405} (\bibinfo {year} {2007})}\BibitemShut {NoStop}%
\bibitem [{\citenamefont {Shilo}\ \emph {et~al.}(2013)\citenamefont {Shilo},
  \citenamefont {Cohen}, \citenamefont {Laikhtman}, \citenamefont {West},
  \citenamefont {Pfeiffer},\ and\ \citenamefont {Rapaport}}]{Shilo2013}%
  \BibitemOpen
  \bibfield  {author} {\bibinfo {author} {\bibfnamefont {Yehiel}\ \bibnamefont
  {Shilo}}, \bibinfo {author} {\bibfnamefont {Kobi}\ \bibnamefont {Cohen}},
  \bibinfo {author} {\bibfnamefont {Boris}\ \bibnamefont {Laikhtman}}, \bibinfo
  {author} {\bibfnamefont {Ken}\ \bibnamefont {West}}, \bibinfo {author}
  {\bibfnamefont {Loren}\ \bibnamefont {Pfeiffer}}, \ and\ \bibinfo {author}
  {\bibfnamefont {Ronen}\ \bibnamefont {Rapaport}},\ }\bibfield  {title}
  {\enquote {\bibinfo {title} {Particle correlations and evidence for dark
  state condensation in a cold dipolar exciton fluid},}\ }\href {\doibase
  10.1038/ncomms3335} {\bibfield  {journal} {\bibinfo  {journal} {Nature
  Communications}\ }\textbf {\bibinfo {volume} {4}},\ \bibinfo {pages} {2335}
  (\bibinfo {year} {2013})}\BibitemShut {NoStop}%
\bibitem [{\citenamefont {Stern}\ \emph {et~al.}(2014)\citenamefont {Stern},
  \citenamefont {Umansky},\ and\ \citenamefont {Bar-Joseph}}]{Stern2014}%
  \BibitemOpen
  \bibfield  {author} {\bibinfo {author} {\bibfnamefont {Michael}\ \bibnamefont
  {Stern}}, \bibinfo {author} {\bibfnamefont {Vladimir}\ \bibnamefont
  {Umansky}}, \ and\ \bibinfo {author} {\bibfnamefont {Israel}\ \bibnamefont
  {Bar-Joseph}},\ }\bibfield  {title} {\enquote {\bibinfo {title} {Exciton
  liquid in coupled quantum wells},}\ }\href {\doibase 10.1126/science.1243409}
  {\bibfield  {journal} {\bibinfo  {journal} {Science}\ }\textbf {\bibinfo
  {volume} {343}},\ \bibinfo {pages} {55--57} (\bibinfo {year} {2014})},\
  \Eprint
  {http://arxiv.org/abs/https://www.science.org/doi/pdf/10.1126/science.1243409}
  {https://www.science.org/doi/pdf/10.1126/science.1243409} \BibitemShut
  {NoStop}%
\bibitem [{\citenamefont {Remeika}\ \emph {et~al.}(2015)\citenamefont
  {Remeika}, \citenamefont {Leonard}, \citenamefont {Dorow}, \citenamefont
  {Fogler}, \citenamefont {Butov}, \citenamefont {Hanson},\ and\ \citenamefont
  {Gossard}}]{Remeika2014}%
  \BibitemOpen
  \bibfield  {author} {\bibinfo {author} {\bibfnamefont {M.}~\bibnamefont
  {Remeika}}, \bibinfo {author} {\bibfnamefont {J.~R.}\ \bibnamefont
  {Leonard}}, \bibinfo {author} {\bibfnamefont {C.~J.}\ \bibnamefont {Dorow}},
  \bibinfo {author} {\bibfnamefont {M.~M.}\ \bibnamefont {Fogler}}, \bibinfo
  {author} {\bibfnamefont {L.~V.}\ \bibnamefont {Butov}}, \bibinfo {author}
  {\bibfnamefont {M.}~\bibnamefont {Hanson}}, \ and\ \bibinfo {author}
  {\bibfnamefont {A.~C.}\ \bibnamefont {Gossard}},\ }\bibfield  {title}
  {\enquote {\bibinfo {title} {Measurement of exciton correlations using
  electrostatic lattices},}\ }\href {\doibase 10.1103/PhysRevB.92.115311}
  {\bibfield  {journal} {\bibinfo  {journal} {Phys. Rev. B}\ }\textbf {\bibinfo
  {volume} {92}},\ \bibinfo {pages} {115311} (\bibinfo {year}
  {2015})}\BibitemShut {NoStop}%
\bibitem [{\citenamefont {Cohen}\ \emph {et~al.}(2016)\citenamefont {Cohen},
  \citenamefont {Shilo}, \citenamefont {West}, \citenamefont {Pfeiffer},\ and\
  \citenamefont {Rapaport}}]{Cohen2016}%
  \BibitemOpen
  \bibfield  {author} {\bibinfo {author} {\bibfnamefont {Kobi}\ \bibnamefont
  {Cohen}}, \bibinfo {author} {\bibfnamefont {Yehiel}\ \bibnamefont {Shilo}},
  \bibinfo {author} {\bibfnamefont {Ken}\ \bibnamefont {West}}, \bibinfo
  {author} {\bibfnamefont {Loren}\ \bibnamefont {Pfeiffer}}, \ and\ \bibinfo
  {author} {\bibfnamefont {Ronen}\ \bibnamefont {Rapaport}},\ }\bibfield
  {title} {\enquote {\bibinfo {title} {Dark high density dipolar liquid of
  excitons},}\ }\href {\doibase 10.1021/acs.nanolett.6b01061} {\bibfield
  {journal} {\bibinfo  {journal} {Nano Letters}\ }\textbf {\bibinfo {volume}
  {16}},\ \bibinfo {pages} {3726--3731} (\bibinfo {year} {2016})}\BibitemShut
  {NoStop}%
\bibitem [{\citenamefont {Misra}\ \emph {et~al.}(2018)\citenamefont {Misra},
  \citenamefont {Stern}, \citenamefont {Joshua}, \citenamefont {Umansky},\ and\
  \citenamefont {Bar-Joseph}}]{Misra2018}%
  \BibitemOpen
  \bibfield  {author} {\bibinfo {author} {\bibfnamefont {Subhradeep}\
  \bibnamefont {Misra}}, \bibinfo {author} {\bibfnamefont {Michael}\
  \bibnamefont {Stern}}, \bibinfo {author} {\bibfnamefont {Arjun}\ \bibnamefont
  {Joshua}}, \bibinfo {author} {\bibfnamefont {Vladimir}\ \bibnamefont
  {Umansky}}, \ and\ \bibinfo {author} {\bibfnamefont {Israel}\ \bibnamefont
  {Bar-Joseph}},\ }\bibfield  {title} {\enquote {\bibinfo {title} {Experimental
  study of the exciton gas-liquid transition in coupled quantum wells},}\
  }\href {\doibase 10.1103/PhysRevLett.120.047402} {\bibfield  {journal}
  {\bibinfo  {journal} {Phys. Rev. Lett.}\ }\textbf {\bibinfo {volume} {120}},\
  \bibinfo {pages} {047402} (\bibinfo {year} {2018})}\BibitemShut {NoStop}%
\bibitem [{\citenamefont {Mazuz-Harpaz}\ \emph {et~al.}(2019)\citenamefont
  {Mazuz-Harpaz}, \citenamefont {Cohen}, \citenamefont {Leveson}, \citenamefont
  {West}, \citenamefont {Pfeiffer}, \citenamefont {Khodas},\ and\ \citenamefont
  {Rapaport}}]{MazuzHarpaz2019}%
  \BibitemOpen
  \bibfield  {author} {\bibinfo {author} {\bibfnamefont {Yotam}\ \bibnamefont
  {Mazuz-Harpaz}}, \bibinfo {author} {\bibfnamefont {Kobi}\ \bibnamefont
  {Cohen}}, \bibinfo {author} {\bibfnamefont {Michael}\ \bibnamefont
  {Leveson}}, \bibinfo {author} {\bibfnamefont {Ken}\ \bibnamefont {West}},
  \bibinfo {author} {\bibfnamefont {Loren}\ \bibnamefont {Pfeiffer}}, \bibinfo
  {author} {\bibfnamefont {Maxim}\ \bibnamefont {Khodas}}, \ and\ \bibinfo
  {author} {\bibfnamefont {Ronen}\ \bibnamefont {Rapaport}},\ }\bibfield
  {title} {\enquote {\bibinfo {title} {Dynamical formation of a strongly
  correlated dark condensate of dipolar excitons},}\ }\href {\doibase
  10.1073/pnas.1903374116} {\bibfield  {journal} {\bibinfo  {journal}
  {Proceedings of the National Academy of Sciences}\ }\textbf {\bibinfo
  {volume} {116}},\ \bibinfo {pages} {18328--18333} (\bibinfo {year} {2019})},\
  \Eprint
  {http://arxiv.org/abs/https://www.pnas.org/doi/pdf/10.1073/pnas.1903374116}
  {https://www.pnas.org/doi/pdf/10.1073/pnas.1903374116} \BibitemShut {NoStop}%
\bibitem [{\citenamefont {Lee}\ \emph {et~al.}(2009)\citenamefont {Lee},
  \citenamefont {Drummond},\ and\ \citenamefont {Needs}}]{Lee2009}%
  \BibitemOpen
  \bibfield  {author} {\bibinfo {author} {\bibfnamefont {R.~M.}\ \bibnamefont
  {Lee}}, \bibinfo {author} {\bibfnamefont {N.~D.}\ \bibnamefont {Drummond}}, \
  and\ \bibinfo {author} {\bibfnamefont {R.~J.}\ \bibnamefont {Needs}},\
  }\bibfield  {title} {\enquote {\bibinfo {title} {Exciton-exciton interaction
  and biexciton formation in bilayer systems},}\ }\href {\doibase
  10.1103/PhysRevB.79.125308} {\bibfield  {journal} {\bibinfo  {journal} {Phys.
  Rev. B}\ }\textbf {\bibinfo {volume} {79}},\ \bibinfo {pages} {125308}
  (\bibinfo {year} {2009})}\BibitemShut {NoStop}%
\bibitem [{\citenamefont {Schinner}\ \emph {et~al.}(2013)\citenamefont
  {Schinner}, \citenamefont {Repp}, \citenamefont {Schubert}, \citenamefont
  {Rai}, \citenamefont {Reuter}, \citenamefont {Wieck}, \citenamefont
  {Govorov}, \citenamefont {Holleitner},\ and\ \citenamefont
  {Kotthaus}}]{Schinner2013}%
  \BibitemOpen
  \bibfield  {author} {\bibinfo {author} {\bibfnamefont {G.~J.}\ \bibnamefont
  {Schinner}}, \bibinfo {author} {\bibfnamefont {J.}~\bibnamefont {Repp}},
  \bibinfo {author} {\bibfnamefont {E.}~\bibnamefont {Schubert}}, \bibinfo
  {author} {\bibfnamefont {A.~K.}\ \bibnamefont {Rai}}, \bibinfo {author}
  {\bibfnamefont {D.}~\bibnamefont {Reuter}}, \bibinfo {author} {\bibfnamefont
  {A.~D.}\ \bibnamefont {Wieck}}, \bibinfo {author} {\bibfnamefont {A.~O.}\
  \bibnamefont {Govorov}}, \bibinfo {author} {\bibfnamefont {A.~W.}\
  \bibnamefont {Holleitner}}, \ and\ \bibinfo {author} {\bibfnamefont {J.~P.}\
  \bibnamefont {Kotthaus}},\ }\bibfield  {title} {\enquote {\bibinfo {title}
  {Confinement and interaction of single indirect excitons in a
  voltage-controlled trap formed inside double ingaas quantum wells},}\ }\href
  {\doibase 10.1103/PhysRevLett.110.127403} {\bibfield  {journal} {\bibinfo
  {journal} {Phys. Rev. Lett.}\ }\textbf {\bibinfo {volume} {110}},\ \bibinfo
  {pages} {127403} (\bibinfo {year} {2013})}\BibitemShut {NoStop}%
\bibitem [{\citenamefont {Tsintzos}\ \emph {et~al.}(2018)\citenamefont
  {Tsintzos}, \citenamefont {Tzimis}, \citenamefont {Stavrinidis},
  \citenamefont {Trifonov}, \citenamefont {Hatzopoulos}, \citenamefont
  {Baumberg}, \citenamefont {Ohadi},\ and\ \citenamefont
  {Savvidis}}]{Tsintzos2018}%
  \BibitemOpen
  \bibfield  {author} {\bibinfo {author} {\bibfnamefont {S.~I.}\ \bibnamefont
  {Tsintzos}}, \bibinfo {author} {\bibfnamefont {A.}~\bibnamefont {Tzimis}},
  \bibinfo {author} {\bibfnamefont {G.}~\bibnamefont {Stavrinidis}}, \bibinfo
  {author} {\bibfnamefont {A.}~\bibnamefont {Trifonov}}, \bibinfo {author}
  {\bibfnamefont {Z.}~\bibnamefont {Hatzopoulos}}, \bibinfo {author}
  {\bibfnamefont {J.~J.}\ \bibnamefont {Baumberg}}, \bibinfo {author}
  {\bibfnamefont {H.}~\bibnamefont {Ohadi}}, \ and\ \bibinfo {author}
  {\bibfnamefont {P.~G.}\ \bibnamefont {Savvidis}},\ }\bibfield  {title}
  {\enquote {\bibinfo {title} {Electrical tuning of nonlinearities in
  exciton-polariton condensates},}\ }\href {\doibase
  10.1103/PhysRevLett.121.037401} {\bibfield  {journal} {\bibinfo  {journal}
  {Phys. Rev. Lett.}\ }\textbf {\bibinfo {volume} {121}},\ \bibinfo {pages}
  {037401} (\bibinfo {year} {2018})}\BibitemShut {NoStop}%
\bibitem [{\citenamefont {Togan}\ \emph {et~al.}(2018)\citenamefont {Togan},
  \citenamefont {Lim}, \citenamefont {Faelt}, \citenamefont {Wegscheider},\
  and\ \citenamefont {Imamoglu}}]{Togan2018}%
  \BibitemOpen
  \bibfield  {author} {\bibinfo {author} {\bibfnamefont {Emre}\ \bibnamefont
  {Togan}}, \bibinfo {author} {\bibfnamefont {Hyang-Tag}\ \bibnamefont {Lim}},
  \bibinfo {author} {\bibfnamefont {Stefan}\ \bibnamefont {Faelt}}, \bibinfo
  {author} {\bibfnamefont {Werner}\ \bibnamefont {Wegscheider}}, \ and\
  \bibinfo {author} {\bibfnamefont {Atac}\ \bibnamefont {Imamoglu}},\
  }\bibfield  {title} {\enquote {\bibinfo {title} {Enhanced interactions
  between dipolar polaritons},}\ }\href {\doibase
  10.1103/PhysRevLett.121.227402} {\bibfield  {journal} {\bibinfo  {journal}
  {Phys. Rev. Lett.}\ }\textbf {\bibinfo {volume} {121}},\ \bibinfo {pages}
  {227402} (\bibinfo {year} {2018})}\BibitemShut {NoStop}%
\bibitem [{\citenamefont {Rosenberg}\ \emph {et~al.}(2018)\citenamefont
  {Rosenberg}, \citenamefont {Liran}, \citenamefont {Mazuz-Harpaz},
  \citenamefont {West}, \citenamefont {Pfeiffer},\ and\ \citenamefont
  {Rapaport}}]{Rosenberg2018dipolar}%
  \BibitemOpen
  \bibfield  {author} {\bibinfo {author} {\bibfnamefont {Itamar}\ \bibnamefont
  {Rosenberg}}, \bibinfo {author} {\bibfnamefont {Dror}\ \bibnamefont {Liran}},
  \bibinfo {author} {\bibfnamefont {Yotam}\ \bibnamefont {Mazuz-Harpaz}},
  \bibinfo {author} {\bibfnamefont {Kenneth}\ \bibnamefont {West}}, \bibinfo
  {author} {\bibfnamefont {Loren}\ \bibnamefont {Pfeiffer}}, \ and\ \bibinfo
  {author} {\bibfnamefont {Ronen}\ \bibnamefont {Rapaport}},\ }\bibfield
  {title} {\enquote {\bibinfo {title} {Strongly interacting
  dipolar-polaritons},}\ }\href {\doibase 10.1126/sciadv.aat8880} {\bibfield
  {journal} {\bibinfo  {journal} {Science Advances}\ }\textbf {\bibinfo
  {volume} {4}},\ \bibinfo {pages} {eaat8880} (\bibinfo {year}
  {2018})}\BibitemShut {NoStop}%
\bibitem [{\citenamefont {Datta}\ \emph {et~al.}(2022)\citenamefont {Datta},
  \citenamefont {Khatoniar}, \citenamefont {Deshmukh}, \citenamefont {Thouin},
  \citenamefont {Bushati}, \citenamefont {De~Liberato}, \citenamefont {Cohen},\
  and\ \citenamefont {Menon}}]{datta2022highly}%
  \BibitemOpen
  \bibfield  {author} {\bibinfo {author} {\bibfnamefont {Biswajit}\
  \bibnamefont {Datta}}, \bibinfo {author} {\bibfnamefont {Mandeep}\
  \bibnamefont {Khatoniar}}, \bibinfo {author} {\bibfnamefont {Prathmesh}\
  \bibnamefont {Deshmukh}}, \bibinfo {author} {\bibfnamefont {F{\'e}lix}\
  \bibnamefont {Thouin}}, \bibinfo {author} {\bibfnamefont {Rezlind}\
  \bibnamefont {Bushati}}, \bibinfo {author} {\bibfnamefont {Simone}\
  \bibnamefont {De~Liberato}}, \bibinfo {author} {\bibfnamefont
  {Stephane~Kena}\ \bibnamefont {Cohen}}, \ and\ \bibinfo {author}
  {\bibfnamefont {Vinod~M}\ \bibnamefont {Menon}},\ }\bibfield  {title}
  {\enquote {\bibinfo {title} {Highly nonlinear dipolar exciton-polaritons in
  bilayer mos2},}\ }\href@noop {} {\bibfield  {journal} {\bibinfo  {journal}
  {Nature communications}\ }\textbf {\bibinfo {volume} {13}},\ \bibinfo {pages}
  {1--7} (\bibinfo {year} {2022})}\BibitemShut {NoStop}%
\bibitem [{\citenamefont {Su\'arez-Forero}\ \emph {et~al.}(2021)\citenamefont
  {Su\'arez-Forero}, \citenamefont {Riminucci}, \citenamefont {Ardizzone},
  \citenamefont {Karpowicz}, \citenamefont {Maggiolini}, \citenamefont
  {Macorini}, \citenamefont {Lerario}, \citenamefont {Todisco}, \citenamefont
  {De~Giorgi}, \citenamefont {Dominici}, \citenamefont {Ballarini},
  \citenamefont {Gigli}, \citenamefont {Lanotte}, \citenamefont {West},
  \citenamefont {Baldwin}, \citenamefont {Pfeiffer},\ and\ \citenamefont
  {Sanvitto}}]{Suarez:PRL2021}%
  \BibitemOpen
  \bibfield  {author} {\bibinfo {author} {\bibfnamefont {D.~G.}\ \bibnamefont
  {Su\'arez-Forero}}, \bibinfo {author} {\bibfnamefont {F.}~\bibnamefont
  {Riminucci}}, \bibinfo {author} {\bibfnamefont {V.}~\bibnamefont
  {Ardizzone}}, \bibinfo {author} {\bibfnamefont {N.}~\bibnamefont
  {Karpowicz}}, \bibinfo {author} {\bibfnamefont {E.}~\bibnamefont
  {Maggiolini}}, \bibinfo {author} {\bibfnamefont {G.}~\bibnamefont
  {Macorini}}, \bibinfo {author} {\bibfnamefont {G.}~\bibnamefont {Lerario}},
  \bibinfo {author} {\bibfnamefont {F.}~\bibnamefont {Todisco}}, \bibinfo
  {author} {\bibfnamefont {M.}~\bibnamefont {De~Giorgi}}, \bibinfo {author}
  {\bibfnamefont {L.}~\bibnamefont {Dominici}}, \bibinfo {author}
  {\bibfnamefont {D.}~\bibnamefont {Ballarini}}, \bibinfo {author}
  {\bibfnamefont {G.}~\bibnamefont {Gigli}}, \bibinfo {author} {\bibfnamefont
  {A.~S.}\ \bibnamefont {Lanotte}}, \bibinfo {author} {\bibfnamefont
  {K.}~\bibnamefont {West}}, \bibinfo {author} {\bibfnamefont {K.}~\bibnamefont
  {Baldwin}}, \bibinfo {author} {\bibfnamefont {L.}~\bibnamefont {Pfeiffer}}, \
  and\ \bibinfo {author} {\bibfnamefont {D.}~\bibnamefont {Sanvitto}},\
  }\bibfield  {title} {\enquote {\bibinfo {title} {Enhancement of parametric
  effects in polariton waveguides induced by dipolar interactions},}\ }\href
  {\doibase 10.1103/PhysRevLett.126.137401} {\bibfield  {journal} {\bibinfo
  {journal} {Phys. Rev. Lett.}\ }\textbf {\bibinfo {volume} {126}},\ \bibinfo
  {pages} {137401} (\bibinfo {year} {2021})}\BibitemShut {NoStop}%
\bibitem [{\citenamefont {Liran}\ \emph {et~al.}(2024)\citenamefont {Liran},
  \citenamefont {Rapaport}, \citenamefont {Hu}, \citenamefont {Lydick},
  \citenamefont {Deng},\ and\ \citenamefont {Pfeiffer}}]{Liran:PRX2024}%
  \BibitemOpen
  \bibfield  {author} {\bibinfo {author} {\bibfnamefont {Dror}\ \bibnamefont
  {Liran}}, \bibinfo {author} {\bibfnamefont {Ronen}\ \bibnamefont {Rapaport}},
  \bibinfo {author} {\bibfnamefont {Jiaqi}\ \bibnamefont {Hu}}, \bibinfo
  {author} {\bibfnamefont {Nathanial}\ \bibnamefont {Lydick}}, \bibinfo
  {author} {\bibfnamefont {Hui}\ \bibnamefont {Deng}}, \ and\ \bibinfo {author}
  {\bibfnamefont {Loren}\ \bibnamefont {Pfeiffer}},\ }\bibfield  {title}
  {\enquote {\bibinfo {title} {Electrically controlled photonic circuits of
  field-induced dipolaritons with huge nonlinearities},}\ }\href {\doibase
  10.1103/PhysRevX.14.031022} {\bibfield  {journal} {\bibinfo  {journal} {Phys.
  Rev. X}\ }\textbf {\bibinfo {volume} {14}},\ \bibinfo {pages} {031022}
  (\bibinfo {year} {2024})}\BibitemShut {NoStop}%
\bibitem [{\citenamefont {Tartakovskii}\ \emph {et~al.}(1998)\citenamefont
  {Tartakovskii}, \citenamefont {Kulakovskii}, \citenamefont {Forchel},\ and\
  \citenamefont {Reithmaier}}]{Tartakovskii:PRB1998}%
  \BibitemOpen
  \bibfield  {author} {\bibinfo {author} {\bibfnamefont {A.~I.}\ \bibnamefont
  {Tartakovskii}}, \bibinfo {author} {\bibfnamefont {V.~D.}\ \bibnamefont
  {Kulakovskii}}, \bibinfo {author} {\bibfnamefont {A.}~\bibnamefont
  {Forchel}}, \ and\ \bibinfo {author} {\bibfnamefont {J.~P.}\ \bibnamefont
  {Reithmaier}},\ }\bibfield  {title} {\enquote {\bibinfo {title}
  {Exciton-photon coupling in photonic wires},}\ }\href {\doibase
  10.1103/PhysRevB.57.R6807} {\bibfield  {journal} {\bibinfo  {journal} {Phys.
  Rev. B}\ }\textbf {\bibinfo {volume} {57}},\ \bibinfo {pages} {R6807--R6810}
  (\bibinfo {year} {1998})}\BibitemShut {NoStop}%
\bibitem [{\citenamefont {Lecomte}\ \emph {et~al.}(2013)\citenamefont
  {Lecomte}, \citenamefont {Ardizzone}, \citenamefont {Abbarchi}, \citenamefont
  {Diederichs}, \citenamefont {Miard}, \citenamefont {Lemaitre}, \citenamefont
  {Sagnes}, \citenamefont {Senellart}, \citenamefont {Bloch}, \citenamefont
  {Delalande}, \citenamefont {Tignon},\ and\ \citenamefont
  {Roussignol}}]{Lecomte:PRB2013}%
  \BibitemOpen
  \bibfield  {author} {\bibinfo {author} {\bibfnamefont {Timoth\'ee}\
  \bibnamefont {Lecomte}}, \bibinfo {author} {\bibfnamefont {Vincenzo}\
  \bibnamefont {Ardizzone}}, \bibinfo {author} {\bibfnamefont {Marco}\
  \bibnamefont {Abbarchi}}, \bibinfo {author} {\bibfnamefont {Carole}\
  \bibnamefont {Diederichs}}, \bibinfo {author} {\bibfnamefont {Audrey}\
  \bibnamefont {Miard}}, \bibinfo {author} {\bibfnamefont {Aristide}\
  \bibnamefont {Lemaitre}}, \bibinfo {author} {\bibfnamefont {Isabelle}\
  \bibnamefont {Sagnes}}, \bibinfo {author} {\bibfnamefont {Pascale}\
  \bibnamefont {Senellart}}, \bibinfo {author} {\bibfnamefont {Jacqueline}\
  \bibnamefont {Bloch}}, \bibinfo {author} {\bibfnamefont {Claude}\
  \bibnamefont {Delalande}}, \bibinfo {author} {\bibfnamefont {Jerome}\
  \bibnamefont {Tignon}}, \ and\ \bibinfo {author} {\bibfnamefont {Philippe}\
  \bibnamefont {Roussignol}},\ }\bibfield  {title} {\enquote {\bibinfo {title}
  {Optical parametric oscillation in one-dimensional microcavities},}\ }\href
  {\doibase 10.1103/PhysRevB.87.155302} {\bibfield  {journal} {\bibinfo
  {journal} {Phys. Rev. B}\ }\textbf {\bibinfo {volume} {87}},\ \bibinfo
  {pages} {155302} (\bibinfo {year} {2013})}\BibitemShut {NoStop}%
\bibitem [{\citenamefont {Levinsen}\ \emph {et~al.}(2019)\citenamefont
  {Levinsen}, \citenamefont {Li},\ and\ \citenamefont
  {Parish}}]{Levinsen2019b}%
  \BibitemOpen
  \bibfield  {author} {\bibinfo {author} {\bibfnamefont {Jesper}\ \bibnamefont
  {Levinsen}}, \bibinfo {author} {\bibfnamefont {Guangyao}\ \bibnamefont {Li}},
  \ and\ \bibinfo {author} {\bibfnamefont {Meera~M.}\ \bibnamefont {Parish}},\
  }\bibfield  {title} {\enquote {\bibinfo {title} {Microscopic description of
  exciton-polaritons in microcavities},}\ }\href {\doibase
  10.1103/PhysRevResearch.1.033120} {\bibfield  {journal} {\bibinfo  {journal}
  {Phys. Rev. Res.}\ }\textbf {\bibinfo {volume} {1}},\ \bibinfo {pages}
  {033120} (\bibinfo {year} {2019})}\BibitemShut {NoStop}%
\bibitem [{\citenamefont {Tan}\ \emph {et~al.}(1990)\citenamefont {Tan},
  \citenamefont {Snider}, \citenamefont {Chang},\ and\ \citenamefont
  {Hu}}]{Tan1990}%
  \BibitemOpen
  \bibfield  {author} {\bibinfo {author} {\bibfnamefont {I.-H.}\ \bibnamefont
  {Tan}}, \bibinfo {author} {\bibfnamefont {Gregory~L.}\ \bibnamefont
  {Snider}}, \bibinfo {author} {\bibfnamefont {L.~D.}\ \bibnamefont {Chang}}, \
  and\ \bibinfo {author} {\bibfnamefont {Evelyn~L.}\ \bibnamefont {Hu}},\
  }\bibfield  {title} {\enquote {\bibinfo {title} {A self‐consistent solution
  of schr{\"o}dinger–poisson equations using a nonuniform mesh},}\
  }\href@noop {} {\bibfield  {journal} {\bibinfo  {journal} {Journal of Applied
  Physics}\ }\textbf {\bibinfo {volume} {68}},\ \bibinfo {pages} {4071--4076}
  (\bibinfo {year} {1990})}\BibitemShut {NoStop}%
\bibitem [{\citenamefont {Lanczos}(1950)}]{Lanczos1950}%
  \BibitemOpen
  \bibfield  {author} {\bibinfo {author} {\bibfnamefont {Cornelius}\
  \bibnamefont {Lanczos}},\ }\bibfield  {title} {\enquote {\bibinfo {title} {An
  iteration method for the solution of the eigenvalue problem of linear
  differential and integral operators},}\ }\href@noop {} {\bibfield  {journal}
  {\bibinfo  {journal} {Journal of research of the National Bureau of
  Standards}\ }\textbf {\bibinfo {volume} {45}},\ \bibinfo {pages} {255--282}
  (\bibinfo {year} {1950})}\BibitemShut {NoStop}%
\bibitem [{\citenamefont {Stewart}(2002)}]{Stewart2002}%
  \BibitemOpen
  \bibfield  {author} {\bibinfo {author} {\bibfnamefont {G.~W.}\ \bibnamefont
  {Stewart}},\ }\bibfield  {title} {\enquote {\bibinfo {title} {A krylov--schur
  algorithm for large eigenproblems},}\ }\href {\doibase
  10.1137/S0895479800371529} {\bibfield  {journal} {\bibinfo  {journal} {SIAM
  Journal on Matrix Analysis and Applications}\ }\textbf {\bibinfo {volume}
  {23}},\ \bibinfo {pages} {601--614} (\bibinfo {year} {2002})}\BibitemShut
  {NoStop}%
\bibitem [{\citenamefont {Tonks}(1936)}]{Tonks36}%
  \BibitemOpen
  \bibfield  {author} {\bibinfo {author} {\bibfnamefont {Lewi}\ \bibnamefont
  {Tonks}},\ }\bibfield  {title} {\enquote {\bibinfo {title} {The complete
  equation of state of one, two and three-dimensional gases of hard elastic
  spheres},}\ }\href {\doibase 10.1103/PhysRev.50.955} {\bibfield  {journal}
  {\bibinfo  {journal} {Phys. Rev.}\ }\textbf {\bibinfo {volume} {50}},\
  \bibinfo {pages} {955--963} (\bibinfo {year} {1936})}\BibitemShut {NoStop}%
\bibitem [{\citenamefont {Girardeau}(1960)}]{girardeau60}%
  \BibitemOpen
  \bibfield  {author} {\bibinfo {author} {\bibfnamefont {M.}~\bibnamefont
  {Girardeau}},\ }\bibfield  {title} {\enquote {\bibinfo {title} {Relationship
  between systems of impenetrable bosons and fermions in one dimension},}\
  }\href {\doibase 10.1063/1.1703687} {\bibfield  {journal} {\bibinfo
  {journal} {Journal of Mathematical Physics}\ }\textbf {\bibinfo {volume}
  {1}},\ \bibinfo {pages} {516--523} (\bibinfo {year} {1960})},\ \Eprint
  {http://arxiv.org/abs/https://doi.org/10.1063/1.1703687}
  {https://doi.org/10.1063/1.1703687} \BibitemShut {NoStop}%
\bibitem [{\citenamefont {Lieb}\ and\ \citenamefont {Liniger}(1963)}]{Lieb63}%
  \BibitemOpen
  \bibfield  {author} {\bibinfo {author} {\bibfnamefont {Elliott~H.}\
  \bibnamefont {Lieb}}\ and\ \bibinfo {author} {\bibfnamefont {Werner}\
  \bibnamefont {Liniger}},\ }\bibfield  {title} {\enquote {\bibinfo {title}
  {Exact analysis of an interacting bose gas. i. the general solution and the
  ground state},}\ }\href {\doibase 10.1103/PhysRev.130.1605} {\bibfield
  {journal} {\bibinfo  {journal} {Phys. Rev.}\ }\textbf {\bibinfo {volume}
  {130}},\ \bibinfo {pages} {1605--1616} (\bibinfo {year} {1963})}\BibitemShut
  {NoStop}%
\bibitem [{\citenamefont {Jaynes}\ and\ \citenamefont
  {Cummings}(1963)}]{JCM63}%
  \BibitemOpen
  \bibfield  {author} {\bibinfo {author} {\bibfnamefont {E.T.}\ \bibnamefont
  {Jaynes}}\ and\ \bibinfo {author} {\bibfnamefont {F.W.}\ \bibnamefont
  {Cummings}},\ }\bibfield  {title} {\enquote {\bibinfo {title} {Comparison of
  quantum and semiclassical radiation theories with application to the beam
  maser},}\ }\href {\doibase 10.1109/PROC.1963.1664} {\bibfield  {journal}
  {\bibinfo  {journal} {Proceedings of the IEEE}\ }\textbf {\bibinfo {volume}
  {51}},\ \bibinfo {pages} {89--109} (\bibinfo {year} {1963})}\BibitemShut
  {NoStop}%
\bibitem [{\citenamefont {Tian}\ and\ \citenamefont
  {Carmichael}(1992)}]{Tian92}%
  \BibitemOpen
  \bibfield  {author} {\bibinfo {author} {\bibfnamefont {L.}~\bibnamefont
  {Tian}}\ and\ \bibinfo {author} {\bibfnamefont {H.~J.}\ \bibnamefont
  {Carmichael}},\ }\bibfield  {title} {\enquote {\bibinfo {title} {Quantum
  trajectory simulations of two-state behavior in an optical cavity containing
  one atom},}\ }\href {\doibase 10.1103/PhysRevA.46.R6801} {\bibfield
  {journal} {\bibinfo  {journal} {Phys. Rev. A}\ }\textbf {\bibinfo {volume}
  {46}},\ \bibinfo {pages} {R6801--R6804} (\bibinfo {year} {1992})}\BibitemShut
  {NoStop}%
\bibitem [{\citenamefont {Pitaevskii}\ and\ \citenamefont
  {Stringari}(2003)}]{PitaevskiiStringari2003book}%
  \BibitemOpen
  \bibfield  {author} {\bibinfo {author} {\bibfnamefont {L.P.}\ \bibnamefont
  {Pitaevskii}}\ and\ \bibinfo {author} {\bibfnamefont {S.}~\bibnamefont
  {Stringari}},\ }\href@noop {} {\emph {\bibinfo {title} {Bose-Einstein
  Condensation}}},\ International Series of Monographs on Physics\ (\bibinfo
  {publisher} {Clarendon Press},\ \bibinfo {year} {2003})\BibitemShut {NoStop}%
\bibitem [{\citenamefont {Born}\ and\ \citenamefont
  {Oppenheimer}(1927)}]{BO1927}%
  \BibitemOpen
  \bibfield  {author} {\bibinfo {author} {\bibfnamefont {M.}~\bibnamefont
  {Born}}\ and\ \bibinfo {author} {\bibfnamefont {R.}~\bibnamefont
  {Oppenheimer}},\ }\bibfield  {title} {\enquote {\bibinfo {title} {Zur
  quantentheorie der molekeln},}\ }\href {\doibase
  https://doi.org/10.1002/andp.19273892002} {\bibfield  {journal} {\bibinfo
  {journal} {Annalen der Physik}\ }\textbf {\bibinfo {volume} {389}},\ \bibinfo
  {pages} {457--484} (\bibinfo {year} {1927})}\BibitemShut {NoStop}%
\bibitem [{\citenamefont {Rochat}\ \emph {et~al.}(2000)\citenamefont {Rochat},
  \citenamefont {Ciuti}, \citenamefont {Savona}, \citenamefont {Piermarocchi},
  \citenamefont {Quattropani},\ and\ \citenamefont
  {Schwendimann}}]{Rochat2000}%
  \BibitemOpen
  \bibfield  {author} {\bibinfo {author} {\bibfnamefont {G.}~\bibnamefont
  {Rochat}}, \bibinfo {author} {\bibfnamefont {C.}~\bibnamefont {Ciuti}},
  \bibinfo {author} {\bibfnamefont {V.}~\bibnamefont {Savona}}, \bibinfo
  {author} {\bibfnamefont {C.}~\bibnamefont {Piermarocchi}}, \bibinfo {author}
  {\bibfnamefont {A.}~\bibnamefont {Quattropani}}, \ and\ \bibinfo {author}
  {\bibfnamefont {P.}~\bibnamefont {Schwendimann}},\ }\bibfield  {title}
  {\enquote {\bibinfo {title} {Excitonic bloch equations for a two-dimensional
  system of interacting excitons},}\ }\href {\doibase
  10.1103/PhysRevB.61.13856} {\bibfield  {journal} {\bibinfo  {journal} {Phys.
  Rev. B}\ }\textbf {\bibinfo {volume} {61}},\ \bibinfo {pages} {13856--13862}
  (\bibinfo {year} {2000})}\BibitemShut {NoStop}%
\bibitem [{Note1()}]{Note1}%
  \BibitemOpen
  \bibinfo {note} {It is interesting to note how the scaling \protect \eqref
  {eq:BOpert_scaling2D} strongly differs from the prediction of the so-called
  capacitor formula, $\Delta E_{\protect \rm int}^{\protect \rm (cap)} =
  (de^2/\epsilon ) n_X$ often used to describe gases of dipolar excitons,
  $n_X=L^{-2}$ being the exciton density~\cite {Liran:PRX2024}. Several
  physical reasons justify this disagreement, including the key role of the
  light-matter interaction and the perturbative nature of the capacitor formula
  in the exciton-exciton interaction potential. In contrast to this, our theory
  assumes in fact from the outset that interaction between excitons are
  strongly repulsive so that $R>d$ and excitons are not able to
  overlap.}\BibitemShut {Stop}%
\bibitem [{\citenamefont {Yu}\ and\ \citenamefont
  {Cardona}(2005)}]{yu-cardona}%
  \BibitemOpen
  \bibfield  {author} {\bibinfo {author} {\bibfnamefont {Peter~Y.}\
  \bibnamefont {Yu}}\ and\ \bibinfo {author} {\bibfnamefont {Manuel}\
  \bibnamefont {Cardona}},\ }\href@noop {} {\emph {\bibinfo {title}
  {Fundamentals of semiconductors}}}\ (\bibinfo  {publisher} {Springer},\
  \bibinfo {year} {2005})\BibitemShut {NoStop}%
\bibitem [{\citenamefont {Agranovich}\ \emph {et~al.}(1998)\citenamefont
  {Agranovich}, \citenamefont {Basko}, \citenamefont {La~Rocca},\ and\
  \citenamefont {Bassani}}]{agranovich1998excitons}%
  \BibitemOpen
  \bibfield  {author} {\bibinfo {author} {\bibfnamefont {VM}~\bibnamefont
  {Agranovich}}, \bibinfo {author} {\bibfnamefont {DM}~\bibnamefont {Basko}},
  \bibinfo {author} {\bibfnamefont {GC}~\bibnamefont {La~Rocca}}, \ and\
  \bibinfo {author} {\bibfnamefont {Franco}\ \bibnamefont {Bassani}},\
  }\bibfield  {title} {\enquote {\bibinfo {title} {Excitons and optical
  nonlinearities in hybrid organic-inorganic nanostructures},}\ }\href@noop {}
  {\bibfield  {journal} {\bibinfo  {journal} {Journal of Physics: Condensed
  Matter}\ }\textbf {\bibinfo {volume} {10}},\ \bibinfo {pages} {9369}
  (\bibinfo {year} {1998})}\BibitemShut {NoStop}%
\bibitem [{\citenamefont {Byrnes}\ \emph {et~al.}(2014)\citenamefont {Byrnes},
  \citenamefont {Kolmakov}, \citenamefont {Kezerashvili},\ and\ \citenamefont
  {Yamamoto}}]{Byrnes2014}%
  \BibitemOpen
  \bibfield  {author} {\bibinfo {author} {\bibfnamefont {Tim}\ \bibnamefont
  {Byrnes}}, \bibinfo {author} {\bibfnamefont {German~V.}\ \bibnamefont
  {Kolmakov}}, \bibinfo {author} {\bibfnamefont {Roman~Ya.}\ \bibnamefont
  {Kezerashvili}}, \ and\ \bibinfo {author} {\bibfnamefont {Yoshihisa}\
  \bibnamefont {Yamamoto}},\ }\bibfield  {title} {\enquote {\bibinfo {title}
  {Effective interaction and condensation of dipolaritons in coupled quantum
  wells},}\ }\href {\doibase 10.1103/PhysRevB.90.125314} {\bibfield  {journal}
  {\bibinfo  {journal} {Phys. Rev. B}\ }\textbf {\bibinfo {volume} {90}},\
  \bibinfo {pages} {125314} (\bibinfo {year} {2014})}\BibitemShut {NoStop}%
\bibitem [{\citenamefont {Barachati}\ \emph {et~al.}(2018)\citenamefont
  {Barachati}, \citenamefont {Fieramosca}, \citenamefont {Hafezian},
  \citenamefont {Gu}, \citenamefont {Chakraborty}, \citenamefont {Ballarini},
  \citenamefont {Martinu}, \citenamefont {Menon}, \citenamefont {Sanvitto},\
  and\ \citenamefont {K{\'e}na-Cohen}}]{barachati2018interacting}%
  \BibitemOpen
  \bibfield  {author} {\bibinfo {author} {\bibfnamefont {F{\'a}bio}\
  \bibnamefont {Barachati}}, \bibinfo {author} {\bibfnamefont {Antonio}\
  \bibnamefont {Fieramosca}}, \bibinfo {author} {\bibfnamefont {Soroush}\
  \bibnamefont {Hafezian}}, \bibinfo {author} {\bibfnamefont {Jie}\
  \bibnamefont {Gu}}, \bibinfo {author} {\bibfnamefont {Biswanath}\
  \bibnamefont {Chakraborty}}, \bibinfo {author} {\bibfnamefont {Dario}\
  \bibnamefont {Ballarini}}, \bibinfo {author} {\bibfnamefont {Ludvik}\
  \bibnamefont {Martinu}}, \bibinfo {author} {\bibfnamefont {Vinod}\
  \bibnamefont {Menon}}, \bibinfo {author} {\bibfnamefont {Daniele}\
  \bibnamefont {Sanvitto}}, \ and\ \bibinfo {author} {\bibfnamefont
  {St{\'e}phane}\ \bibnamefont {K{\'e}na-Cohen}},\ }\bibfield  {title}
  {\enquote {\bibinfo {title} {Interacting polariton fluids in a monolayer of
  tungsten disulfide},}\ }\href@noop {} {\bibfield  {journal} {\bibinfo
  {journal} {Nature nanotechnology}\ }\textbf {\bibinfo {volume} {13}},\
  \bibinfo {pages} {906--909} (\bibinfo {year} {2018})}\BibitemShut {NoStop}%
\bibitem [{\citenamefont {Fieramosca}\ \emph {et~al.}(2019)\citenamefont
  {Fieramosca}, \citenamefont {Polimeno}, \citenamefont {Ardizzone},
  \citenamefont {De~Marco}, \citenamefont {Pugliese}, \citenamefont {Maiorano},
  \citenamefont {De~Giorgi}, \citenamefont {Dominici}, \citenamefont {Gigli},
  \citenamefont {Gerace} \emph {et~al.}}]{fieramosca2019two}%
  \BibitemOpen
  \bibfield  {author} {\bibinfo {author} {\bibfnamefont {A}~\bibnamefont
  {Fieramosca}}, \bibinfo {author} {\bibfnamefont {L}~\bibnamefont {Polimeno}},
  \bibinfo {author} {\bibfnamefont {V}~\bibnamefont {Ardizzone}}, \bibinfo
  {author} {\bibfnamefont {L}~\bibnamefont {De~Marco}}, \bibinfo {author}
  {\bibfnamefont {M}~\bibnamefont {Pugliese}}, \bibinfo {author} {\bibfnamefont
  {V}~\bibnamefont {Maiorano}}, \bibinfo {author} {\bibfnamefont
  {M}~\bibnamefont {De~Giorgi}}, \bibinfo {author} {\bibfnamefont
  {L}~\bibnamefont {Dominici}}, \bibinfo {author} {\bibfnamefont
  {G}~\bibnamefont {Gigli}}, \bibinfo {author} {\bibfnamefont {D}~\bibnamefont
  {Gerace}},  \emph {et~al.},\ }\bibfield  {title} {\enquote {\bibinfo {title}
  {Two-dimensional hybrid perovskites sustaining strong polariton interactions
  at room temperature},}\ }\href@noop {} {\bibfield  {journal} {\bibinfo
  {journal} {Science advances}\ }\textbf {\bibinfo {volume} {5}},\ \bibinfo
  {pages} {eaav9967} (\bibinfo {year} {2019})}\BibitemShut {NoStop}%
\bibitem [{\citenamefont {Frérot}\ \emph {et~al.}(2023)\citenamefont
  {Frérot}, \citenamefont {Vashisht}, \citenamefont {Morassi}, \citenamefont
  {Lemaître}, \citenamefont {Ravets}, \citenamefont {Bloch}, \citenamefont
  {Minguzzi},\ and\ \citenamefont {Richard}}]{frerot2023bogoliubov}%
  \BibitemOpen
  \bibfield  {author} {\bibinfo {author} {\bibfnamefont {Irénée}\
  \bibnamefont {Frérot}}, \bibinfo {author} {\bibfnamefont {Amit}\
  \bibnamefont {Vashisht}}, \bibinfo {author} {\bibfnamefont {Martina}\
  \bibnamefont {Morassi}}, \bibinfo {author} {\bibfnamefont {Aristide}\
  \bibnamefont {Lemaître}}, \bibinfo {author} {\bibfnamefont {Sylvain}\
  \bibnamefont {Ravets}}, \bibinfo {author} {\bibfnamefont {Jacqueline}\
  \bibnamefont {Bloch}}, \bibinfo {author} {\bibfnamefont {Anna}\ \bibnamefont
  {Minguzzi}}, \ and\ \bibinfo {author} {\bibfnamefont {Maxime}\ \bibnamefont
  {Richard}},\ }\href@noop {} {\enquote {\bibinfo {title} {Bogoliubov
  excitations driven by thermal lattice phonons in a quantum fluid of light},}\
  } (\bibinfo {year} {2023}),\ \Eprint {http://arxiv.org/abs/2304.08677}
  {arXiv:2304.08677 [cond-mat.quant-gas]} \BibitemShut {NoStop}%
\bibitem [{\citenamefont {Sun}\ \emph {et~al.}(2017)\citenamefont {Sun},
  \citenamefont {Wen}, \citenamefont {Yoon}, \citenamefont {Liu}, \citenamefont
  {Steger}, \citenamefont {Pfeiffer}, \citenamefont {West}, \citenamefont
  {Snoke},\ and\ \citenamefont {Nelson}}]{SunY2017}%
  \BibitemOpen
  \bibfield  {author} {\bibinfo {author} {\bibfnamefont {Yongbao}\ \bibnamefont
  {Sun}}, \bibinfo {author} {\bibfnamefont {Patrick}\ \bibnamefont {Wen}},
  \bibinfo {author} {\bibfnamefont {Yoseob}\ \bibnamefont {Yoon}}, \bibinfo
  {author} {\bibfnamefont {Gangqiang}\ \bibnamefont {Liu}}, \bibinfo {author}
  {\bibfnamefont {Mark}\ \bibnamefont {Steger}}, \bibinfo {author}
  {\bibfnamefont {Loren~N.}\ \bibnamefont {Pfeiffer}}, \bibinfo {author}
  {\bibfnamefont {Ken}\ \bibnamefont {West}}, \bibinfo {author} {\bibfnamefont
  {David~W.}\ \bibnamefont {Snoke}}, \ and\ \bibinfo {author} {\bibfnamefont
  {Keith~A.}\ \bibnamefont {Nelson}},\ }\bibfield  {title} {\enquote {\bibinfo
  {title} {Bose-einstein condensation of long-lifetime polaritons in thermal
  equilibrium},}\ }\href {\doibase 10.1103/PhysRevLett.118.016602} {\bibfield
  {journal} {\bibinfo  {journal} {Phys. Rev. Lett.}\ }\textbf {\bibinfo
  {volume} {118}},\ \bibinfo {pages} {016602} (\bibinfo {year}
  {2017})}\BibitemShut {NoStop}%
\bibitem [{\citenamefont {Peyronel}\ \emph {et~al.}(2012)\citenamefont
  {Peyronel}, \citenamefont {Firstenberg}, \citenamefont {Liang}, \citenamefont
  {Hofferberth}, \citenamefont {Gorshkov}, \citenamefont {Pohl}, \citenamefont
  {Lukin},\ and\ \citenamefont {Vuleti{\'c}}}]{peyronel2012quantum}%
  \BibitemOpen
  \bibfield  {author} {\bibinfo {author} {\bibfnamefont {Thibault}\
  \bibnamefont {Peyronel}}, \bibinfo {author} {\bibfnamefont {Ofer}\
  \bibnamefont {Firstenberg}}, \bibinfo {author} {\bibfnamefont {Qi-Yu}\
  \bibnamefont {Liang}}, \bibinfo {author} {\bibfnamefont {Sebastian}\
  \bibnamefont {Hofferberth}}, \bibinfo {author} {\bibfnamefont {Alexey~V}\
  \bibnamefont {Gorshkov}}, \bibinfo {author} {\bibfnamefont {Thomas}\
  \bibnamefont {Pohl}}, \bibinfo {author} {\bibfnamefont {Mikhail~D}\
  \bibnamefont {Lukin}}, \ and\ \bibinfo {author} {\bibfnamefont {Vladan}\
  \bibnamefont {Vuleti{\'c}}},\ }\bibfield  {title} {\enquote {\bibinfo {title}
  {Quantum nonlinear optics with single photons enabled by strongly interacting
  atoms},}\ }\href@noop {} {\bibfield  {journal} {\bibinfo  {journal} {Nature}\
  }\textbf {\bibinfo {volume} {488}},\ \bibinfo {pages} {57--60} (\bibinfo
  {year} {2012})}\BibitemShut {NoStop}%
\bibitem [{\citenamefont {Gorshkov}\ \emph {et~al.}(2011)\citenamefont
  {Gorshkov}, \citenamefont {Otterbach}, \citenamefont {Fleischhauer},
  \citenamefont {Pohl},\ and\ \citenamefont {Lukin}}]{Gorshkov2011}%
  \BibitemOpen
  \bibfield  {author} {\bibinfo {author} {\bibfnamefont {Alexey~V.}\
  \bibnamefont {Gorshkov}}, \bibinfo {author} {\bibfnamefont {Johannes}\
  \bibnamefont {Otterbach}}, \bibinfo {author} {\bibfnamefont {Michael}\
  \bibnamefont {Fleischhauer}}, \bibinfo {author} {\bibfnamefont {Thomas}\
  \bibnamefont {Pohl}}, \ and\ \bibinfo {author} {\bibfnamefont {Mikhail~D.}\
  \bibnamefont {Lukin}},\ }\bibfield  {title} {\enquote {\bibinfo {title}
  {Photon-photon interactions via rydberg blockade},}\ }\href {\doibase
  10.1103/PhysRevLett.107.133602} {\bibfield  {journal} {\bibinfo  {journal}
  {Phys. Rev. Lett.}\ }\textbf {\bibinfo {volume} {107}},\ \bibinfo {pages}
  {133602} (\bibinfo {year} {2011})}\BibitemShut {NoStop}%
\bibitem [{\citenamefont {Firstenberg}\ \emph {et~al.}(2013)\citenamefont
  {Firstenberg}, \citenamefont {Peyronel}, \citenamefont {Liang}, \citenamefont
  {Gorshkov}, \citenamefont {Lukin},\ and\ \citenamefont
  {Vuleti{\'c}}}]{firstenberg2013attractive}%
  \BibitemOpen
  \bibfield  {author} {\bibinfo {author} {\bibfnamefont {Ofer}\ \bibnamefont
  {Firstenberg}}, \bibinfo {author} {\bibfnamefont {Thibault}\ \bibnamefont
  {Peyronel}}, \bibinfo {author} {\bibfnamefont {Qi-Yu}\ \bibnamefont {Liang}},
  \bibinfo {author} {\bibfnamefont {Alexey~V}\ \bibnamefont {Gorshkov}},
  \bibinfo {author} {\bibfnamefont {Mikhail~D}\ \bibnamefont {Lukin}}, \ and\
  \bibinfo {author} {\bibfnamefont {Vladan}\ \bibnamefont {Vuleti{\'c}}},\
  }\bibfield  {title} {\enquote {\bibinfo {title} {Attractive photons in a
  quantum nonlinear medium},}\ }\href@noop {} {\bibfield  {journal} {\bibinfo
  {journal} {Nature}\ }\textbf {\bibinfo {volume} {502}},\ \bibinfo {pages}
  {71--75} (\bibinfo {year} {2013})}\BibitemShut {NoStop}%
\bibitem [{\citenamefont {Walther}\ \emph {et~al.}(2018)\citenamefont
  {Walther}, \citenamefont {Johne},\ and\ \citenamefont {Pohl}}]{Walther2018}%
  \BibitemOpen
  \bibfield  {author} {\bibinfo {author} {\bibfnamefont {Valentin}\
  \bibnamefont {Walther}}, \bibinfo {author} {\bibfnamefont {Robert}\
  \bibnamefont {Johne}}, \ and\ \bibinfo {author} {\bibfnamefont {Thomas}\
  \bibnamefont {Pohl}},\ }\bibfield  {title} {\enquote {\bibinfo {title} {Giant
  optical nonlinearities from rydberg excitons in semiconductor
  microcavities},}\ }\href {\doibase 10.1038/s41467-018-03742-7} {\bibfield
  {journal} {\bibinfo  {journal} {Nature Communications}\ }\textbf {\bibinfo
  {volume} {9}},\ \bibinfo {pages} {1309} (\bibinfo {year} {2018})}\BibitemShut
  {NoStop}%
\bibitem [{\citenamefont {Orfanakis}\ \emph {et~al.}(2022)\citenamefont
  {Orfanakis}, \citenamefont {Rajendran}, \citenamefont {Walther},
  \citenamefont {Volz}, \citenamefont {Pohl},\ and\ \citenamefont
  {Ohadi}}]{Orfanakis2022}%
  \BibitemOpen
  \bibfield  {author} {\bibinfo {author} {\bibfnamefont {Konstantinos}\
  \bibnamefont {Orfanakis}}, \bibinfo {author} {\bibfnamefont {Sai~Kiran}\
  \bibnamefont {Rajendran}}, \bibinfo {author} {\bibfnamefont {Valentin}\
  \bibnamefont {Walther}}, \bibinfo {author} {\bibfnamefont {Thomas}\
  \bibnamefont {Volz}}, \bibinfo {author} {\bibfnamefont {Thomas}\ \bibnamefont
  {Pohl}}, \ and\ \bibinfo {author} {\bibfnamefont {Hamid}\ \bibnamefont
  {Ohadi}},\ }\bibfield  {title} {\enquote {\bibinfo {title} {Rydberg
  exciton--polaritons in a cu2o microcavity},}\ }\href {\doibase
  10.1038/s41563-022-01230-4} {\bibfield  {journal} {\bibinfo  {journal}
  {Nature Materials}\ }\textbf {\bibinfo {volume} {21}},\ \bibinfo {pages}
  {767--772} (\bibinfo {year} {2022})}\BibitemShut {NoStop}%
\bibitem [{\citenamefont {Kn{\"o}rzer}\ \emph {et~al.}(2024)\citenamefont
  {Kn{\"o}rzer}, \citenamefont {O{\l}dziejewski}, \citenamefont {Murthy},\ and\
  \citenamefont {Amelio}}]{knorzer2024fermionization}%
  \BibitemOpen
  \bibfield  {author} {\bibinfo {author} {\bibfnamefont {Johannes}\
  \bibnamefont {Kn{\"o}rzer}}, \bibinfo {author} {\bibfnamefont {Rafa{\l}}\
  \bibnamefont {O{\l}dziejewski}}, \bibinfo {author} {\bibfnamefont {Puneet~A}\
  \bibnamefont {Murthy}}, \ and\ \bibinfo {author} {\bibfnamefont {Ivan}\
  \bibnamefont {Amelio}},\ }\bibfield  {title} {\enquote {\bibinfo {title}
  {Fermionization and collective excitations of 1d polariton lattices},}\
  }\href@noop {} {\bibfield  {journal} {\bibinfo  {journal} {arXiv preprint
  arXiv:2405.02251}\ } (\bibinfo {year} {2024})}\BibitemShut {NoStop}%
\bibitem [{\citenamefont {Hu}\ \emph {et~al.}(2024)\citenamefont {Hu},
  \citenamefont {Lorchat}, \citenamefont {Chen}, \citenamefont {Watanabe},
  \citenamefont {Taniguchi}, \citenamefont {Heinz}, \citenamefont {Murthy},\
  and\ \citenamefont {Chervy}}]{hu2024quantum}%
  \BibitemOpen
  \bibfield  {author} {\bibinfo {author} {\bibfnamefont {Jenny}\ \bibnamefont
  {Hu}}, \bibinfo {author} {\bibfnamefont {Etienne}\ \bibnamefont {Lorchat}},
  \bibinfo {author} {\bibfnamefont {Xueqi}\ \bibnamefont {Chen}}, \bibinfo
  {author} {\bibfnamefont {Kenji}\ \bibnamefont {Watanabe}}, \bibinfo {author}
  {\bibfnamefont {Takashi}\ \bibnamefont {Taniguchi}}, \bibinfo {author}
  {\bibfnamefont {Tony~F}\ \bibnamefont {Heinz}}, \bibinfo {author}
  {\bibfnamefont {Puneet~A}\ \bibnamefont {Murthy}}, \ and\ \bibinfo {author}
  {\bibfnamefont {Thibault}\ \bibnamefont {Chervy}},\ }\bibfield  {title}
  {\enquote {\bibinfo {title} {Quantum control of exciton wave functions in 2d
  semiconductors},}\ }\href@noop {} {\bibfield  {journal} {\bibinfo  {journal}
  {Science Advances}\ }\textbf {\bibinfo {volume} {10}},\ \bibinfo {pages}
  {eadk6369} (\bibinfo {year} {2024})}\BibitemShut {NoStop}%
\bibitem [{\citenamefont {Thureja}\ \emph {et~al.}(2024)\citenamefont
  {Thureja}, \citenamefont {Yazici}, \citenamefont {Smolenski}, \citenamefont
  {Kroner}, \citenamefont {Norris},\ and\ \citenamefont
  {Imamoglu}}]{thureja2024electrically}%
  \BibitemOpen
  \bibfield  {author} {\bibinfo {author} {\bibfnamefont {Deepankur}\
  \bibnamefont {Thureja}}, \bibinfo {author} {\bibfnamefont {F~Emre}\
  \bibnamefont {Yazici}}, \bibinfo {author} {\bibfnamefont {Tomasz}\
  \bibnamefont {Smolenski}}, \bibinfo {author} {\bibfnamefont {Martin}\
  \bibnamefont {Kroner}}, \bibinfo {author} {\bibfnamefont {David~J}\
  \bibnamefont {Norris}}, \ and\ \bibinfo {author} {\bibfnamefont {Atac}\
  \bibnamefont {Imamoglu}},\ }\bibfield  {title} {\enquote {\bibinfo {title}
  {Electrically defined quantum dots for bosonic excitons},}\ }\href@noop {}
  {\bibfield  {journal} {\bibinfo  {journal} {arXiv preprint arXiv:2402.19278}\
  } (\bibinfo {year} {2024})}\BibitemShut {NoStop}%
\end{thebibliography}%

\clearpage

\appendix

\section{Coupled wire model}
\label{2Dapp}

In this Appendix, we validate the use of a BO approach in higher dimensions by comparing its predictions with the result of a full numerical calculation for a simplified model including a pair of coupled wires. In particular, we consider two kinds of excitons sitting on a pair of parallel wires, with the possibility of tunneling between them and, in this way, of passing next one another at a not excessive energy cost. Given the spatial separation $\Delta z$ between the wires, dipolar interactions are in fact stronger for excitons located on the same wire.

\begin{widetext}
This model can be formalized in the following Hamiltonian
\begin{multline}
    \hat{H} = \sum_{k=1,2} \int d{x_k} \!\!\!\sum_{i,j=\{C,X_1,X_2\}}\!\! \hat{\Psi}_i^\dagger ({x_k}) h^{0}_{ij}({x_k}) \hat{\Psi}_j ({x_k}) 
    +\frac{1}{2} \sum_{i=1,2} \int d{x_1}\, d{x_2} \,\big[ \hat{\Psi}^\dagger_{X_i}({x_1})\hat{\Psi}^\dagger_{X_i}({x_2}) V_{\text{dd}}(|{x_1-x_2}|) \hat{\Psi}_{X_i}({x_1})\hat{\Psi}_{X_i}({x_2})\big] \\
    + 
    \int d{x_1}\, d{x_2}\, \big[ \hat{\Psi}^\dagger_{X_1}({x_1})\hat{\Psi}^\dagger_{X_2}({x_2}) V_{\text{dd}}\left(\sqrt{|{x_1-x_2}|^2+(\Delta z)^2}\right) \hat{\Psi}_{X_1}({x_1})\hat{\Psi}_{X_2}({x_2})\big] 
    ,
    \label{Hamiltonian2dnew}
\end{multline}
where h.c. means the hermitian conjugate of the term just before, and $h^0_{ij}$ are elements of the matrix $h^0$ given as 
\begin{align}
    h^0({x_k}) = \begin{pmatrix}
     -\frac{\hbar^2}{2m_C}\frac{\partial^2}{\partial x_k^2}+\Delta+\delta_C& \Omega& \Omega \\
     \Omega & -\frac{\hbar^2}{2m_{X}}\frac{\partial^2}{\partial x_k^2}& -J\\ \Omega& -J& -\frac{\hbar^2}{2m_{X} }\frac{\partial^2}{\partial x_k^2} \end{pmatrix},
     \label{Hamiltonian2dsmall}
\end{align}
\end{widetext}
where the masses of the excitons on the two wires are taken to be equal. Here, $\Delta = \epsilon_{X,1}-J-\epsilon_{C,1}$ where $\epsilon_{X,1}$ and $\epsilon_{C,1}$ are the lowest excitonic- and photonic single particle states in each wire in the presence of the confinement potential, and the tunneling amplitude gives a further red-shift $-J$.  

\begin{figure}[htb]
    \centering
    \includegraphics[width=0.95\columnwidth]{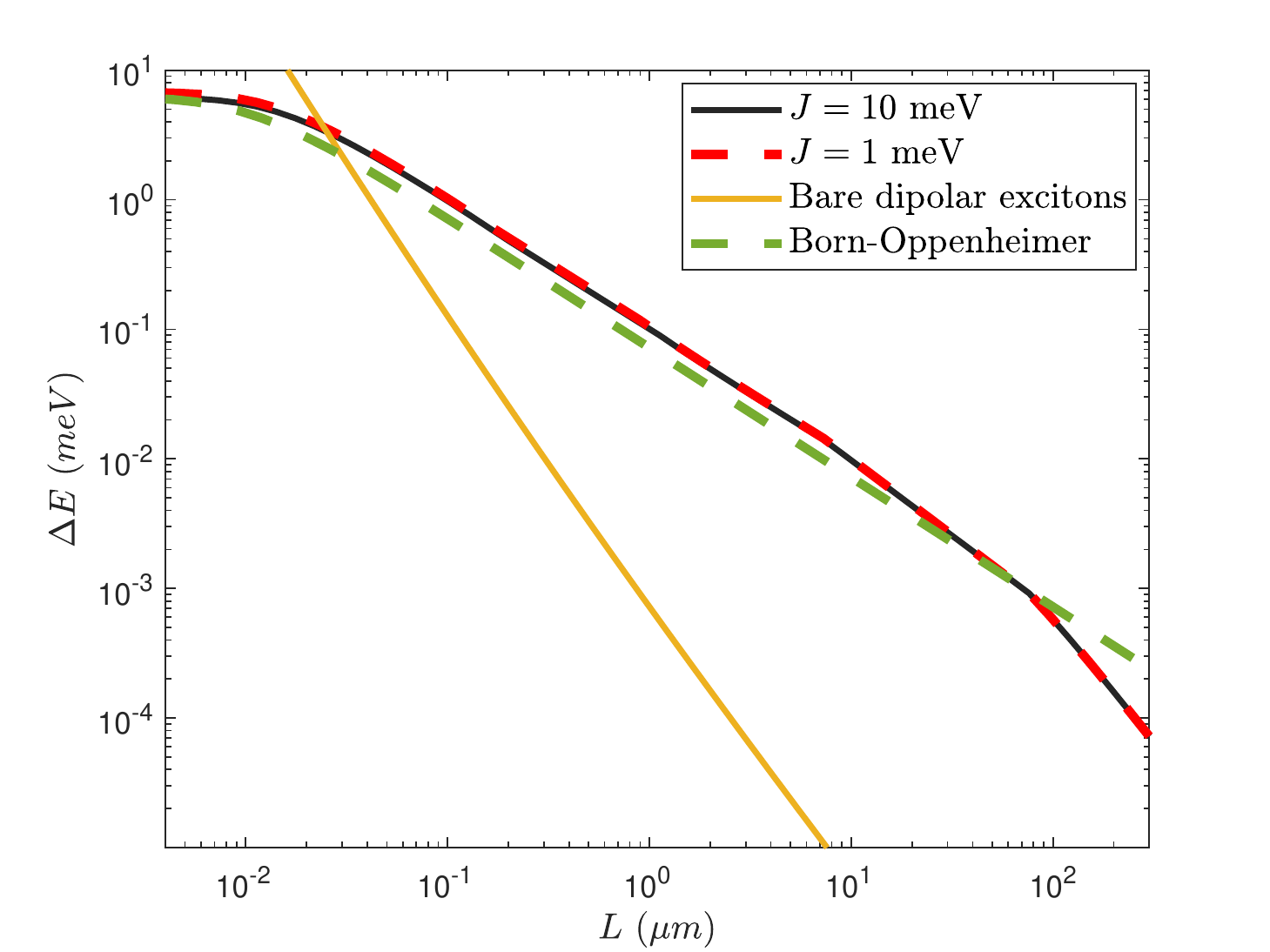}
    \caption{Numerical calculation of the interaction energy in a coupled wire model. The solid black and blue curves correspond to different values of tunneling strength $J$. The dashed green line is the prediction of a generalized perturbative BO theory.  The Rabi coupling is set here to $\Omega=10$ meV. Other parameters as in the figures of the main text. }
    \label{pseudo2d}
\end{figure}

In Fig.~\ref{pseudo2d} we show the numerical results for the interaction energy as a function of the wire length $L$ for two values of the tunneling amplitude $J$ and a fixed lateral spacing of $\Delta z = 5 $ nm. It is seen that for large $L$ the interaction energy is approximately equal in the two cases. As $L$ is decreased, one sees a slight dependence on $J$, a smaller $J$ yielding a slightly larger interaction energy. This is explained by the fact that at small values of $J$, the system does not decouple completely from the other single particle state.
This slight dependence on $J$ is more pronounced when $\Delta z$ is increased. It shows the importance of choosing a large enough $J$ to obtain a realistic quasi-1D model. The interaction energy is again $J$-independent in the quantum dot limit where light is decoupled completely from the states that feature dipolar repulsion. 

Most interestingly, the BO prediction (green dashed line) accurately captures  the essential physics for intermediate $L$-values and the quantum dot limit, as for the strict 1D case.

\section{Model with coupled direct and indirect excitons}
\label{app:DXIX}



In order to validate the BO approach for the model involving coupled direct and indirect excitons, in this Appendix we report full numerical calculations for the interaction energy for this system in a one-dimensional configuration. 

\begin{widetext}
The system can be described by the Hamiltonian 
\begin{align}
    \hat{H} & = \sum_{k=1,2} \int d{x_k} \!\!\! \sum_{i,j=\{C,X,IX\}} \!\!\! \hat{\Psi}_i^\dagger ({x_k}) h^{0}_{ij}({x_k}) \hat{\Psi}_j ({x_k}) 
    +\frac{1}{2} \int d{x_1}\,d{x_2}\, \big[ \hat{\Psi}^\dagger_{IX}({x_1})\hat{\Psi}^\dagger_{IX}({x_2}) V_{\text{dd}}(|{x_1-x_2}|) \hat{\Psi}_{IX}({x_1})\hat{\Psi}_{IX}({x_2})\big],
    \label{Hamiltoniancix}
\end{align}
where IX (X) stand for indirect (direct) exciton. Here $h^0_{ij}$ are elements of the matrix 
\begin{align}
    h^0({x_k}) = \begin{pmatrix}
     -\frac{\hbar^2}{2m_C}\frac{\partial^2}{\partial x_k^2}+\Delta+\delta_C& \Omega& 0 \\
     \Omega & -\frac{\hbar^2}{2m_{X}}\frac{\partial^2}{\partial x_k^2}& -J\\ 0& -J& -\frac{\hbar^2}{2m_{IX} }\frac{\partial^2}{\partial x_k^2} \end{pmatrix},
     \label{Hamiltoniancixsmall}
\end{align}
and equal masses are taken for the two types of excitons $m_X=m_{IX}$. The detunings $\Delta$ and $\delta_C$ follow the same definitions as in the main text. Interactions are assumed to only occur between pairs of indirect excitons.
\end{widetext}

\begin{figure}[htb]
    \centering
    \includegraphics[width=0.95\columnwidth]{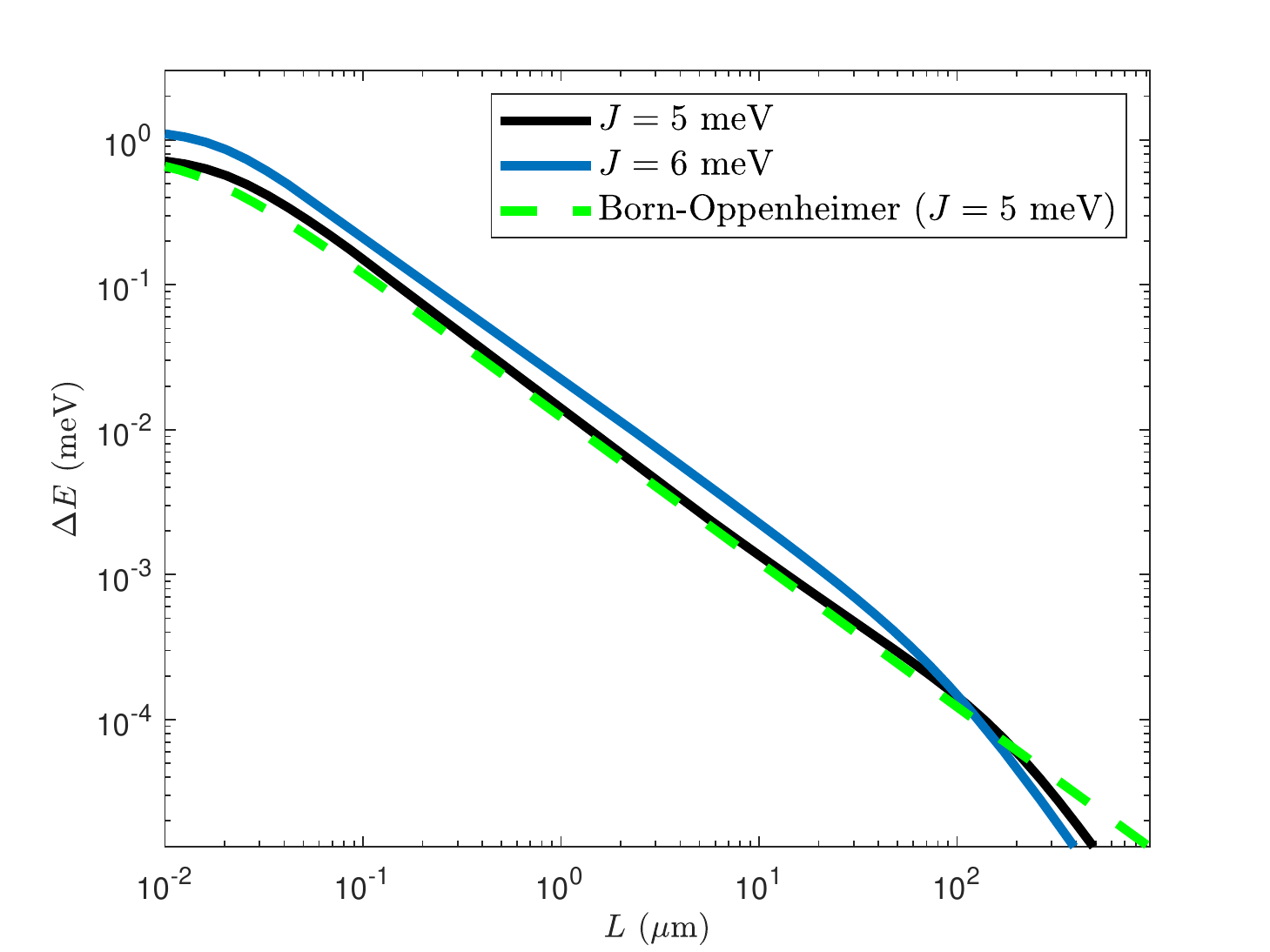}
    \caption{Numerical calculation for the interaction energy for a model with both direct- and indirect excitons. The solid black and blue curves correspond to different values of the inter-exciton coupling strength $J$. The green dashed curve is the prediction of a generalized perturbative BO theory. Other parameters as in the figures of the main text. 
    }
    \label{CIXX}
\end{figure}

In Fig.~\ref{CIXX} we show the numerical predictions for the interaction energy as a function of $L$ for $\Omega = 5 $ meV and two values of the inter-exciton coupling $J=5$ and 6 meV. 
For large $L$, the system is in the Tonks-Girardeau (TG) regime as indicated by the $1/L^2$ scaling. To explain why the prefactor is lower for a higher $J$, one can notice that, at the single-particle level, a larger $J$ gives a larger excitonic component in the lower polariton which in turn increases the effective mass. 
As the confinement length is decreased, the system enters the intermediate regime with a $1/L$ scaling. In this regime, increasing $J$ results in a larger interaction energy for a fixed Rabi coupling. In the small $L$ limit, the two-particle state consisting of two indirect excitons gets decoupled, and Jaynes-Cummings physics is realized. As a result, the interaction energy is given as $2 f_1(\Omega,J)-f_2(\Omega,J)$, where $f_{1,2}$ are the generalized Rabi couplings with subscript indicating either single- or two-particle system, and is a increasing function of $J$.

Most interestingly, also in this case the perturbative BO prediction (green dashed line) accurately captures  the essential physics for intermediate $L$-values, as it happened for the simpler model with a single exciton species.

\end{document}